\documentclass[usenatbib]{emulateapj}
\usepackage{epsf}
\usepackage{psfig}
\usepackage{graphicx}

\begin{document}

\def\mpch {$h^{-1}$ Mpc} 
\def\kpch {$h^{-1}$ kpc}
\def\kms {km s$^{-1}$} 
\def\lcdm {$\Lambda$CDM } 
\def\etal {et al.}

\def\kms {km s$^{-1}$} 
\def\eg{{\it e.g.}}
\def\etal{{\it et al.}}
\def\lir{\hbox{$L_{IR}$}}
\def\lsun{\hbox{$L_{\odot}$}}
\def\msun{\hbox{$M_{\odot}$}}
\def\msunyr{\hbox{$M_{\odot}~{\rm yr}^{-1}$}}
\def\deg{\hbox{$^{\circ}$}}
\def\ujy{\hbox{$\mu$Jy}}
\def\um{\hbox{$\mu$m}}

\def\mgtwol{\ion{Mg}{2} 2796, 2803 \AA}
\def\mgonel{\ion{Mg}{1} 2852 \AA}
\def\mgtwo{\ion{Mg}{2}}
\def\mgone{\ion{Mg}{1}}
\def\fetwo{\ion{Fe}{2}}
\def\nethree{[\ion{Ne}{3}]}

\title{Outflowing Galactic Winds in Post-starburst and AGN Host Galaxies 
at $0.2<z<0.8$}

\author{
Alison L. Coil\altaffilmark{1,2}, 
Benjamin J. Weiner\altaffilmark{3},
Daniel E. Holz\altaffilmark{4},
Michael C. Cooper\altaffilmark{5,6},
Renbin Yan\altaffilmark{7},
James Aird\altaffilmark{1}}

\altaffiltext{1}{Department of Physics, Center for Astrophysics and Space Sciences, University of California at San Diego, 9500 Gilman Dr., La Jolla, San Diego, CA 92093}
\altaffiltext{2}{Alfred P. Sloan Foundation Fellow}
\altaffiltext{3}{Steward Observatory, The University of Arizona, 933 N. Cherry Ave., Tucson, AZ 85721}
\altaffiltext{4}{Theoretical Division, Los Alamos National Laboratory, Los Alamos, NM 87545}
\altaffiltext{5}{Department of Physics and Cosmology, University of California at Irvine, 4129 Frederick Reines Hall, Irvine, CA 92697}
\altaffiltext{6}{Hubble Fellow, Center for Galaxy Evolution fellow}
\altaffiltext{7}{Center for Cosmology and Particle Physics, Department of Physics, New York University, 4 Washington Place, New York, NY 10003}

\begin{abstract}

We present Keck/LRIS-B spectra for a sample of ten AEGIS X-ray AGN
host galaxies and thirteen post-starburst galaxies from SDSS and DEEP2
at $0.2<z<0.8$ in order to investigate the presence, properties, and
influence of outflowing galactic winds at intermediate redshifts.  We
focus on galaxies that either host a low-luminosity AGN or have
recently had their star formation quenched to test whether these
galaxies have winds of sufficient velocity to potentially clear gas
from the galaxy.  We find, using absorption features of \fetwo,
\mgtwo, and \mgone, that six of the ten (60\%) X-ray AGN host galaxies 
and four of the thirteen (31\%) post-starburst galaxies have outflowing 
galactic winds, with 
typical velocities of $\sim200$ \kms.  We additionally find that most
of the galaxies in our sample show line emission, possibly from the
wind, in either \fetwo* or \mgtwo.  A total of 100\% of our X-ray AGN
host sample (including four red sequence galaxies) and 77\% of our
post-starburst sample has either blueshifted absorption or line
emission.  Several K+A galaxies have small amounts of 
cool gas absorption at the systemic velocity, indicating that not all
of the cool gas has been expelled.
We conclude that while outflowing galactic
winds are common in both X-ray low-luminosity AGN host
galaxies and post-starburst galaxies at intermediate redshifts, the
winds are likely driven by supernovae (as opposed to AGN) and do not
appear to have sufficiently high velocities to quench star formation
in these galaxies.

\end{abstract}


\section{Introduction} \label{sec:introduction}

Galaxy redshift surveys have revealed that the optical restframe color
distribution of galaxies is bimodal in both the local and distant
Universe, beyond $z=2$ \citep[e.g.,][]{Strateva01, Blanton03, Faber07, 
Kriek08}.  Galaxies predominantly lie either along the ``red
sequence'', which is composed of mostly quiescent, early-type
galaxies, or in the ``blue cloud'' of star-forming, late-type
galaxies.  From $z=1$ to today, the red sequence has roughly doubled
in mass \citep{Bell04, Bundy06, Brown07, Faber07}, presumably as galaxies have
moved from the blue cloud to the red sequence as a result of ending
star formation and passively evolving.  It is not yet understood what
causes galaxies to stop forming stars, nor why their star formation remains
quenched, i.e. why they do not continue to accrete additional gas 
and resume star formation.

One popular proposed mechanism for quenching star formation is gas
blowout by a starburst and/or AGN, resulting from a merger event
\citep[e.g.][]{Sanders88, Hopkins05}. In this picture, the starburst
and/or AGN is activated by gas inflow during the merger, and this 
subsequently drives an outflowing wind that fully clears gas from the
galaxy, leading to the creation of a quiescent elliptical.  A
remaining low-luminosity AGN may provide a second mode of feedback and 
deter further star formation by
heating any accreted gas, keeping the galaxy red 
\citep[e.g.][]{Binney01,Ostriker05,Croton06,Bower06,Best06,Cattaneo07}.  
While this proposed
scenario is attractive in that it solves many outstanding questions
relating to the formation of elliptical galaxies, 
it has yet to be confirmed observationally.
Understanding what quenches star
formation is critical to understanding how the red sequence is built
up over time.

Post-starburst galaxies are potentially an ideal population with which
to study the processes that quench star formation.  Post-starburst
galaxies are passing through a brief phase of galaxy evolution: they
are seen just after abruptly ceasing star formation.  
These galaxies are also known as ``E+A'' or ``K+A''
galaxies from their spectral types, which show a mixture of older
stars plus younger A stars \citep[e.g.][]{Dressler83, Zabludoff96,
  Quintero04}. Their spectra show strong Balmer absorption from A
stars but no nebular emission, which indicates that they are not
currently forming stars. However, the presence of A stars, which have
lifetimes $\lesssim 1$ Gyr, show that star formation has only recently
stopped. As these galaxies age, the A stars will disappear and their
spectra will look like that of an early-type galaxy.  Their
morphologies and metallicities also imply that they are the immediate
progenitors of early-type, red, spheroidal galaxies \citep{Yang08,
  Goto07}.  Post-starburst galaxies are therefore likely caught in
the act of moving from the blue cloud to the red sequence.

Observationally, there are a variety of ways in which to find evidence 
for the star formation quenching mechanism in these galaxies.  One method 
is to search for outflowing galactic winds using either blueshifted 
interstellar medium (ISM) 
absorption lines (see Veilleux et al. 2005 for a recent review)
\nocite{Veilleux05} or emission lines \citep{Rubin11}.
Detecting winds through blueshifted absorption has the advantage of 
providing kinematic information on the wind speed along the line of 
sight.
The AGN gas blowout picture discussed above recently gained support
with the discovery by \citet{Tremonti07} of extreme velocity
outflows in very luminous post-starburst galaxies at $z\sim0.6$
selected from the Sloan Digital Sky Survey \citep[SDSS, ][]{York00}.
These galaxies are extremely rare, have low space densities, and are
bright, blue, and massive ($\geq 10^{11}$ \msun).  The outflowing
winds are detected in blueshifted \mgtwol \ interstellar absorption
lines in ten of the fourteen galaxies observed.  The outflow
velocities seen in these galaxies are $\sim 1000$ \kms\ or more, which
the authors interpret as galaxy-scale AGN winds launched at the epoch
of cessation of the starburst.  The motivation for AGN activity as the driver
of these winds is the high velocity of the outflows; the velocities
seen are intermediate between winds in local starburst galaxies and
winds in broad absorption line quasars.


\begin{table*}[]
\tablewidth{0pt}
\begin{center}
\label{tab:properties}
\small
\footnotesize
\caption{\small Properties of Galaxies in Sample}
\begin{tabular}{cccccccccc}
\cr
\colrule
\colrule
\vspace{-3 mm} \cr
\vspace{-3 mm} \cr
Object  & RA      & Dec     & redshift & $B$ & $R$ & $I$ & $U-B$ & $M_B$ &   \cr
        & (J2000) & (J2000) &          &     &     &     &         &       &  \cr
\vspace{-3 mm} \cr
\colrule 
\colrule 
\vspace{-3 mm} \cr
\vspace{-3 mm} \cr
\multicolumn{10}{c}{X-ray AGN Host Galaxies} \cr
\vspace{-3 mm} \cr
\colrule 
11046507 \ & 14:16:04.970 & +52:18:27.971 & \ 0.4514 & \ 21.3 & 20.6 & 20.5 & 0.31 & -20.12 &   \cr
12016790 \ & 14:17:32.498 & +52:34:41.209 & \ 0.4645 & \ 20.9 & 19.9 & 19.6 & 0.56 & -20.91 &   \cr
13004312 \ & 14:19:10.463 & +52:48:31.240 & \ 0.3458 & \ 21.4 & 19.0 & 18.3 & 1.27 & -20.73 &   \cr
13025528 \ & 14:20:38.780 & +52:57:27.729 & \ 0.2004 & \ 19.9 & 18.1 & 17.6 & 1.10 & -20.12 &   \cr
13041622 \ & 14:21:12.279 & +53:06:22.173 & \ 0.2017 & \ 20.0 & 18.7 & 18.2 & 0.77 & -19.77 &   \cr
13043681 \ & 14:20:00.785 & +53:06:43.926 & \ 0.2014 & \ 20.0 & 18.0 & 17.4 & 1.25 & -20.20 &   \cr
13051909 \ & 14:19:49.857 & +53:08:05.335 & \ 0.2334 & \ 20.3 & 18.4 & 17.8 & 1.11 & -20.27 &   \cr
13063597 \ & 14:22:17.219 & +53:14:27.123 & \ 0.3029 & \ 19.9 & 18.4 & 17.8 & 0.84 & -21.15 &   \cr
13063920 \ & 14:21:41.788 & +53:15:06.358 & \ 0.5589 & \ 21.7 & 20.7 & 20.3 & 0.59 & -20.65 &   \cr
22029058 \ & 16:51:10.627 & +34:54:53.390 & \ 0.3407 & \ 20.3 & 19.3 & 18.9 & 0.49 & -20.61 &   \cr
\vspace{-3 mm} \cr
\colrule
\colrule
\vspace{-3 mm} \cr
\vspace{-3 mm} \cr
\multicolumn{10}{c}{DEEP2 K+A Galaxies} \cr
\vspace{-3 mm} \cr
\colrule 
\vspace{-3 mm} \cr
31046744 \ & 23:27:12.737 & +00:17:16.822 & 0.8542 & 23.9 & 22.4 & 21.2 & 1.03 & -20.99 \cr
32003698 \ & 23:30:14.824 & +00:01:29.920 & 0.7960 & 23.8 & 21.3 & 20.1 & 1.10 & -21.82 \cr
32008909 \ & 23:30:29.216 & +00:03:54.400 & 0.7894 & 24.2 & 22.1 & 21.1 & 1.01 & -20.85 \cr
41057700 \ & 02:27:28.413 & +00:48:04.204 & 0.7200 & 22.9 & 21.5 & 20.8 & 0.76 & -20.85 \cr
42020386 \ & 02:31:02.398 & +00:33:12.964 & 0.7757 & 23.1 & 21.6 & 20.8 & 0.81 & -21.10 \cr
42021012 \ & 02:30:54.429 & +00:32:50.010 & 0.7482 & 22.7 & 21.6 & 21.0 & 0.60 & -20.82 \cr
43030800 \ & 02:32:03.446 & +00:39:47.795 & 0.8445 & 23.7 & 22.8 & 21.7 & 1.02 & -20.50 \cr
\vspace{-3 mm} \cr
\colrule
\colrule
\end{tabular}
\begin{tabular}{ccccccccccc}
\vspace{-3 mm} \cr
\vspace{-3 mm} \cr
Object  & RA      & Dec     & redshift & $u$ & $g$ & $r$ & $i$ & $z$ & $U-B$ & $M_B$  \cr
        & (J2000) & (J2000) & & & & & &  \cr
\vspace{-3 mm} \cr
\colrule 
\colrule 
\vspace{-3 mm} \cr
\vspace{-3 mm} \cr
\multicolumn{11}{c}{SDSS K+A Galaxies} \cr
\vspace{-3 mm} \cr
\colrule 
\vspace{-3 mm} \cr
J022743.21$-$001523.0  & 02:27:43.211 & -00:15:23.077 & 0.2191 & 19.0 & 17.7 & 17.0 & 16.7 & 16.5 & 0.91 & -21.92 \cr
J210025.41+011319.2  & 21:00:25.415 & +01:13:19.267 & 0.2079 & 20.5 & 18.9 & 17.9 & 17.6 & 17.4 & 1.13 & -21.00 \cr
J212043.73+114345.4  & 21:20:32.739 & +11:43:45.430 & 0.2435 & 20.4 & 18.6 & 17.6 & 17.2 & 17.0 & 1.25 & -21.80 \cr
J215518.35$-$071010.6  & 21:55:18.354 & -07:10:10.697 & 0.2767 & 21.0 & 19.0 & 17.8 & 17.4 & 17.2 & 1.14 & -21.62 \cr
J224603.64$-$000918.8  & 22:46:03.640 & -00:09:18.804 & 0.2052 & 20.1 & 18.4 & 17.3 & 17.0 & 16.7 & 1.13 & -21.51 \cr
J225656.77+130402.6  & 22:56:56.770 & +13:04:02.609 & 0.2089 & 19.8 & 18.1 & 17.3 & 17.0 & 16.7 & 1.10 & -21.61 \cr
\vspace{-3 mm} \cr
\colrule
\colrule
\end{tabular}
\end{center}
\end{table*}

Galactic-scale outflows can also be driven by supernovae (SNe) resulting
from high star formation rates.
At low redshift, outflows of $\sim$-100 to -600 \kms\
are observed in infrared-luminous galaxies (SFR $\gtrsim$ 20-50$~\msunyr$)
in the Na I D 5890, 5896 \AA\ doublet \citep{Heckman00,Rupke05, Martin05}.
At $z=1.4$, where much of the blue galaxy population is forming
stars with high SFR, $\sim20-100~\msunyr$,  \cite{Weiner09} found 
outflowing galactic winds detected as blueshifted absorption 
in the \mgtwol \ doublet 
by stacking spectra of star-forming galaxies in the 
DEEP2 redshift survey, with typical velocities of $\sim$-300 \kms.  
Similar results were found by \citet{Rubin10a} in
star-forming galaxies at $z\sim1$.
At an even higher redshift of $z\sim2-3$, Lyman-break galaxies also exhibit
strong winds with speeds of $\sim$-200 -- -500 \kms\ \citep{Shapley03, Steidel10}. 

The winds seen in these
galaxies are likely SNe-driven, as opposed to AGN-driven, as most of these
galaxies do not have detected AGN, and the fraction of star-forming 
galaxies with winds is high ($\gtrsim50\%$).
These SNe-driven winds raise the questions of whether AGN-driven outflows
are necessary and/or sufficient to cause the quenching of star formation. 
High-luminosity AGN are expected to drive faster winds than star-formation 
powered objects \citep{Thacker06} and may be more effective at clearing the 
ISM of the host galaxy. 
The wind velocity can therefore potentially 
distinguish the cause of the galactic wind: starburst or AGN. 

In the \citet{Thacker06} model, which assumes that AGN drive outflows 
with an energy output equal to 5\% of the bolometric luminosity, typical 
outflow velocities are $\sim$1000 \kms \ and depend on the bulge (and 
therefore AGN) mass.  For example, galaxies 
with bulge masses $>10^{12}$ \msun \ (with corresponding AGN masses 
$>10^9$ \msun) drive outflows with velocities $>$1000 \kms, 
while galaxies with bulge masses $>3 \ 10^{11}$ \msun \ 
have typical velocities of 1000 \kms \ but can have velocities of $\sim$
700-800 \kms.  They 
further show that winds in excess of 1000 \kms \ are difficult to produce 
with SNe-driven winds and would require extremely efficient star formation, 
with a ratio of gas to stars of $<$25\%.  It is worth noting that the 
$\sim$1000 \kms \ winds predicted in this model are for luminous AGN.

The \cite{Tremonti07} sample included only a very specific and rare 
type of galaxy with low space density; 
it is not clear whether most red galaxies have followed 
this path to the red sequence.  Their result therefore may 
not be applicable to the bulk of the red sequence.
In this paper we present Keck/LRIS-B restframe UV spectra of a sample of 
ten X-ray selected narrow-line 
AGN host galaxies from the AEGIS survey and thirteen post-starburst galaxies at 
intermediate redshifts ($0.2<z<0.8$) from the SDSS and DEEP2 surveys.  
We observe galaxies with lower-luminosity AGN, 
both in star-forming and quiescent galaxies, as well as more 
common post-starburst galaxies, in order to measure the 
frequency and velocity of outflowing winds 
via blueshifted absorption in several UV ISM lines 
(\ion{Fe}{2} 2343, 2374, 2382, 2586, 2599 \AA, \mgtwol, and \mgonel).  
Our aim is to determine
(1) whether winds are 
frequently found in galaxies that host low-level (non-quasar) AGN, 
(2) whether winds are commonly associated with the truncation of star 
formation at intermediate redshift, 
(3) whether wind presence correlates with star formation activity, 
(4) whether the velocities 
indicate a star formation- or AGN-driven wind, and 
(5) whether the velocities
are high enough to potentially clear the ISM of the galaxy and halt star 
formation.
The galaxies in our study should represent typical paths to the
red sequence; therefore our results are potentially relevant for the 
formation of much of the quiescent population.

Until very recently, the 
\fetwo, \mgtwo, and \mgone \ 
lines have been little studied due to their
location in the near-UV. Most $z<1$ searches for absorption have used
the \ion{Na}{1} D doublet, however it is harder to interpret as 
(1) the velocity separation is low (304 \kms, as compared to the 
770 \kms \ separation of \mgtwo) such that it is often blended,
(2) it has a lower ionization potential (5.1 eV for \ion{Na}{1} D, 
compared to 15.0 eV for \mgtwo) and therefore traces dense gas while
\mgtwo \ more faithfully traces cool, photoionized gas, and 
(3) the stellar contribution to \ion{Na}{1} D is 
quite significant (~80\% in Chen et al.) and requires extensive 
modeling to separate
the outflow and stellar components \citep[e.g.][] {Rupke05, Chen10}.  
We choose here to focus on the 
\fetwo, \mgtwo, and \mgone \ lines.

The outline of the paper is as follows: \S 2 describes our galaxy samples,
LRIS spectra, and data reduction.  In \S 3 we
discuss the methods used in this paper to obtain absorption line fits
to the UV lines of interest, including removing absorption due to stars or
gas in the galaxy.  We present results on winds detected in absorption in 
\S 4 and winds detected in emission in \S 5.  We discuss our 
results and conclude
in \S 6. In this paper all magnitudes are AB.  Restframe 
magnitudes are $M-5 ~{\rm log}(h)$ with $h=1$. 

\begin{table*}[]
\tablewidth{0pt}
\begin{center}
\label{tab:xray}
\small
\footnotesize
\caption{\small Properties of X-ray AGN Host Galaxies}
\begin{tabular}{ccccccc}
\cr
\colrule
\colrule
\vspace{-3 mm} \cr
\vspace{-3 mm} \cr
Object & log SFR\tablenotemark{1} & log stellar  &  log $L_{\rm X}$   &  log $L_{\rm X}$     & log $L_{\rm X}$        &  HR   \cr
       &                         & mass         &  est. from SF     &  from $f_{(0.5-2 keV)}$ & from $f_{(2-10 keV)}$ &  \cr
       & (\msun \ $yr^{-1}$)       & (\msun)      &  ($erg \ s^{-1}$) &  ($erg \ s^{-1}$)    & ($erg \ s^{-1}$)      &        \cr
\vspace{-3 mm} \cr
\colrule 
\colrule 
\vspace{-3 mm} \cr
\vspace{-3 mm} \cr
\multicolumn{7}{c}{DEEP2 X-ray AGN Host Galaxies} \cr
\vspace{-3 mm} \cr
\colrule 
11046507 \ & 1.10 & 9.82  & 40.50 & 41.60 $^{+0.21}_{-0.24}$ & $<42.15$               & $-0.74 ^{+0.06}_{-0.26}$ \cr
12016790 \ & 1.41 & 10.97 & 41.02 & 41.93 $^{+0.16}_{-0.18}$ & $<42.44$               & $-0.59 ^{+0.11}_{-0.41}$ \cr
13004312 \ & ...  & 11.15 & 40.10 & 41.41 $^{+0.19}_{-0.21}$ & $<41.83$               & $-0.68 ^{+0.08}_{-0.32}$ \cr
13025528 \ & 0.40 & 11.10 & 40.55 & 40.69 $^{+0.25}_{-0.30}$ & $41.63 ^{+0.18}_{-0.21}$ & $+0.26 ^{+0.31}_{-0.24}$ \cr
13041622 \ & 1.09 & 10.74 & 40.73 & 41.47 $^{+0.08}_{-0.08}$ & $41.66 ^{+0.18}_{-0.26}$ & $-0.48 ^{+0.20}_{-0.17}$ \cr
13043681 \ & ...  & 10.52 & 39.47 & 41.39 $^{+0.11}_{-0.12}$ & $<41.58$               & $-0.72 ^{+0.10}_{-0.25}$ \cr
13051909 \ & 1.08 & 11.40 & 41.01 & 42.11 $^{+0.08}_{-0.08}$ & $43.55 ^{+0.03}_{-0.03}$ & $+0.71 ^{+0.05}_{-0.04}$ \cr
13063597 \ & 1.21 & 10.53 & 40.75 & 41.50 $^{+0.19}_{-0.23}$ & $<42.13$               & $-0.24 ^{+0.35}_{-0.34}$ \cr
13063920 \ & 1.62 & 11.18 & 41.23 & 41.79 $^{+0.26}_{-0.29}$ & $<42.38$               & $-0.66 ^{+0.08}_{-0.34}$ \cr
\vspace{-3 mm} \cr
\colrule
\colrule
\end{tabular}
\tablenotetext{1}{The columns list the star formation rate derived from MIPS 24\micron \ flux, the stellar mass derived following \citet{Lin07}, the X-ray luminosity estimated from star formation alone, the X-ray luminosity at 2-10 keV derived from the observed X-ray flux at either 0.5-2 keV or 2-10 keV, and the hardness ratio (see text for details).}
\end{center}
\end{table*}

\vspace{2cm}

\section{Data} \label{sec:data}

\subsection{Galaxy Samples}

For this study we selected two samples of intermediate redshift
galaxies to study their outflowing wind properties: 1) X-ray AGN host
galaxies, and 2) post-starburst (``K+A'') galaxies.
Properties of the galaxies in our samples are given in Table~\ref{tab:properties}. The redshifts listed are derived from our LRIS 
data.  
K-corrections, absolute $M_B$ magnitudes and restframe $(U-B)$ 
colors for DEEP2 objects have been derived as described in \cite{Willmer06} 
and for SDSS objects were derived using the kcorrect package 
\citep{Blanton07}. We 
do not include luminosity evolution in the K-corrections for $M_B$. 

X-ray AGN host galaxies were identified using deep 200\,ks {\it
  Chandra} data \citep{Laird09} obtained as part of the AEGIS survey
\citep{Davis07}.  The {\it Chandra} data reduction, source detection
and flux estimation are described in detail by \citet{Laird09}.
Initial optical spectroscopy of these sources was provided by the
DEEP2 survey \citep{Davis00,Davis03} using DEIMOS on Keck and by
\citet{Coil09} using Hectospec on the MMT.  Here we present follow-up
LRIS-B spectroscopy of nine X-ray AGN host galaxies with redshifts
$0.20<z<0.56$ and $B<21.7$.  We selected sources that are bright
in the $B$ band, such that we would be able to detect 
continuum levels in their spectra with relatively short exposures 
(under an hour), and we did not target objects with broad \mgtwo \ emission 
in their DEEP2 or MMT spectra, as strong broad emission may have affected
our ability to detect blueshifted absorption features. 

\begin{figure*}[tbp]
\plottwo{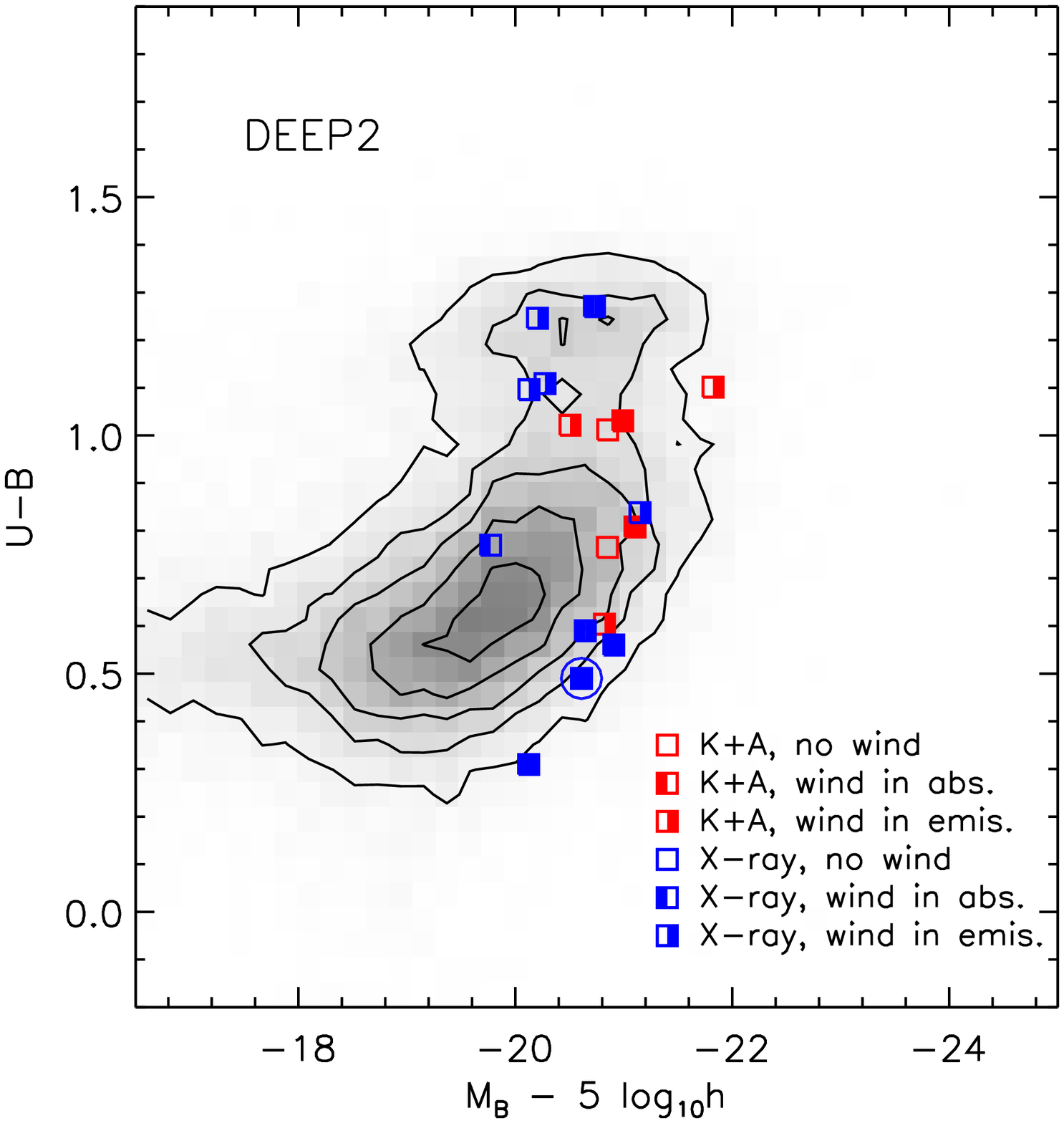}{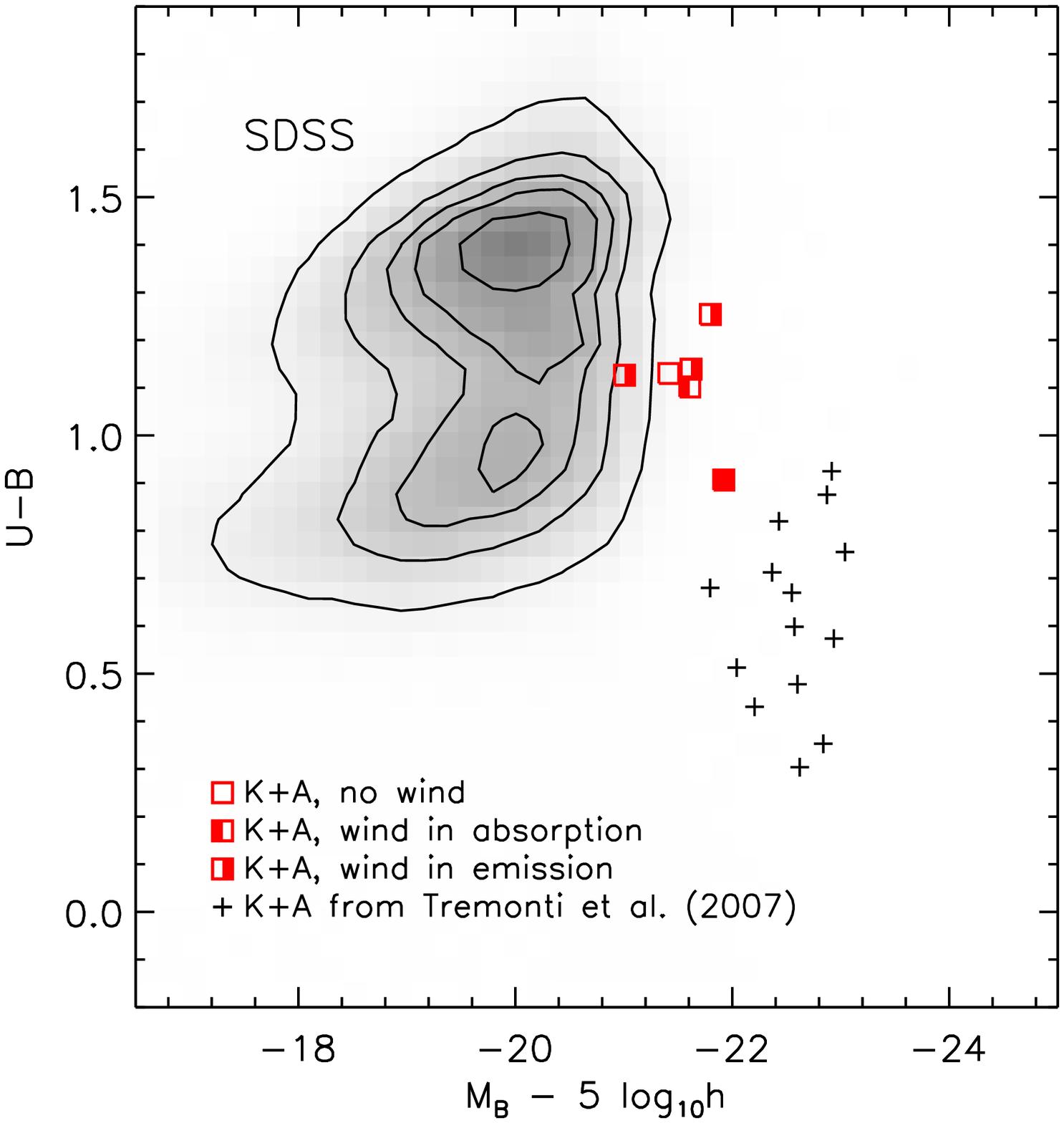}
\caption{\label{fig:colmag}
\small Left: 
Restframe $U-B$ color versus $M_B$ magnitude of galaxies in the DEEP2 
parent galaxy sample 
with $0.2<z<1.2$ (grey shaded contours) compared to galaxies in our DEEP2  
K+A sample (red squares) and AEGIS X-ray AGN host galaxies (blue squares).  
Filled red and blue squares are those galaxies in our sample
with winds detected either by blueshifted \mgtwo \ absorption (left side of
square filled) or by \mgtwo \ or \fetwo* emission (right side of square
filled).  Objects with winds detected both in absorption and emission
are shown with completely filled squares.  The LIRG is indicated with a blue
circle.
Right: Restframe color-magnitude diagram of galaxies in the SDSS main 
galaxy sample with 
$0.02<z<0.20$ (grey shaded contours) compared to galaxies in our SDSS 
K+A sample with $0.20<z<0.28$ (red squares).  
Symbols are filled according to how a wind is detected, as in the left panel.
K+A galaxies from \citet{Tremonti07} are shown as black crosses.
}
\end{figure*}

Table~\ref{tab:xray} lists additional properties of these X-ray sources.  Shown are 
the SFRs derived from MIPS 24\micron \ observations of the AEGIS field 
\citep{Davis07}, following the prescription of \citet{Rieke09}.  These
SFR estimates are upper limits as there may be some contribution to the
 IR luminosity from an AGN.  All but one of the sources are detected in
the MIPS data; for the non-detection the SFR should be extremely low. 
We list stellar masses derived following
the prescription of \citet{Lin07} for DEEP2 galaxies.  Using the stellar
mass and SFR, we estimate the X-ray luminosity expected from X-ray 
binaries, following \citet{Lehmer10}.  
We list the X-ray luminosity (at 2-10 keV rest-frame) 
that is estimated from the observed
{\it Chandra} flux in either the soft or hard bands, 
assuming $\Gamma=$1.9, corrected for Galactic absorption but assuming
no intrinsic absorption.  If the source was not detected in the hard band,
an upper limit is given based on the 99\% upper limit on the observed flux.
We also list the hardness ratio, estimated using
Bayesian techniques as described in \citet{Laird09}, following \citet{Park06}.
Two of our nine X-ray sources have positive hardness ratios, indicative of
moderately obscured AGN.  One of those also has log 
$L_{\rm X}>42 \ \mbox{erg} \ \mbox{s}^{-1}$, 
above the traditional cutoff often used to define AGN.
The other seven sources have 
log $L_{\rm X}\sim 41-42 \ \mbox{erg} \ \mbox{s}^{-1}$. The log $L_{\rm X}$ values estimated
from the observed X-ray fluxes are typically $\sim$10 times higher
than the upper limits on the 
contribution to log $L_{\rm X}$ estimated from their SFRs.  We 
therefore conclude that these galaxies have low luminosity, relatively 
unobscured AGN.
 We note that while these are relatively low luminosity AGN, they should 
be fairly representative of the broader population of low luminosity 
AGN at these redshifts.

During our observations of the X-ray AGN host galaxies we 
also observed one luminous infrared galaxy (LIRG) in the DEEP2 16hr 
field that was selected as a {\it Spitzer}/MIPS 24 $\micron$
source with high $L_{IR}$.
The DEEP2 16hr field was observed with MIPS by the MIPS GTO team
in August 2007.  The observations were reduced with the MIPS
GTO pipeline and fluxes cataloged using DAOphot PSF fitting.
For 24 $\micron$ sources without DEEP2 redshifts, follow-up spectra
were obtained with MMT/Hectospec.  A full catalog will be
published in a forthcoming paper.
We computed the total IR luminosity from the 24 $\micron$ flux, K-correcting
with a template from \citet{Dale02}.
This source has a total IR luminosity of log $L_{IR}>11.7$ $L_\Sun$.
We selected a source at 
intermediate redshift ($z=0.34$) that was bright at observed blue
wavelengths ($B=20.3$) and did not have broad optical emission lines.
It is not detected in relatively shallow {\it XMM} and {\it Chandra} 
data of this field and has an upper limit of log $L_{\rm X}<44 \ \mbox{erg} \ \mbox{s}^{-1}$.
While we do not know if this source
contains an AGN, we include it in our `X-ray AGN host galaxy'
sample for the purposes of this paper, as it was selected to be IR-bright 
(and is not a post-starburst galaxy). 

Post-starburst (K+A) galaxies were identified in both the DEEP2 and SDSS 
surveys from the optical spectroscopy.  We decomposed spectra in each 
survey into two components: a young and an old stellar population.
The templates were constructed using \citet{Bruzual03} 
models with a Salpeter IMF and solar metallicity. The young population 
template is a model taken 0.3 Gyr after a starburst, which had a constant star
formation rate and a duration of 0.1 Gyr. The old stellar population template is
a 7 Gyr old simple stellar population. Using a linear decomposition, 
we measure the fractional contribution of the young component 
between 4450--4550 \AA \ in the restframe, denoted as the 
A-star fraction ($f_A$). Post-starburst galaxies were identified
as having a strong young stellar component ($f_A > 0.25$) and a weak or 
non-detectable H$\beta$ emission equivalent width (EW; 
Eq. 2 in Yan et al. 2009). \nocite{Yan09}
We restricted the SDSS sample to $z>0.2$ so that the observed \mgtwol \  
doublet falls redward of 3360 \AA, 
for better sensitivity. In the DEEP2 sample we can only identify post-starbursts
to $z=0.9$, where the spectral coverage includes the Balmer lines.
We observed a total of seven K+A galaxies in
DEEP2 with redshifts $0.77<z<0.85$ and $B<24.2$ and six K+A galaxies 
in SDSS with redshifts $0.21<z<0.28$ and $u<21$.

Restframe color-magnitude diagrams for the DEEP2 ($0.2<z<1.2$) 
and SDSS main galaxy ($0.02<z<0.20$)
samples are shown in Figure~1, with our targets marked.  Our 
sample of X-ray AGN host galaxies (shown as blue squares in the left panel 
of Figure~1)
spans both the blue cloud and red sequence.  We purposely targeted both
blue and red host galaxies in order to study the prevalence of 
winds in both AGN host populations.  Visual inspection of HST/ACS imaging
of AEGIS AGN host galaxies in the blue cloud shows that the blue light 
in these objects is galaxy light and not dominated by a central blue point 
source; therefore our AGN host sample includes both star-forming and 
quiescent galaxies.

Galaxies in our DEEP2 K+A sample (shown as red squares in the left panel of
Figure~1)
lie either at the bright end of the blue cloud or in the ``green valley''
between the blue cloud and red sequence.  This reflects the fact that these
galaxies recently stopped forming stars and are in transition, moving towards
the red sequence.  
Our SDSS K+A sample, with $0.20<z<0.28$, is compared in 
the right panel of Figure~1 
to the main SDSS galaxy sample at lower redshifts, $0.02<z<0.20$.
At $z\sim0.2$ SDSS targeted only very bright galaxies, with $M_r^{0.1}<-21.2$.
Therefore, compared to the main SDSS sample, our SDSS K+A sample is composed 
of brighter galaxies.  All but one of the galaxies in our SDSS K+A sample lie 
 in the green valley; the other K+A galaxy is in the blue cloud.  
For comparison, we also plot the K+A sample of \citet{Tremonti07}, which is 
at $0.5<z<0.7$, in the right panel.  The \citet{Tremonti07} 
sample is brighter and bluer than our lower redshift SDSS 
K+A galaxies.  This likely indicates that most galaxies in 
our SDSS K+A sample are at a slightly later evolutionary stage than the 
galaxies in the \citet{Tremonti07} sample.  

\subsection{Observations}

\begin{table*}[]
\tablewidth{0pt}
\begin{center}
\label{tab:observations}
\footnotesize
\caption{\small LRIS Observational Setup for Each Sample}
\begin{tabular}{lccc}
\cr
\colrule
\colrule
\vspace{-3 mm} \cr
\vspace{-3 mm} \cr
 & X-ray AGN & DEEP2 K+A  & SDSS K+A \cr
\vspace{-3 mm} \cr
\vspace{-3 mm} \cr
\colrule 
\colrule 
\vspace{-3 mm} \cr
\vspace{-3 mm} \cr
grism (line mm$^{-1}$)  & 400  & 600  & 1200 \cr
dichroic                & d560 & d560 & d460 \cr
grating (line mm$^{-1}$) & 600 & 600 & 600 \cr
red central wave. (\AA) & 6800 & 6900 & 6200 \cr
blue wave. range (\AA)  & $\sim$1900--5600 & $\sim$3300--5600 & $\sim$2900--3900 \cr
blue dispersion (\AA \ pix$^{-1}$)   & $\sim$2.1 & $\sim$1.2 & $\sim$0.5 \cr
blue FWHM  (\AA)        & $\sim$6.8 & $\sim$4.3 & $\sim$1.7 \cr
red wave. range (\AA)   & $\sim$5600--8200 & $\sim$5600-8200 & $\sim$4900-7500\cr
red dispersion (\AA \ pix$^{-1}$)  & $\sim$1.3 & $\sim$1.3 & $\sim$1.3\cr
red FWHM (\AA)          & $\sim$5.8 & $\sim$5.8 & $\sim$5.8 \cr
\vspace{-3 mm} \cr
\colrule
\colrule
\end{tabular}
\end{center}
\end{table*}

\begin{figure*}[tbp]
\epsscale{1.05}
\plottwo{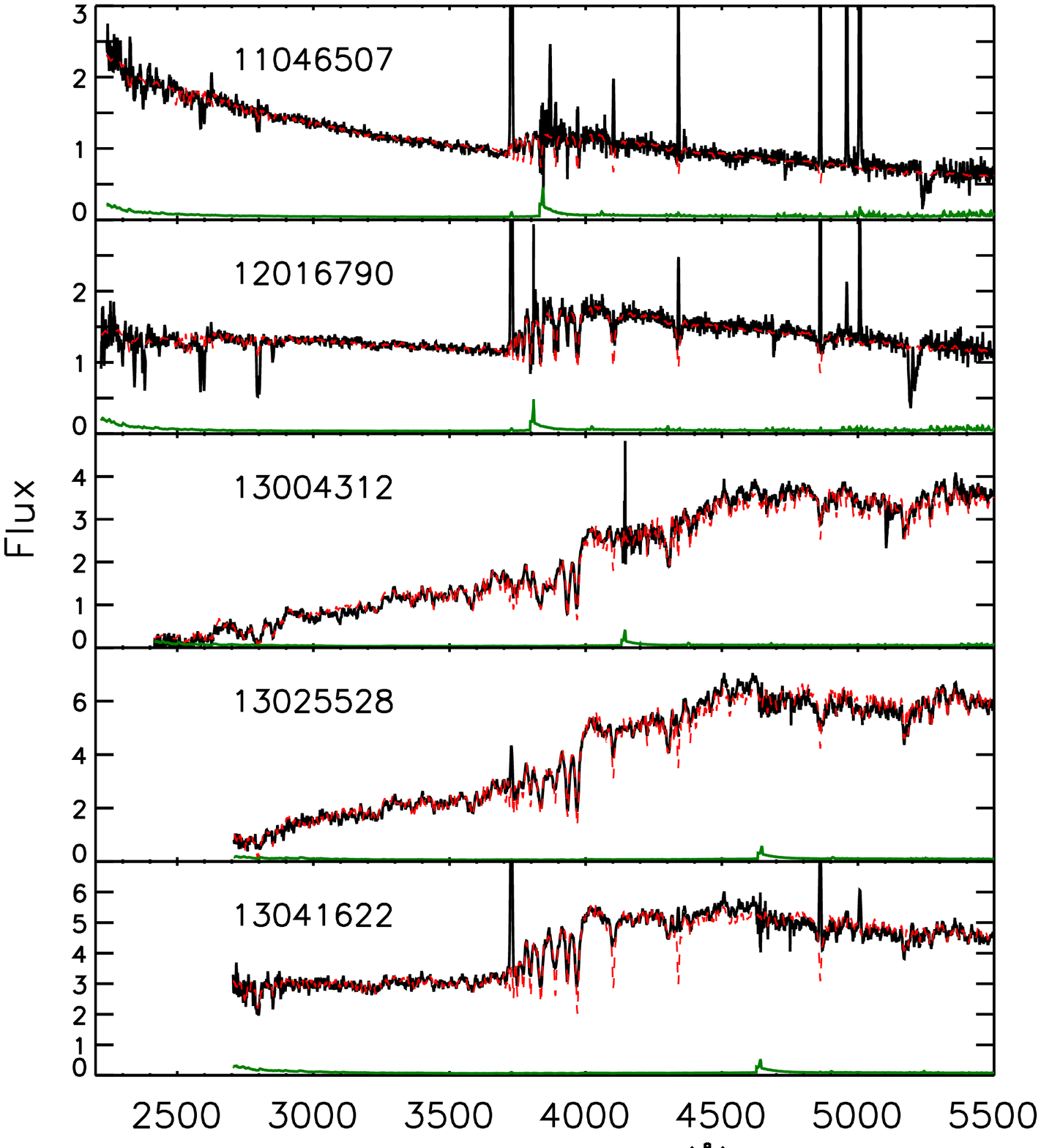}{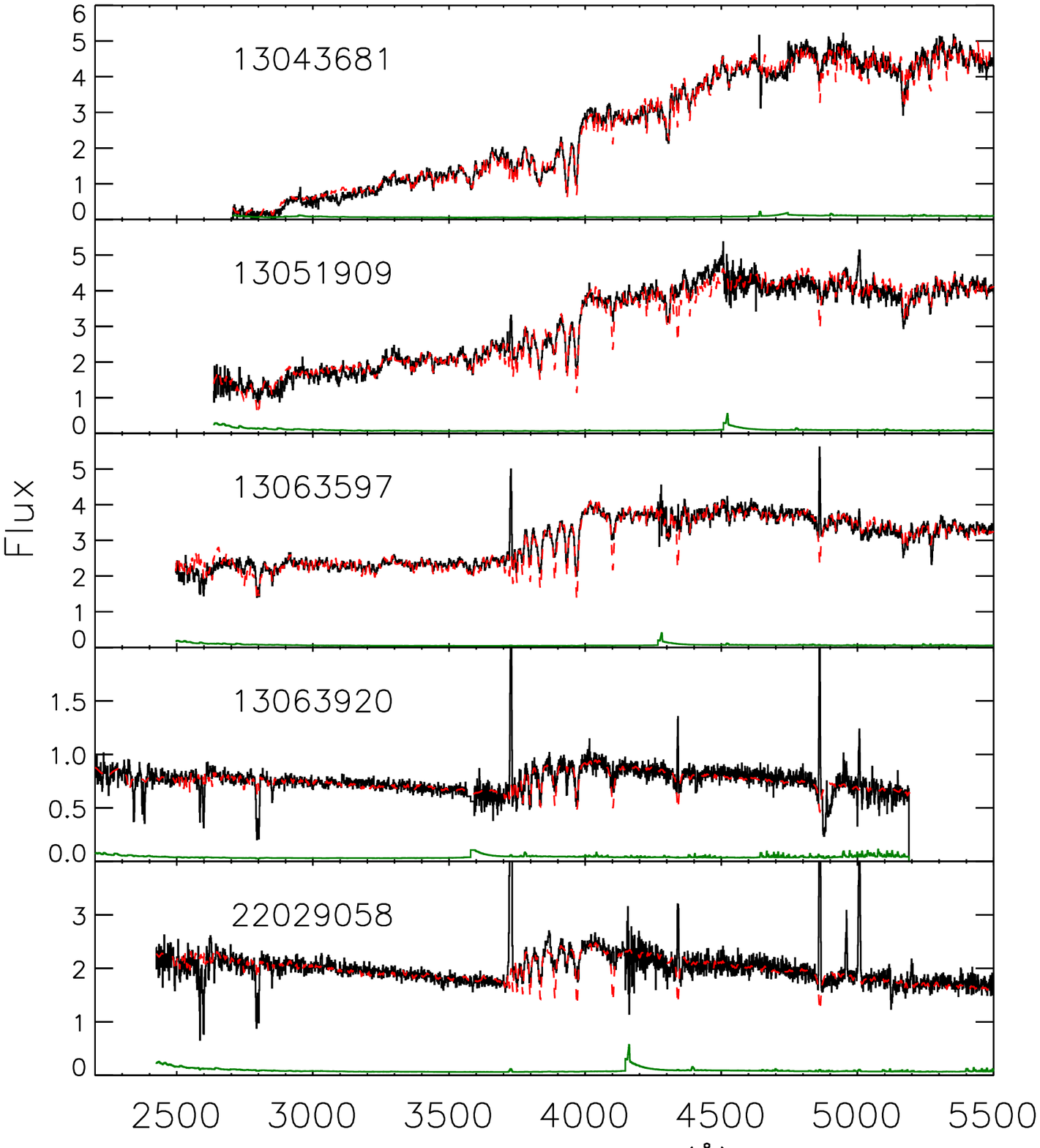}
\caption{\label{fig:fitsxray}
\small Observed spectra (black) and stellar continuum fits (red) for all
objects in the X-ray AGN host galaxy sample. Error spectra are shown
in green. } 
\end{figure*}

Observations of both samples were performed with the Low Resolution 
Imaging Spectrograph \citep[LRIS, ][]{Oke95} on the Keck I 10m telescope.  
The high UV/blue throughput of LRIS opens up the possibility of observing 
restframe UV lines at intermediate redshift.
The X-ray AGN host sample was observed on April 23, 2009
with clear skies and 1-1.5\arcsec \ seeing. 
The details of the spectrograph setup are given in Table~3.
All objects were observed with the 1.0\arcsec \ longslit.
When possible two objects were
placed on the longslit, otherwise the slit position angle was set to 
parallactic.  LRIS has an atmospheric dispersion corrector 
\citep[ADC, ][]{Phillips08}, which 
was used. The blue side data were binned both spatially
and spectrally. Total exposure times varied between 20 to 60 minutes
per object.  Standard stars were taken at the beginning and end of the
night and flats were taken in twilight.

The post-starburst sample was observed August 28 and 29, 2008, with clear
skies and $\sim$0.7-1.2\arcsec \ seeing.  
The SDSS objects were observed with the 1.0\arcsec \ longslit.
Again, when possible two objects were
placed on the longslit, otherwise the slit position angle was set to 
parallactic, and the ADC was used.  
Total exposure times varied between 30 to 60 minutes per object.
The DEEP2 post-starburst galaxies 
were observed on slitmasks with 1\arcsec \ slit widths. 
Total exposure times varied between 1 to 3 hours per slitmask.
The blue side data were binned both spatially and spectrally.
Standard stars were taken at the beginning and end of the night and flats 
were taken in twilight.  

\subsection{Data Reduction}

The data were reduced using the XIDL LowRedux 
\footnote[1]{http://www.ucolick.org/$\sim$xavier/LowRedux/}
data reduction pipeline. The pipeline includes bias subtraction
and flat fielding, wavelength calibration, object identification, 
sky subtraction, cosmic ray rejection, and flux calibration.

The wavelength solutions (which are extremely important for measuring
velocity shifts to detect 
outflowing winds, as we do here) were checked by eye both in terms of
the scatter in the arc solution (the resulting rms is 
$\sim$0.8 \AA \ in the blue and 0.1 \AA \ in
the red for the X-ray AGN host galaxies and $\sim$0.2 \AA \ in the blue and 
0.05 \AA \ in the red for the post-starburst galaxies) and by checking the
wavelengths of prominent sky lines across the observed wavelength 
range (at 3910 \AA, 5577 \AA, and 7341 \AA).  
 The final spectra are in air wavelengths.
We flux calibrated our spectra using standard stars that were observed
with the longslit using the same observational setup.

Redshifts for all of the sources were measured in the LRIS 
spectra using IDL code adapted from the DEEP2 data reduction pipeline.
Using the redshift derived from the pre-existing DEIMOS or MMT spectra as an
initial guess, we performed a $\chi^2$-minimization between the 
observed red-side LRIS data (to not be affected by the \fetwo,
 \mgtwo, or \mgone \ absorption lines on the blue side) and a linear combination 
of three galaxy templates: an artificial emission-line galaxy spectrum, 
an early-type spectrum, and a post-starburst spectrum.  The resulting
best-fit redshifts were checked by eye and are used here to define 
the systemic velocity of each galaxy.

The typical redshift error from the $\chi^2$ fit is $<$1e$^{-5}$ 
and is therefore negligible.  However, as any systemic shifts in the 
wavelength solutions between the blue and red side data would affect the e
stimated systemic 
velocity, we also measure redshifts for the blue side data for the X-ray AGN 
objects with \ion{O}{2} (3727 \AA) emission and/or Balmer absorption lines and 
find that the redshift difference is typically $\sim10$ \kms.  
For the K+A objects
the lines of interest all fall on the red side, so this test can not be 
performed.

\section{Analysis} \label{sec:analysis}

\begin{figure*}[tbp]
\epsscale{1.05}
\plottwo{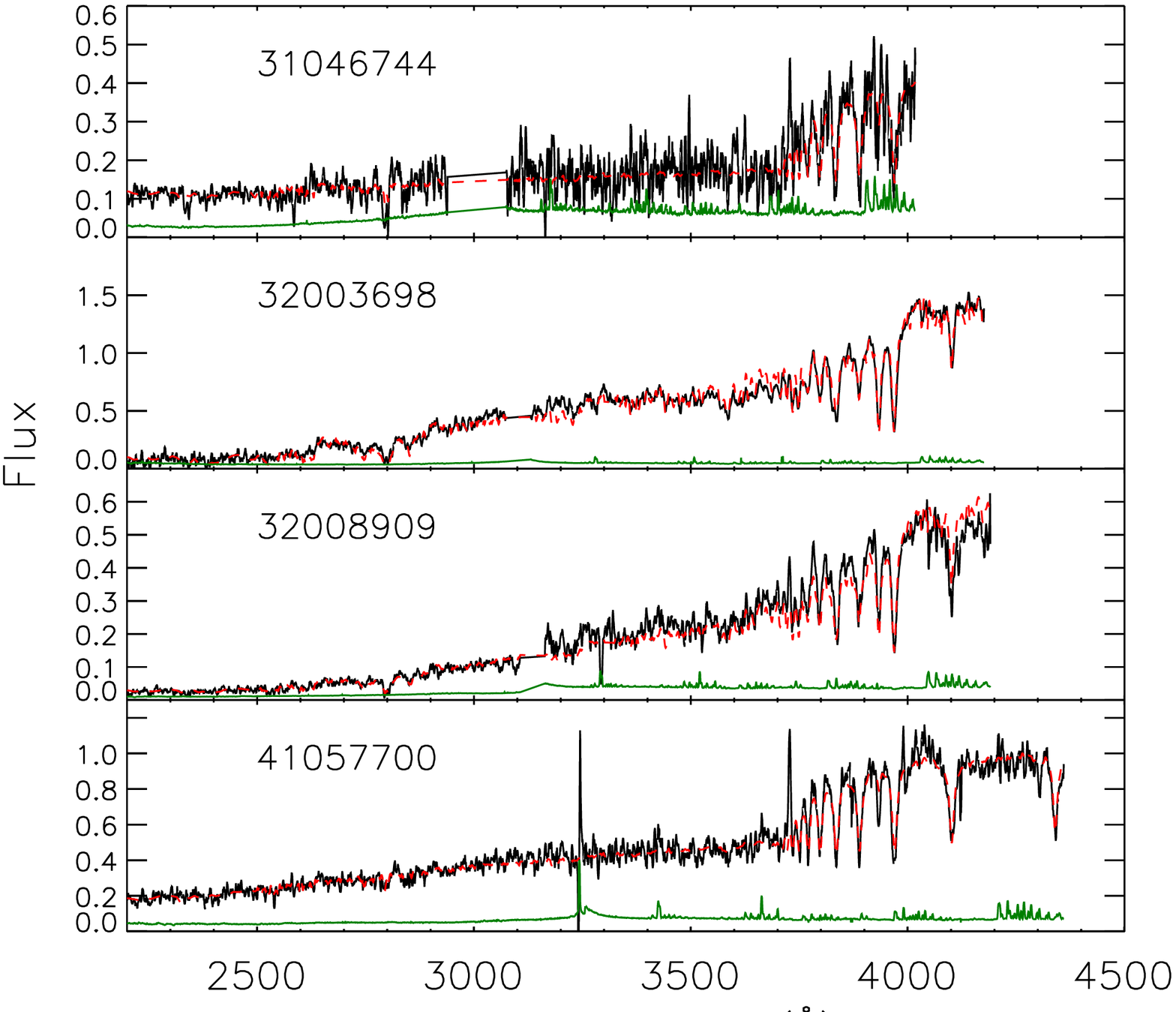}{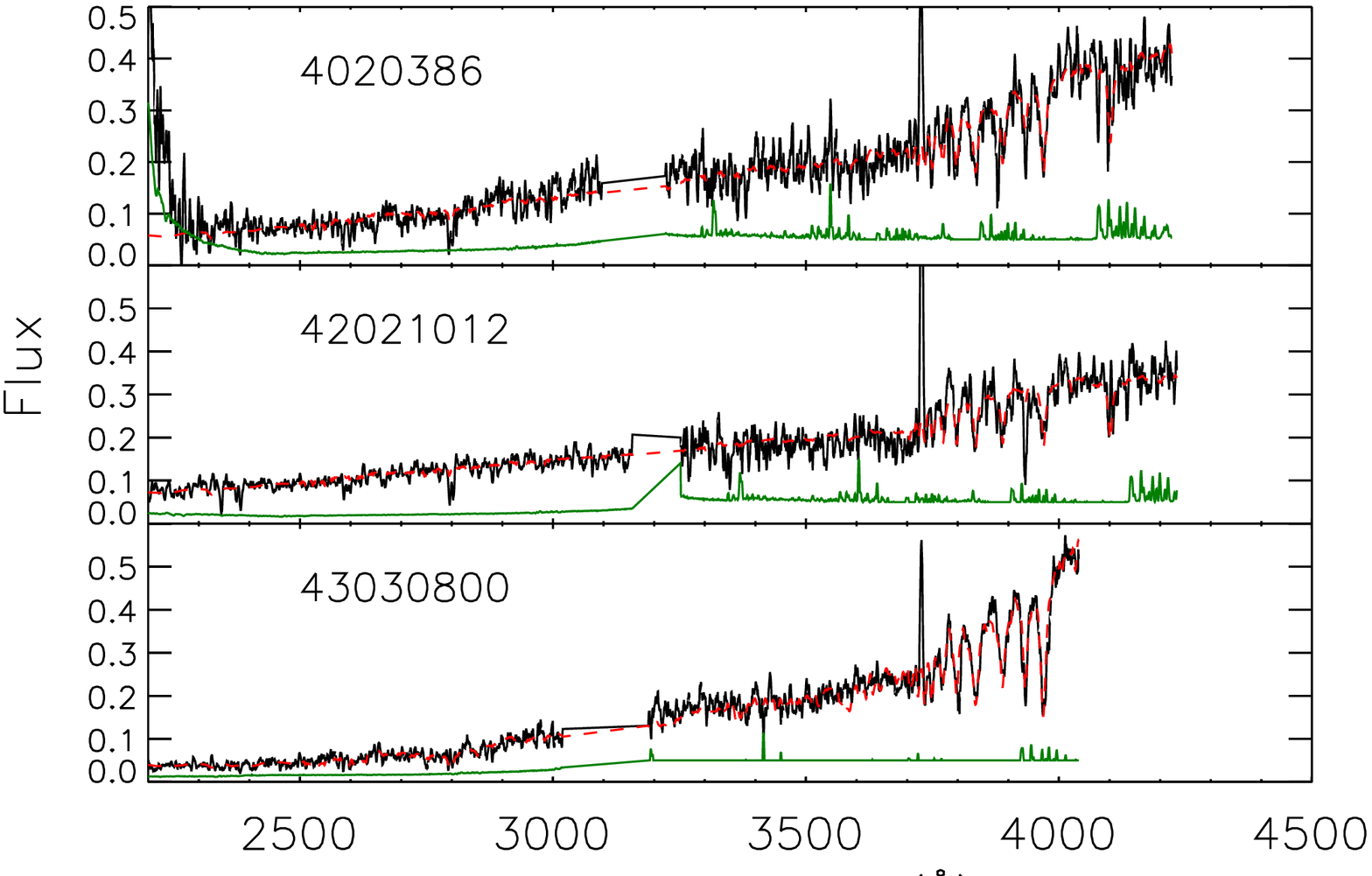}
\caption{\label{fig:fitsdeep2}
\small Observed spectra (black) and stellar continuum fits (red) for all
DEEP2 K+A galaxies in our sample. Error spectra are shown in green. 
The spectra have been smoothed by a boxcar 
of width five pixels for this figure.  The data have not been smoothed for
any of the analysis presented. }
\end{figure*}

\begin{figure*}[tbp]
\epsscale{1.05}
\plottwo{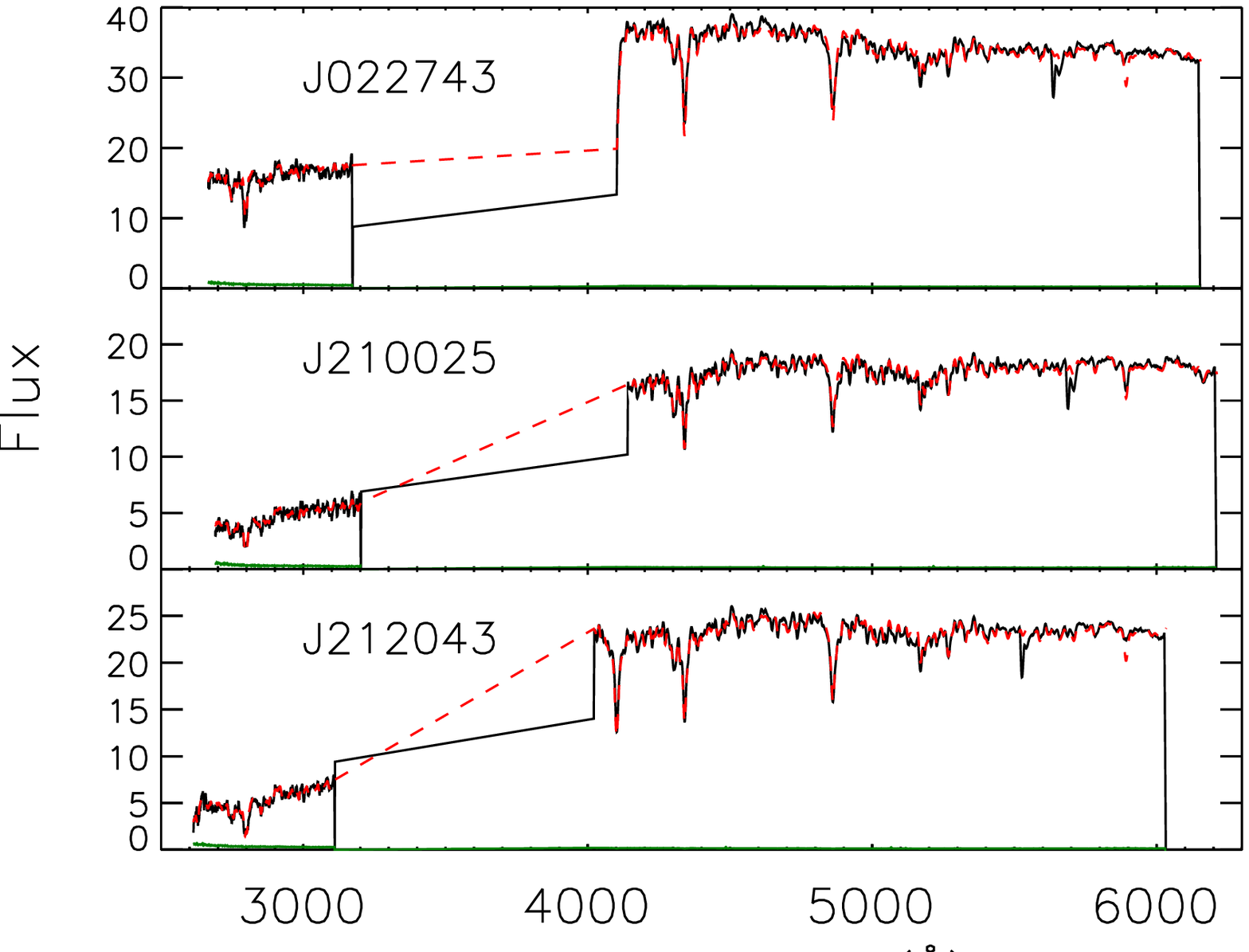}{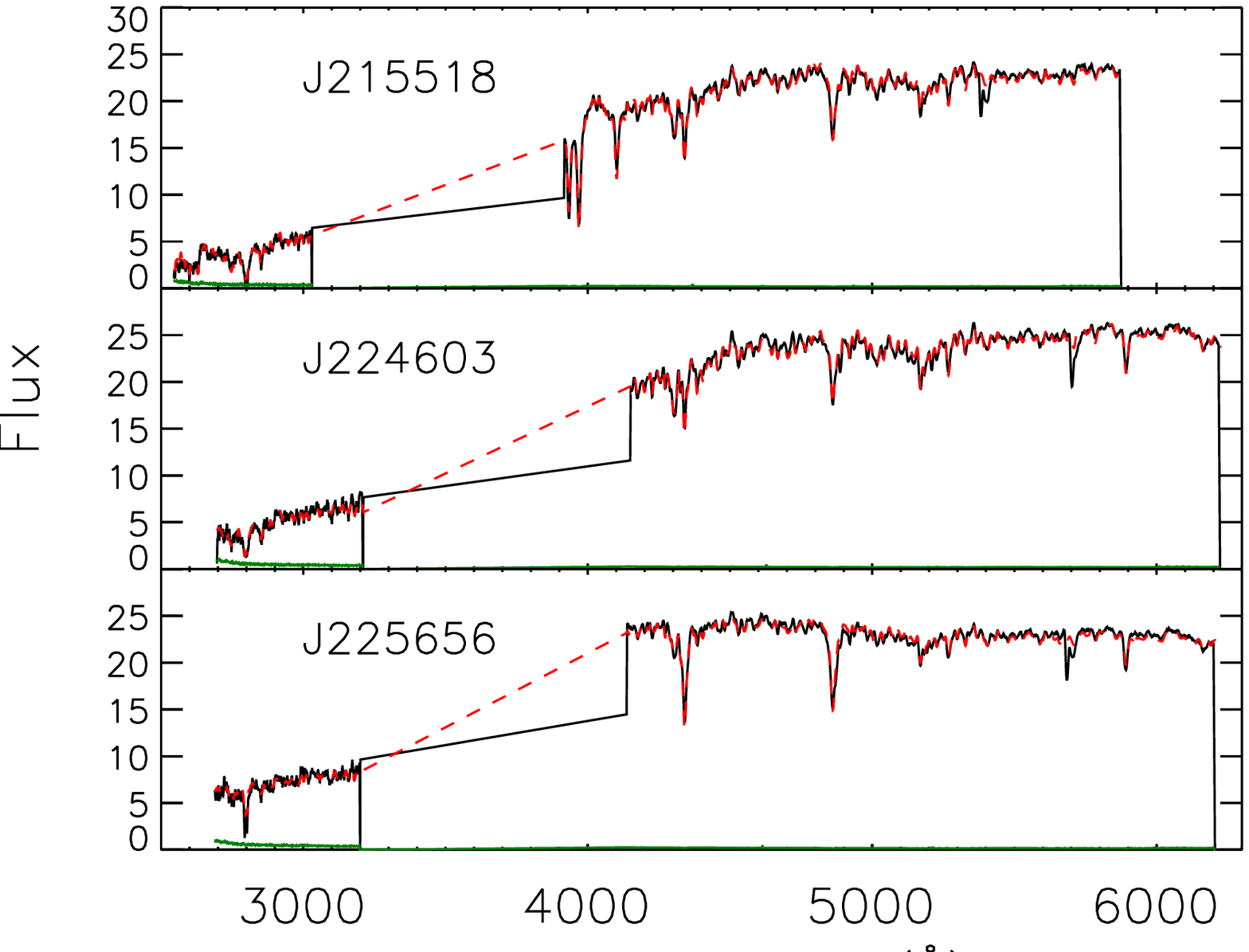}
\caption{\label{fig:fitssdss}
\small Observed spectra (black) and stellar continuum fits (red) for all
SDSS K+A galaxies in our sample.  As in Figure~\ref{fig:fitsdeep2}, the spectra
have been smoothed by a boxcar of width five pixels for this figure. }
\end{figure*}


\begin{table}[]
\tablewidth{0pt}
\begin{center}
\label{tab:age}
\footnotesize
\caption{\small Light-Weighted Stellar Age Estimates}
\begin{tabular}{lccr}
\cr
\colrule
\colrule
\vspace{-3 mm} \cr
\vspace{-3 mm} \cr
Object   &  Light-Weighted  & $<$2 Gyr Light & $<$2 Gyr  \cr
         &  Age (Gyr)       &  Fraction      & Age (Myr)        \cr
\vspace{-3 mm} \cr
\vspace{-3 mm} \cr
\colrule 
\colrule 
\vspace{-3 mm} \cr
\vspace{-3 mm} \cr
\multicolumn{4}{c}{X-ray AGN Host Galaxies} \cr
\vspace{-3 mm} \cr
\colrule 
11046507 & 0.35 $\pm$0.25 & 0.9 & 36 $\pm$8  \cr
12016790 & 0.06 $\pm$0.00 & 1.0 & 61 $\pm$2 \cr
13004312 & 6.03 $\pm$1.10 & 0.2 & 640 $\pm$12 \cr
13025528 & 5.48 $\pm$1.79 & 0.4 & 640 $\pm$9 \cr
13041622 & 1.03 $\pm$0.04 & 0.7 & 505 $\pm$15 \cr
13043681 & 4.37 $\pm$0.16 & 0.2 & 1,434 $\pm$88 \cr
13051909 & 3.07 $\pm$1.82 & 0.3 & 850 $\pm$245 \cr
13063597 & 1.45 $\pm$0.42 & 0.6 & 753 $\pm$346 \cr
13063920 & 0.06 $\pm$0.00 & 1.0 & 59  $\pm$3 \cr
22029058 & 0.52 $\pm$2.08 & 0.9 & 127 $\pm$375 \cr
\vspace{-3 mm} \cr
\colrule
\colrule
\vspace{-3 mm} \cr
\vspace{-3 mm} \cr
\multicolumn{4}{c}{DEEP2 K+A Galaxies} \cr
\vspace{-3 mm} \cr
\colrule 
31046744 & 0.37 $\pm$0.06 & 1.0 & 365 $\pm$64 \cr  
32003698 & 4.54 $\pm$0.25 & 0.6 & 696 $\pm$147 \cr  
32008909 & 6.55 $\pm$0.34 & 0.4 & 161 $\pm$31 \cr  
41057700 & 0.54 $\pm$0.12 & 1.0 & 543 $\pm$116 \cr  
42020386 & 2.16 $\pm$12.0 & 0.6 & 732 $\pm$5940 \cr  
42021012 & 0.49 $\pm$0.07 & 1.0 & 488 $\pm$65 \cr  
43030800 & 0.59 $\pm$0.06 & 1.0 & 592 $\pm$56 \cr  
\vspace{-3 mm} \cr
\colrule
\colrule
\vspace{-3 mm} \cr
\vspace{-3 mm} \cr
\multicolumn{4}{c}{SDSS K+A Galaxies} \cr
\vspace{-3 mm} \cr
\colrule 
J022743 & 0.80 $\pm$1.07 & 1.0 & 801 $\pm$1074 \cr  
J210025 & 2.54 $\pm$8.48 & 0.7 & 805 $\pm$8483 \cr  
J212043 & 0.96 $\pm$1.23 & 1.0 & 957 $\pm$1225 \cr  
J215518 & 2.17 $\pm$0.75 & 0.7 & 813 $\pm$754 \cr  
J224603 & 1.89 $\pm$12.0 & 0.3 & 686 $\pm$12.0 \cr  
J225656 & 0.78 $\pm$0.20 & 1.0 & 781 $\pm$198 \cr  
\vspace{-3 mm} \cr
\colrule
\colrule
\end{tabular}
\end{center}
\end{table}

The observed spectra are the product of a stellar population spectrum, 
ISM absorption occurring at the systemic velocity, and blueshifted 
absorption from outflowing gas, if it exists.  To decompose these we 
first model the stellar population (sec 3.1) and then the systemic 
absorption (sec 3.2).

\subsection{Stellar Continuum Fits}

Older stellar populations can show significant stellar 
\mgtwo \ and \mgone \ absorption.  While this absorption will 
likely not be relevant for our galaxies 
where the UV continuum is dominated by younger stars,
it is potentially important for the red 
X-ray AGN host galaxies and the K+A galaxies in our sample.
To estimate and account for stellar \mgtwo \ and \mgone \ absorption
 in our spectra, following \citet{Tremonti07} we model the stellar 
continuum in our LRIS spectra 
by fitting each spectrum with a linear combination of 
\citet{Bruzual03} single stellar population (SSP) models 
and adopt the model with the minimum $\chi^2$.  
A total of ten solar metallicity models are used, spanning
a range of ages from 5 Myr to 10 Gyr, and 
reddening is treated as a free parameter.
As discussed in \citet{Tremonti07}, the \citet{Bruzual03} models use the
\citet{Pickles98} stellar library which has a spectral resolution
of 10 \AA \ at wavelengths less than 3300 \AA, which is too low to model the
\mgtwol \ doublet at the resolution required.  Therefore 
theoretical spectra from the UVBLUE stellar library \citep{Merino05}
are used for wavelengths 2600-–3300 \AA.  While this does not include the 
wavelength range of the various \fetwo \ lines studied here, 
in the best fit models to our data there is essentially
no \fetwo \ absorption. 
We are mainly interested in 
modeling the \mgtwo \ and \mgone \ stellar absorption.
In deriving these fits we do not include the region of the spectra 
with rest wavelengths between 2760 \AA \ and 2870 \AA, to avoid the \mgtwo \
and \mgone \ absorption features in our data.

The fluxed spectra (black) and resulting stellar continuum fits (red) 
are shown for
each sample in Figures~2-4.  Close inspection of the Ca H+K and Balmer 
absorption features in the fits and data show that the fits reproduce 
the data well redward of 3900 \AA \ and should therefore be reasonably
 good estimates of the stellar absorption at \mgtwo \ and \mgone.

\begin{figure*}[tbp]
\plotone{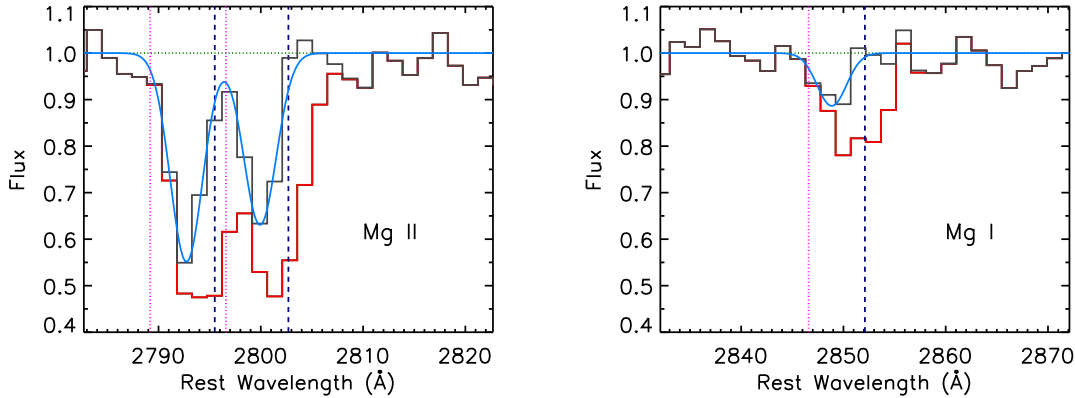}
\caption{\label{fig:method}
\small Absorption profiles for \mgtwo \ and \mgone \ for one of the X-ray AGN
host galaxies, object 12016790.  The red line is the observed
continuum-normalized flux, the dark grey is the continuum-normalized 
flux after removing the systemic absorption, and the best-fit Gaussian to
the grey data is shown in light blue.
The continuum level is shown as a dotted green line, and 
the \mgtwo \ and \mgone \ lines at rest are shown as dashed dark blue lines, 
indicating the systemic velocity of each line.  
The maximum interval over which the equivalent width is measured is 
shown as a dotted pink line; this also corresponds to the maximum velocity
that we report for each object with a wind observed in absorption.
}
\end{figure*}

The best-fit continuum model can be used to provide a rough estimate of the age 
of the stars in each galaxy.  For the post-starburst sample in particular 
we are interested in the time since the last episode of star formation.
For the X-ray AGN host galaxy sample, the ages estimated from these fits
are generally secure, as the LRIS spectra taken for this study 
cover red enough wavelengths to constrain the older stars and cover the 
Balmer absorption lines.  For the DEEP2 K+A sample, our LRIS spectra do not
extend redward of $\sim$4000 \AA, and so are not sensitive to older stellar
populations.  We therefore fit the DEEP2/DEIMOS spectra of these sources to
estimate the light-weighted age.  The ages derived from the DEIMOS and LRIS
spectra for this sample are similar, such that the stellar continuum fit used
to normalize the spectra is similar.  For the SDSS K+A sample, our LRIS spectra
extend to red wavelengths but do not cover the spectra region that contains 
Balmer absorption features.  We therefore fit the SDSS spectra to estimate
the ages of these galaxies.  

Table~4 lists the light-weighted mean of the ages of the SSPs in the
best-fit model for each object in our sample,
where we include a rough estimate of the 
light-weighted age, the fraction of the light that is due to relatively 
young ($<$2 Gyr) stars, and the light-weighted age of the young ($<$2 Gyr) 
stars only.  We note that there is an error of 
at least 1 Gyr on the light-weighted age of the stars and an error of 
at least 200 Myr on the 
age of the young population due to fitting degeneracies. We use
these age estimates merely to establish relative ages between galaxies in
our sample.
Comparing the DEEP2 and SDSS K+A samples, we find that stars in 
the younger stellar population of the SDSS K+A galaxies is older, 
on average, which may not be surprising given that the SDSS galaxies 
are at lower redshift.

We use the stellar continuum fits to correct any small residual fluxing 
errors in our data below rest wavelengths 2900 \AA, by ensuring that the 
ratio of the fit to the continuum level in our spectra is close
to unity in regions around the \fetwo, \mgtwo, and \mgone \ lines of interest. 
We then divide the data by the fit to produce continuum-normalized 
spectra 
where the continuum fit includes the stellar absorption, so that the 
normalized spectra show the excess or non-stellar absorption.

Errors on the stellar continuum model will propagate to the 
continuum-normalized spectra.  This may be particularly important for 
\mgtwo, as there is substantial \mgtwo \ absorption in older stellar 
populations.  We estimate the error on the \mgtwo \ stellar absorption 
EW by creating 25 Monte Carlo realizations of continuum spectra, drawn 
from the error distributions of the individual SSP fit amplitudes.  For the 
X-ray AGN host galaxies, the median \mgtwo \ EW error is 3\% and therefore 
negligible.  There is one galaxy, 13051909, for which the error is 20\%; 
the relevance of this is discussed further in Section 5.1.
 
For the DEEP2 K+A galaxies, the median stellar continuum \mgtwo \ EW error is 
9\% and is subdominant.  There are two objects that have larger \mgtwo \ EW 
errors: 42020386 (32\%) and 43030800 (26\%); these are discussed further in 
Sections 4.1 and 5.1.  For the SDSS K+A galaxies, the continuum errors are 
negligible ($<2$\%) using the best fits to the LRIS spectra.  
However, for the DEEP2 and SDSS K+A galaxies, there may be additional 
systematic errors in the stellar continuum, as the LRIS spectra do not 
include red wavelengths for the DEEP2 sources or the Balmer series for 
the SDSS sources.  We may therefore be underestimating the \mgtwo \ 
photospheric absorption in the DEEP2 stellar continua and overestimating the \mgtwo \ 
photospheric absorption in the SDSS stellar continua.  This would imply that the SDSS 
wind absorption EWs that we measure are a conservative lower limit; however, the 
DEEP2 wind absorption EWs may be overestimated.  The importance of this can be 
determined by comparing the \mgtwo \ EW in the LRIS stellar continuum best-fit 
models and those from the DEIMOS and SDSS spectra.  The EWs are generally 
similar between the LRIS and DEIMOS/SDSS spectra and therefore our 
measurements should not be significantly affected.  This is discussed further 
in Section 5.1 below.

In addition, we produce separate continuum normalized spectra that 
that are normalized by a smooth continuum not including stellar absorption.
These spectra are used to estimate
how much our results are biased if stellar absorption is {\it not}
explicitly accounted for.  To create these spectra we use the best fit
model above to define the broad shape of the continuum level.  We
divide the model fit into bins $\sim$50 \AA \ wide and use the maximum
value of the fit in that bin as an estimate of the continuum level.
This ensures that we use the continuum level of the fit but do not
include any absorption lines.  The width of 50 \AA \ is large enough
to ensure that no absorption line features contribute but is small
enough to retain the shape and wiggles in the overall continuum
level.  We then interpolate this coarse continuum estimate onto the 
wavelength grid in the data, and divide the observed spectrum by the
continuum estimate.  Investigating the continuum-normalized spectra by eye, 
we conclude that this method works well for our purposes.

\subsection{Absorption Line Fits}


\begin{figure*}[tbp]
\epsscale{1.1}
\plottwo{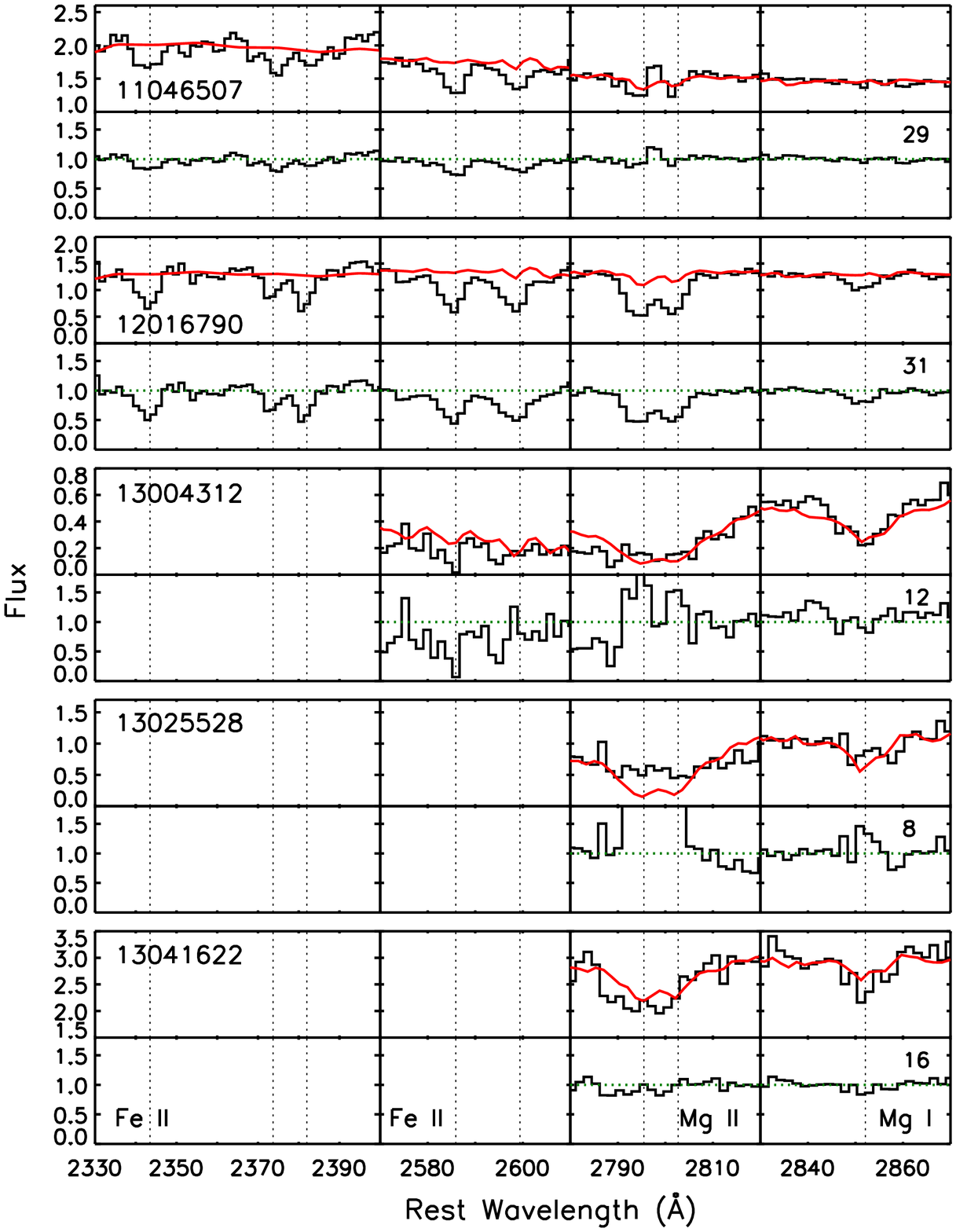}{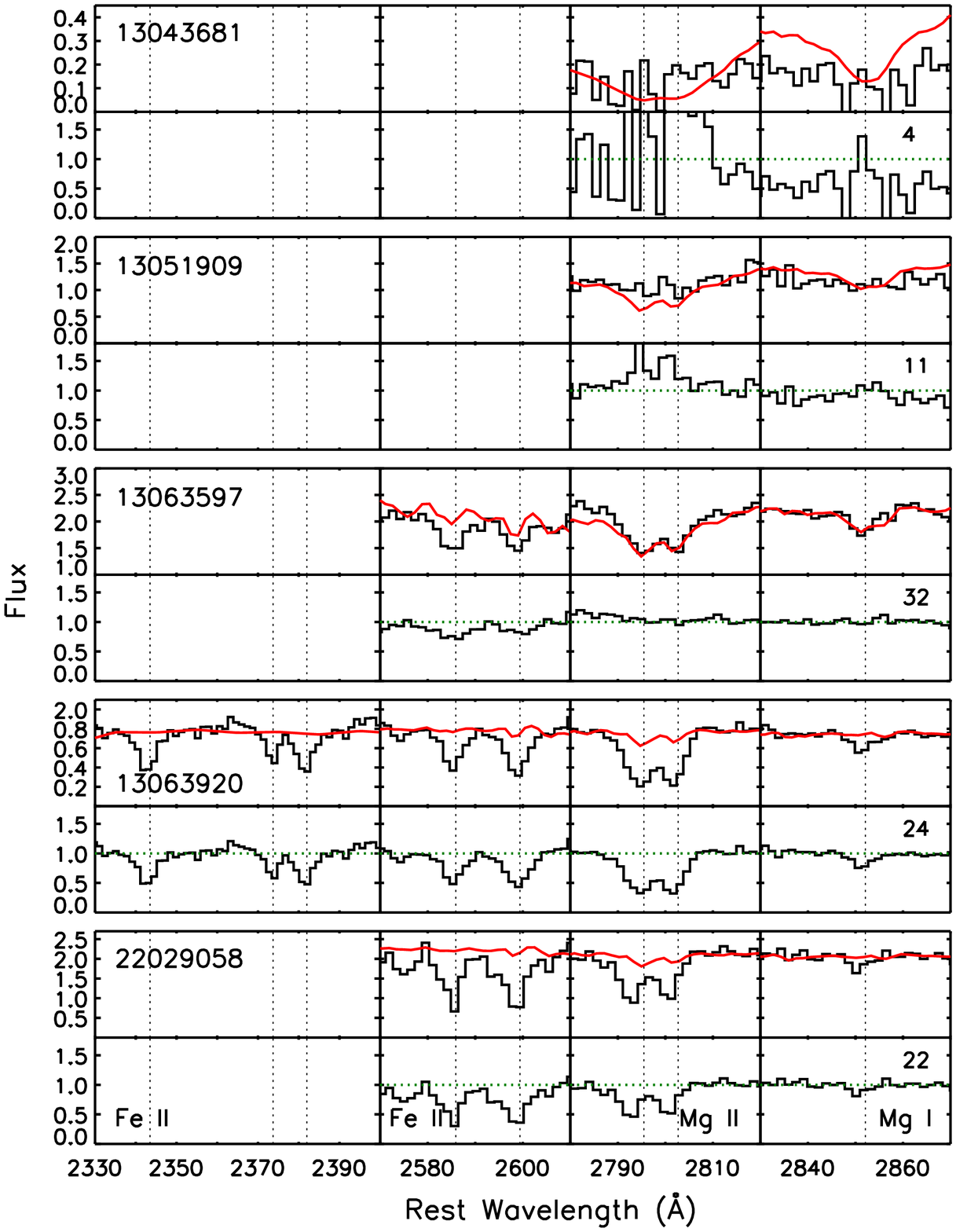}
\caption{\label{fig:linesxray}
\small Regions of the observed and continuum-normalized spectra around 
\fetwo, \mgtwo, and \mgone \ for objects in the DEEP2 X-ray AGN host galaxy
sample.  For each object we plot two rows of spectra; the top row shows the 
observed spectrum (black) and stellar continuum fit (red), while the bottom
row shows the continuum-normalized spectrum (black) and continuum level 
(green). We do not show spectra with the systemic component removed.
In the lower right panel for each galaxy we list the median S/N/pixel 
averaged between two windows near \mgtwo \ and \mgone: 2650--2750 \AA \ and 
2900--3000 \AA.
}
\end{figure*}

We model the \fetwo, \mgtwo, and \mgone \ absorption lines in our data by
fitting a Gaussian to the continuum-normalized spectra, from which we
can estimate the velocity offset of the center of the absorption
profile, the velocity width of the profile, and the lower limit on the 
covering fraction (equal to the observed absorption depth).

First, following \citet{Weiner09} we attempt to separate absorption due
to outflowing gas from absorption due to the ISM of the galaxy (i.e. not
in an outflow component) with the following model:

\begin{equation}
F_{obs}(\lambda) = C(\lambda) (1-A_{sym}) (1-A_{flow}),
\end{equation}

\noindent
where $F_{obs}(\lambda)$ is the observed flux density, $C(\lambda)$ is the 
underlying stellar continuum fit, and $A_{sym}$ and $A_{flow}$
are the absorption features from the intrinsic (symmetric) and outflow (blueshifted)
absorption.  To estimate the intrinsic ISM absorption of the galaxy,
if any, we first fit a Gaussian to the continuum-normalized
flux and use the result to model the absorption {\it redward} of
the systemic velocity for each line as an estimate of the systemic
absorption.  We assume that this absorption is symmetric about the
systemic velocity of the line and divide out this model fit to the
data both blueward and redward of the systemic velocity.  We then fit
the resulting data (where the remaining absorption should be due only
to an outflowing component) with a new Gaussian.  This step of
removing the systemic absorption is performed only for those objects
and lines in which the data redward of the systemic velocity was
negative (i.e. showed absorption).  Only galaxies in the blue cloud
show this absorption, including four X-ray AGN host galaxies, 
two DEEP2 K+A galaxies, and one SDSS K+A galaxy, not all of which 
have detected blueshifted absorption. 

For the \mgtwol \ and \fetwo \ 2586, 2599 \AA \ lines 
we model the symmetric absorption as the product
of two Gaussians $G(v)$ centered on the wavelengths $\lambda_1,\lambda_2$, 
with velocity dispersion $\sigma$ and intensities $A_1,A_2$:

\begin{equation}
A_{sym}(\lambda) = A_1 G(v,\lambda_1,\sigma) + A_2 G(v,\lambda_2,\sigma)
\end{equation}
 For the \fetwo \ 2586, 2599 \AA \ lines 
we allow $A_1$ and $A_2$ to be fit independently.

For the \mgtwol \ doublet the systemic component of the bluer line may be
impacted by absorption from the outflowing wind component of the redder
line.  We therefore use the measured systemic component from
the redder line alone for both lines in the doublet, assuming that the systemic
component is the same for both, which would be true if the 
systemic absorption in the \mgtwo \ line is saturated.
Given that the ratio of the minimum absorption in the two 
lines is close to unity, this assumption of $\tau>1$ is well justified.  
We then remove this systemic absorption from 
both lines in the doublet and fit the remaining data with a double
Gaussian, where we restrict both lines to have the same velocity width but 
allow the relative depths of the lines to vary.

\begin{figure*}[tbp]
\epsscale{1.1}
\plottwo{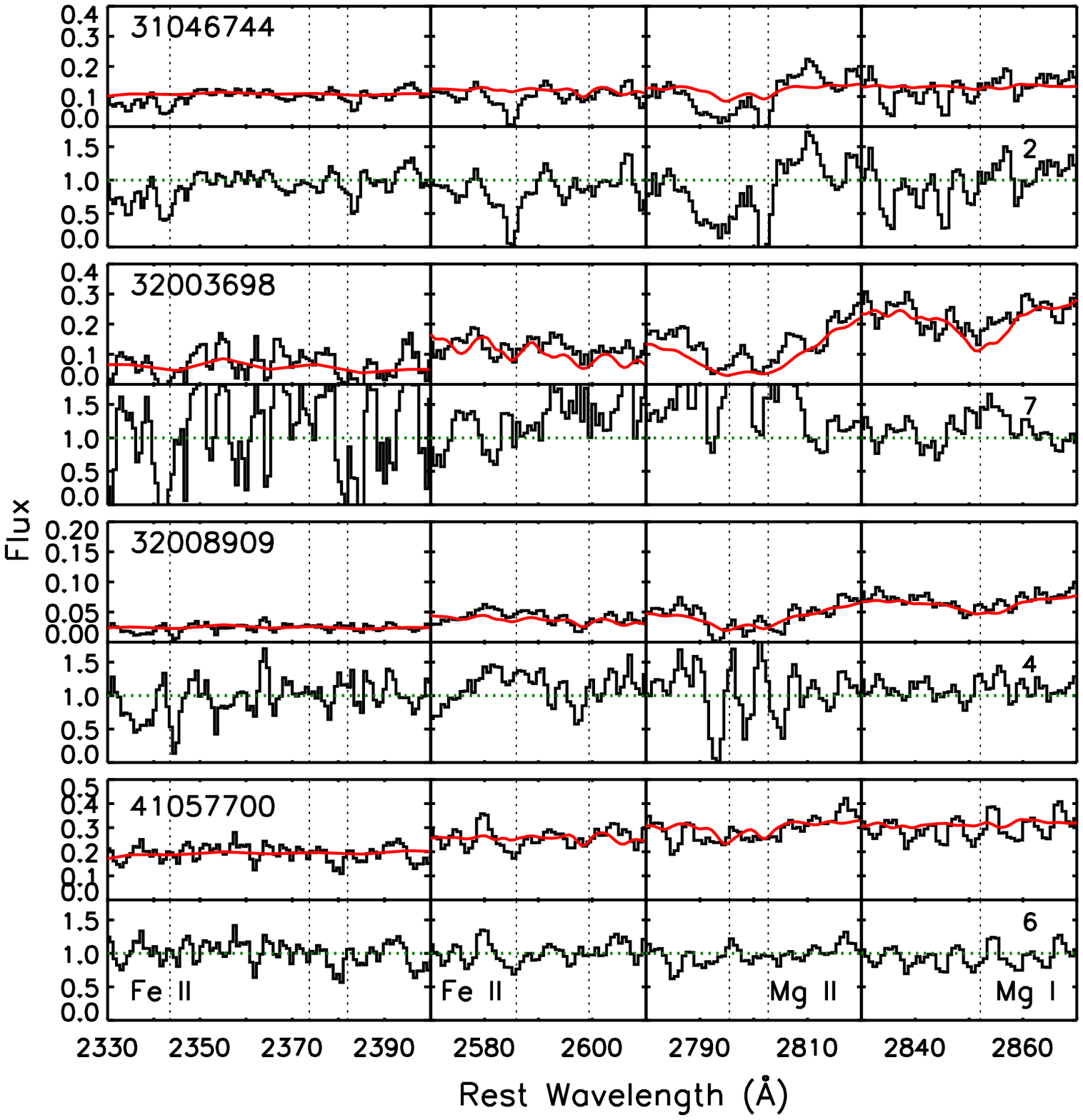}{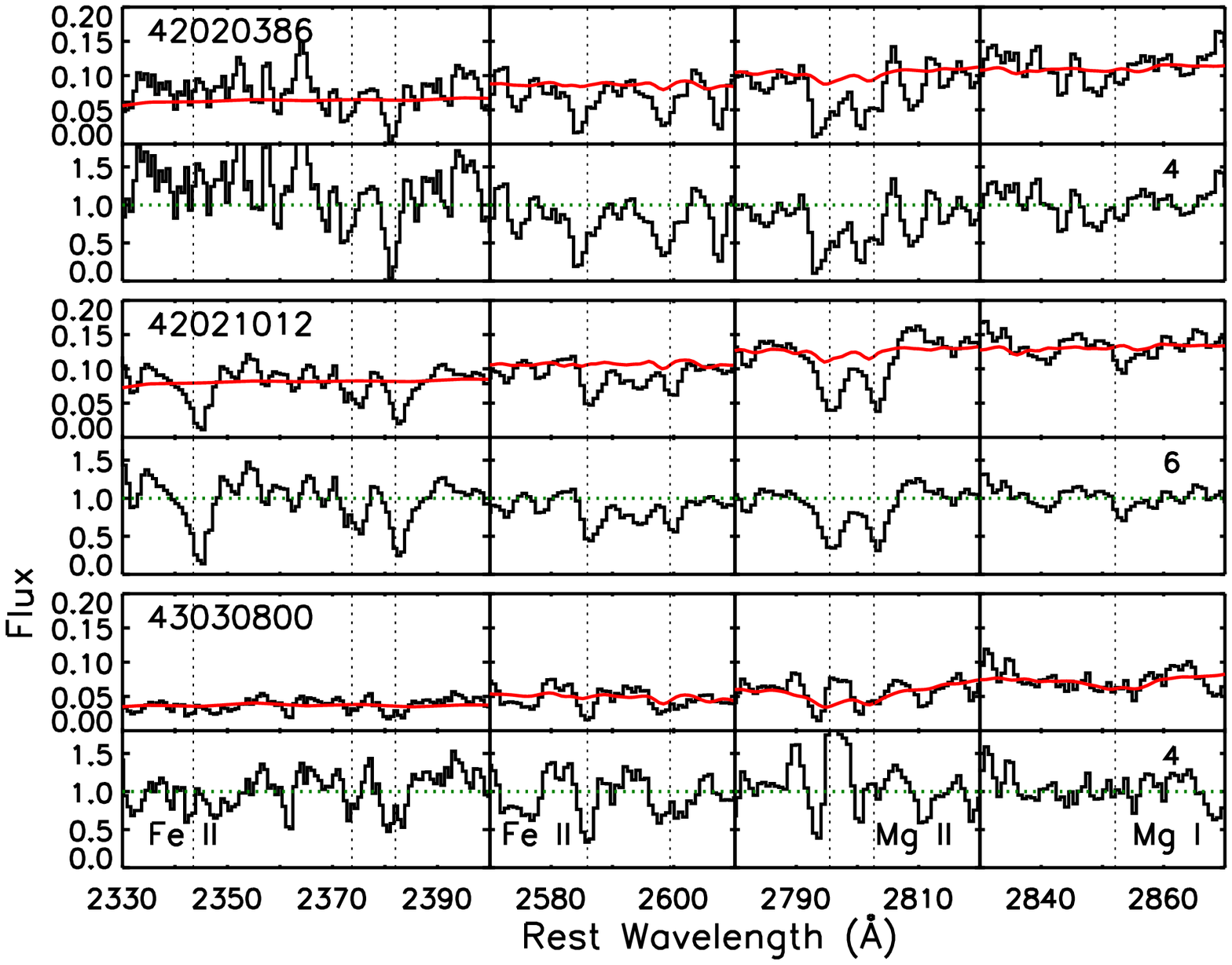}
\caption{\label{fig:linesdeep2}
\small Regions of the observed and continuum-normalized spectra around 
\fetwo, \mgtwo, and \mgone \ for objects in the DEEP2 K+A galaxy 
sample, similar to Figure~6.
}
\end{figure*}

The Gaussian fit to the remaining absorption blueward of systemic provides 
an estimate of the velocity centroid of the outflowing wind, the 
velocity width of the wind, and the  minimum line depth.  
The covering fraction is simply 1 - the minimum line depth (or $A_{flow}$) 
if the data are high resolution and the line is optically thick; here we 
can only obtain a lower limit on the covering fraction due to our 
resolution.

As the various AGN and K+A samples were observed using different blue side 
grisms, the velocity resolution varies.  The resolution at 2800 \AA \ restframe 
for the X-ray AGN sample is $\sim$540 \kms, while for the DEEP2 K+A sample it 
is $\sim$250 \kms, and for the SDSS K+A sample it is $\sim$150 \kms.  
These differences will affect measurements of the velocity width and line 
depth, which should be interpreted with caution.  However, they should not 
significantly affect the velocity centroid or our ability to detect outflows, 
which is dominated by the S/N of the spectra. 
As discussed above, only a few galaxies in our sample with blueshifted 
absorption have systemic absorption 
removed.  While in theory this, when combined with the different velocity 
resolution of our samples, could affect the velocity centroid of the outflow in
these objects, in one X-ray AGN host galaxy where systemic absorption is 
removed from the \fetwo \ lines but not the \mgtwo \ doublet (due to the 
presence of \mgtwo \ emission), the outflow velocity centroids of the \fetwo \ 
and \mgtwo \ lines are consistent with each other.

From the Gaussian fit we can estimate where the blueward flux of the fit is
within 1$\sigma$ of the continuum; we call this the `maximum velocity' of 
the wind.  We integrate the continuum-normalized flux 
from the systemic velocity to this maximum velocity to measure the
EW and associated error.  The velocity range used 
to measure the EW therefore varies from object to object. 
This approach is summarized in Figure~5, which shows the results 
of fitting the \mgtwo \ doublet and \mgone \ for an X-ray AGN host galaxy.

The absorption EW is measured after removing any systemic component, 
such that it is the EW of the outflow component alone.
Without subtracting systemic absorption due to ISM in the galaxy, 
any estimates of the outflow kinematics or EW 
may be systematically biased.  For example, for the \mgone \ line 
shown in Figure~5, the EW is 2.3 times greater, the velocity width is 1.9
times wider, and the velocity centroid is 0.3 times smaller if the
systemic absorption is not removed.  For the \mgtwo \ doublet, the
EW is 1.7 times greater, the velocity width is 1.5 times wider, and
the velocity centroid is 0.5 times smaller if the systemic absorption
is not removed.  Therefore, quantitative results can be {\it up to a factor 
of 2 to 3 different} if systemic absorption is not accounted for.

For lines without detectable absorption, we estimate the 2$\sigma$ 
upper limit on the
EW that could have been measured.  This is complicated due to \mgtwo \ emission 
in many cases; \mgtwo \ emission is discussed further in Section 4.3 below.

We further test whether our measured \mgtwo \ absorption parameters are 
sensitive to the stellar absorption accounted for in our continuum fits by 
measuring \mgtwo \ absorption in continuum-normalized spectra both with and 
without stellar absorption included.  We test the higher S/N X-ray AGN 
host galaxies and find that when \mgtwo \ emission is not present, the 
results are not sensitive to stellar absorption, as our method above 
accounts for all absorption at systemic, whether from stars or the ISM.
However, the stellar absorption included in our continuum fits can 
affect the detected \mgtwo \ emission, in that one can underestimate the 
\mgtwo \ emission if stellar absorption is not accounted for (see Section 4.3).

\begin{figure*}[tbp]
\epsscale{1.1}
\plottwo{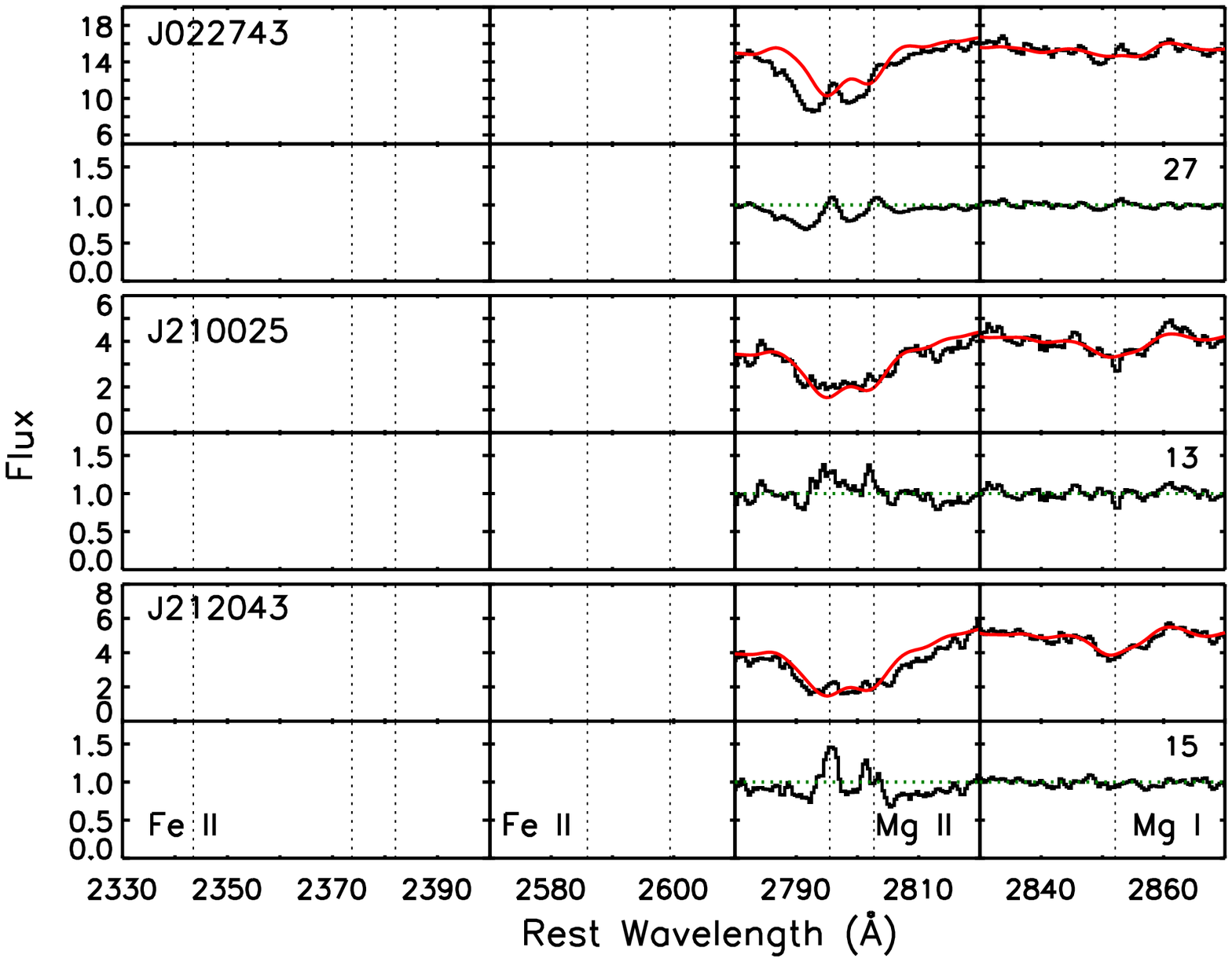}{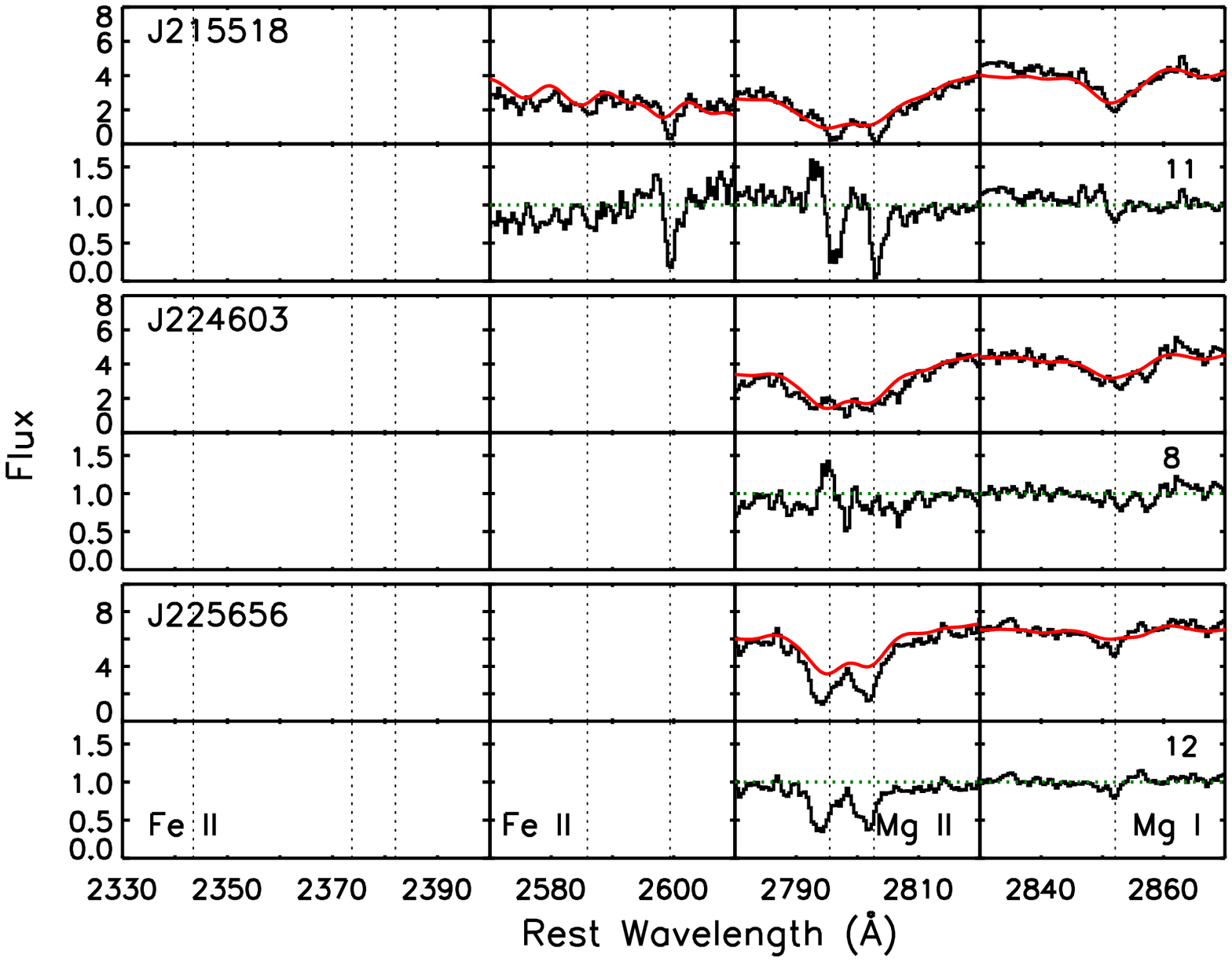}
\caption{\label{fig:linessdss}
\small Regions of the observed and continuum-normalized spectra around 
\fetwo, \mgtwo, and \mgone \ for objects in the SDSS K+A galaxy 
sample, similar to Figures~6 and 7.
}
\end{figure*}

\section{Results on Winds Detected in Absorption} \label{sec:results}

In this section we first present results of fitting our spectra 
for \fetwo, \mgtwo, and \mgone \ absorption, investigating which objects
have outflowing winds as seen in blueshifted absorption 
and what the wind properties are.   We then compare the different 
absorption lines.  In the following section we  
present measurements of \mgtwo \ and \fetwo* emission in 
each object, which likely originates in the wind as well 
\citep{Weiner09,Rubin10a,Prochaska11}.


\begin{table*}[]
\tablewidth{0pt}
\begin{center}
\label{tab:ews}
\setlength{\tabcolsep}{4.5pt}
\scriptsize
\caption{\small \fetwo, \mgtwo, and \mgone \ 
Absorption Line Measurements and Equivalent Widths}
\vspace{0.25cm}
\resizebox{\columnwidth}{!} {
	\begin{tabular*}{1.0\columnwidth}{lrrrcc}
          \colrule
          \colrule
          \vspace{-3 mm} \cr
          Line \  & Velocity &   Velocity   & Max.   & $A_{flow}$\tablenotemark{4} & EW  \cr
         & centroid\tablenotemark{1} &  width\tablenotemark{2}  & vel.\tablenotemark{3} &  & \cr
          (\AA)& (km/s) &(km/s) &(km/s) &          & (\AA) \cr
          \vspace{-3 mm} \cr
          \colrule 
          \colrule 
          \vspace{-3 mm} \cr
          \multicolumn{6}{c}{DEEP2 X-ray AGN Host Galaxies} \cr
          \vspace{-3 mm} \cr
          \colrule 
          \vspace{-3 mm} \cr
          \multicolumn{6}{c}{11046507} \cr
          \vspace{-3 mm} \cr
          2852.1  &   X\tablenotemark{5}         &   X        &       X   &    X          &     X            \cr
          2802.7  & $-$141 $\pm$103 &   69 $\pm$46 &      $-$225   &  0.11 $\pm$0.08 &     0.18 $\pm$0.08 \cr
          2795.5  & $-$279 $\pm$44  &  115 $\pm$45 &      $-$505   &  0.13 $\pm$0.04 &     0.42 $\pm$0.11 \cr
          2599.4  & $-$312 $\pm$71  &  327 $\pm$66 &      $-$729   &  0.08 $\pm$0.03 &     0.45 $\pm$0.12 \cr
          2585.9  & $-$312 $\pm$71  &  327 $\pm$66 &      $-$782   &  0.11 $\pm$0.02 &     0.64 $\pm$0.13 \cr
          2382.0  &   X         &   X        &       X   &    X          &     X            \cr
          2373.7  &   X         &   X        &       X   &    X          &     X            \cr
          2343.5  & $-$360 $\pm$71  &  151 $\pm$76 &      $-$561   &  0.14 $\pm$0.06 &     0.39 $\pm$0.13 \cr
          \colrule 
          \vspace{-3 mm} \cr
          \multicolumn{6}{c}{12016790} \cr
          \vspace{-3 mm} \cr
          2852.1  & $-$329 $\pm$42  &   145 $\pm$43  &    $-$558  &    0.12 $\pm$0.03   &    0.39 $\pm$0.10 \cr
          2802.7  & $-$297 $\pm$10  &   170 $\pm$9   &    $-$653  &    0.37 $\pm$0.10   &    1.40 $\pm$0.09 \cr
          2795.5  & $-$297 $\pm$10  &   170 $\pm$9   &    $-$677  &    0.45 $\pm$0.03   &    1.80 $\pm$0.11 \cr
          2599.4  & $-$255 $\pm$16  &   264 $\pm$16  &    $-$822  &    0.38 $\pm$0.14   &    2.20 $\pm$0.13 \cr
          2585.9  & $-$255 $\pm$16  &   264 $\pm$16  &    $-$805  &    0.33 $\pm$0.02   &    1.63 $\pm$0.12 \cr
          2382.0  & $-$177 $\pm$18  &   138 $\pm$16  &    $-$451  &    0.52 $\pm$0.07   &    1.48 $\pm$0.17 \cr
          2373.7  & $-$177 $\pm$18  &   138 $\pm$16  &    $-$416  &    0.33 $\pm$0.06   &    0.82 $\pm$0.15 \cr
          2343.5  & $-$253 $\pm$41  &   209 $\pm$42  &    $-$612  &    0.36 $\pm$0.06   &    1.26 $\pm$0.17 \cr
          \colrule 
          \vspace{-3 mm} \cr
          \multicolumn{6}{c}{13004312} \cr
          \vspace{-3 mm} \cr
          2852.1 &    X      &    X      &     X  &     X        &      X   \cr
          2802.7 &    X      &    X      &     X  &     X        &      X   \cr
          2795.5 &  $-$1225 $\pm$161 &   496 $\pm$156  & $-$2136  &   0.47 $\pm$0.13  &      3.07 $\pm$0.77  \cr
          2599.4 &    X      &    X      &     X  &     X        &      X   \cr
          2585.9 &    X      &    X      &     X  &     X        &      X   \cr
          \colrule 
          \vspace{-3 mm} \cr
          \multicolumn{6}{c}{13041622} \cr
          \vspace{-3 mm} \cr
          2852.1 &    X      &    X      &     X  &     X        &      X   \cr
          2802.7 &  $-$576 $\pm$106 &   271 $\pm$82  &    $-$717  &   0.18 $\pm$0.08  &      0.63 $\pm$0.27  \cr
          2795.5 &  $-$576 $\pm$106 &   271 $\pm$82  &    $-$913  &   0.17 $\pm$0.05  &      1.21 $\pm$0.28  \cr
          \colrule 
          \vspace{-3 mm} \cr
          \multicolumn{6}{c}{13063920} \cr
          \vspace{-3 mm} \cr
          2852.1 &  $-$136 $\pm$48  &    92 $\pm$38   &   $-$263   &  0.13 $\pm$0.05   &     0.25 $\pm$0.08  \cr
          2802.7 &  $-$289 $\pm$11  &   204 $\pm$9    &   $-$739   &  0.51 $\pm$0.13   &     2.52 $\pm$0.12  \cr 
          2795.5 &  $-$289 $\pm$11  &   204 $\pm$9    &   $-$741   &  0.49 $\pm$0.03   &     2.17 $\pm$0.13  \cr
          2599.4 &  $-$217 $\pm$20  &   185 $\pm$20   &   $-$591   &  0.41 $\pm$0.04   &     1.58 $\pm$0.13  \cr
          2585.9 &  $-$152 $\pm$15  &   126 $\pm$18   &   $-$409   &  0.42 $\pm$0.05   &     1.10 $\pm$0.11  \cr
          2382.0 &  $-$177 $\pm$27  &   184 $\pm$27   &   $-$526   &  0.43 $\pm$0.05   &     1.55 $\pm$0.16  \cr
          2373.7 &  $-$102 $\pm$33  &   133 $\pm$32   &   $-$326   &  0.31 $\pm$0.07   &     0.80 $\pm$0.15  \cr
          2343.5 &  $-$207 $\pm$24  &   164 $\pm$24   &   $-$535   &  0.48 $\pm$0.06   &     1.37 $\pm$0.15  \cr
          \colrule 
          \vspace{-3 mm} \cr
          \multicolumn{6}{c}{22029058} \cr
          \vspace{-3 mm} \cr
          2852.1  & $-$249 $\pm$56  &    84 $\pm$54 &   $-$389 &    0.18 $\pm$0.06 &         0.37 $\pm$0.11  \cr
          2802.7  & $-$314 $\pm$11  &   169 $\pm$10 &   $-$675 &    0.49 $\pm$0.14 &         1.87 $\pm$0.14  \cr
          2795.5  & $-$314 $\pm$11  &   169 $\pm$10 &   $-$677 &    0.58 $\pm$0.04 &         2.19 $\pm$0.14  \cr
          2599.4  & $-$141 $\pm$14  &   175 $\pm$15 &   $-$499 &    0.55 $\pm$0.19 &         1.89 $\pm$0.15  \cr
          2585.9  & $-$141 $\pm$14  &   175 $\pm$15 &   $-$503 &    0.45 $\pm$0.04 &         1.58 $\pm$0.14  \cr
          \vspace{-3 mm} \cr
          \colrule 
          \colrule 
        \end{tabular*}
        }	
\setlength{\tabcolsep}{4.5pt}
\resizebox{\columnwidth}{!} {
	\begin{tabular*}{1.0\columnwidth}{lrrrcc}
          \colrule
          \colrule
          \vspace{-3 mm} \cr
          \ \ Line \ \ &  Velocity    &   Velocity    & Max.    & $A_{flow}$ & EW  \cr
          & centroid                 &  width                       & vel.           &  & \cr
          \ \ (\AA)& (km/s) &(km/s) &(km/s) &          & (\AA) \cr
          \vspace{-3 mm} \cr
          \colrule 
          \colrule 
          \vspace{-3 mm} \cr
          \multicolumn{6}{c}{DEEP2 K+A Galaxies} \cr
          \vspace{-3 mm} \cr
          \colrule 
          \vspace{-3 mm} \cr
          \multicolumn{6}{c}{31046744} \cr
          \vspace{-3 mm} \cr
          2852.1 &  X             & X              &  X       &    X              &    X \cr
          2802.7 &  $-$82  $\pm$30  &     103 $\pm$30  &    $-$225  &    1.45 $\pm$0.37   &    2.48 $\pm$0.59 \cr
          2795.5 &  $-$433 $\pm$86  &     265 $\pm$83  &    $-$633  &    0.73 $\pm$0.20  &     3.45 $\pm$0.91 \cr
          2599.4 &  X             & X              &  X       &     X             &     X \cr
          2585.9 &  $-$133 $\pm$50  &     249 $\pm$50  &    $-$480  &    0.79 $\pm$0.14  &     3.41 $\pm$0.54 \cr
          2382.0 &  X             & X              &  X       &     X             &     X  \cr
          2373.7 &  X             & X              &  X       &    X              &    X  \cr
          2343.5 &  $-$161 $\pm$38  &    135 $\pm$38   &   $-$356   &    0.73 $\pm$0.17  &    1.76 $\pm$0.36 \cr
          \colrule 
          \vspace{-3 mm} \cr
          \multicolumn{6}{c}{42021012\tablenotemark{6}} \cr
          \vspace{-3 mm} \cr
          2852.1  &   115 $\pm$34 &  73 $\pm$35 & 200 & 0.40 $\pm$0.16 & 0.72 $\pm$0.28 \cr
          2802.7  &    46 $\pm$23 & 166 $\pm$20 & 268 & 0.66 $\pm$0.39 & 2.13 $\pm$0.39 \cr
          2795.5  &    46 $\pm$23 & 166 $\pm$20 & 311 & 0.75 $\pm$0.10 & 2.58 $\pm$0.40 \cr
          2599.4  &    76 $\pm$25 & 136 $\pm$25 & 217 & 0.48 $\pm$0.34 & 1.28 $\pm$0.32 \cr
          2585.9  &    76 $\pm$25 & 136 $\pm$25 & 286 & 0.65 $\pm$0.12 & 2.03 $\pm$0.37 \cr
          2382.0  &    93 $\pm$34 & 170 $\pm$34 & 305 & 0.81 $\pm$0.53 & 2.39 $\pm$0.46 \cr
          2373.7  &    93 $\pm$34 & 170 $\pm$34 & 292 & 0.47 $\pm$0.13 & 1.47 $\pm$0.44 \cr
          2343.5  &   158 $\pm$35 & 176 $\pm$35 & 848 & 0.93 $\pm$0.16 & 3.15 $\pm$0.54 \cr
          \colrule 
          \vspace{-3 mm} \cr
          \multicolumn{6}{c}{42020386} \cr
          2852.1  & X              &   X          &    X     &     X           &     X  \cr
          2802.7  & $-$258 $\pm$23   &   69 $\pm$22   &    $-$353  &    0.97 $\pm$0.29 &    1.02 $\pm$0.49 \cr
          2795.5  & $-$236 $\pm$21   &   70 $\pm$22   &    $-$333  &    1.10 $\pm$0.29 &    1.34 $\pm$0.45 \cr
          2599.4  & $-$167 $\pm$17   &   81 $\pm$18   &    $-$268  &    0.70 $\pm$0.20 &    1.65 $\pm$0.43 \cr
          2585.9  & $-$167 $\pm$17   &   81 $\pm$18   &    $-$294  &    1.17 $\pm$0.26 &    1.92 $\pm$0.42 \cr
          2382.0  & $-$109 $\pm$32   &   89 $\pm$32   &    $-$224  &    1.29 $\pm$0.40 &    2.29 $\pm$0.62 \cr
          2373.7  & X              &   X          &    X     &     X           &     X   \cr
          2343.5  & X              &   X          &    X     &     X           &     X   \cr
          \colrule 
          \colrule 
          \vspace{-3 mm} \cr
          \vspace{-3 mm} \cr
          \multicolumn{6}{c}{SDSS K+A Galaxies} \cr
          \vspace{-3 mm} \cr
          \colrule 
          \vspace{-3 mm} \cr
          \multicolumn{6}{c}{J022743} \cr
          \vspace{-3 mm} \cr
          2852.1 &  $-$286 $\pm$32  &    94 $\pm$32   &    $-$389   &    0.07 $\pm$0.02  &   0.14 $\pm$0.04 \cr
          2802.7 &  $-$386 $\pm$15  &   138 $\pm$15   &    $-$632   &    0.23 $\pm$0.02  &   0.71 $\pm$0.07 \cr
          2795.5 &  $-$509 $\pm$20  &   371 $\pm$20   &    $-$1192  &    0.27 $\pm$0.02  &   2.00 $\pm$0.08 \cr
          \colrule  
          \vspace{-3 mm} \cr
          \multicolumn{6}{c}{J215518\tablenotemark{5}} \cr
          \vspace{-3 mm} \cr
          2852.1  &  X & X & X & X & X \cr
          2802.7  &    75 $\pm$15 & 100 $\pm$14 & 225 & 0.92 $\pm$0.30 & 1.92 $\pm$0.25 \cr
          2795.5  &    75 $\pm$15 & 100 $\pm$14 & 203 & 0.82 $\pm$0.16 & 1.53 $\pm$0.33 \cr
          2599.4  &    30 $\pm$24 &  89 $\pm$23 & 102 & 0.81 $\pm$0.39 & 1.64 $\pm$0.40 \cr
          2585.9  &   X & X & X & X & X \cr
          \colrule  
          \vspace{-3 mm} \cr
          \multicolumn{6}{c}{J225656} \cr
          \vspace{-3 mm} \cr
          2852.1  & $-$224 $\pm$66 &    130 $\pm$65  &     $-$305   &  0.10 $\pm$0.04   &    0.22 $\pm$0.09 \cr
          2802.7  & $-$213 $\pm$14 &    142 $\pm$13  &     $-$439   &  0.53 $\pm$0.19   &	 1.60 $\pm$0.16 \cr
          2795.5  & $-$213 $\pm$14 &    142 $\pm$13  &     $-$441   &  0.60 $\pm$0.06   &	 1.71 $\pm$0.18 \cr
          \vspace{-3 mm} \cr
          \colrule 
          \colrule 
          \vspace{0.01 mm} \cr
	\end{tabular*}
}
\tablenotetext{1}{Velocity offset of center of Gaussian fit from systemic velocity}
\tablenotetext{2}{Velocity width of Gaussian fit}
\tablenotetext{3}{Maximum velocity used in EW measurement}
\tablenotetext{4}{$A_{flow}$ is 1 - the minimum line depth of the outflow component and corresponds to a lower limit on the covering fraction for optically thick lines}
\tablenotetext{5}{An `X' indicates spectral coverage of this line but no significant outflowing or inflowing absorption is detected}
\tablenotetext{6}{Possible `inflow' object; EW measured for the observed absorption without a systemic component removed}
\end{center}
\end{table*}

\subsection{Outflowing Winds Detected in Absorption}

Figures~6-8 show details of the observed and continuum-normalized spectra
for each galaxy in our sample centered on the \fetwo, \mgtwo, and \mgone \ 
absorption lines.  For each object we present two rows of plots: the upper
row is the fluxed spectrum (black) and stellar continuum fit (red), while
the lower panel is the continuum-normalized spectrum (black), with the 
continuum level (equal to unity) 
shown in green. These figures clearly show the variance
seen among different objects in our sample in terms of the absorption line
strengths and velocity profiles.  The figures also illuminate the importance 
of accounting for stellar absorption at \mgtwo \ and \mgone \ in objects 
with older stars, as the continuum-normalized spectra often have features, 
such as \mgtwo \ in emission, that are not clearly seen in the observed
spectra.

Table~5 
 lists the results of the absorption line fits and 
EWs of the \fetwo, \mgtwo, and \mgone \ lines 
for each galaxy in our sample that has an 
outflowing or possibly inflowing wind 
with an EW detected at the $\geq3\sigma$ level.  
We do not include objects in this table for which we did 
not measure a significant wind in absorption (after removing absorption
at systemic), 
and we list for each object only those lines
for which we had spectral coverage.  For line doublets that were fit 
simultaneously, the velocity width of both lines is constrained to be identical.
The quoted error on the \mgtwo \ outflow EW for galaxy 42020386 may be 
underestimated, as the error on the \mgtwo \ stellar continuum EW was 32\%.
However, this object also has significant absorption in \fetwo, which is 
not affected by the stellar continuum fit.

Five of nine X-ray AGN host galaxies, plus the LIRG (object 22029058),
 have detected outflowing winds detected in absorption, for a total of six
out of ten objects in this sample.
Five of these galaxies are in the blue cloud, and their spectra (Figure~2) 
show strong nebular emission lines.  The 
velocity centroids of the absorption features in these five galaxies are 
$\sim-$200 -- $-$500 \kms,
similar to the velocities found by \cite{Weiner09} and \cite{Rubin10a} for
star-forming galaxies at $z\sim1$.

\begin{figure}[tbp]
\epsscale{1.05}
\plotone{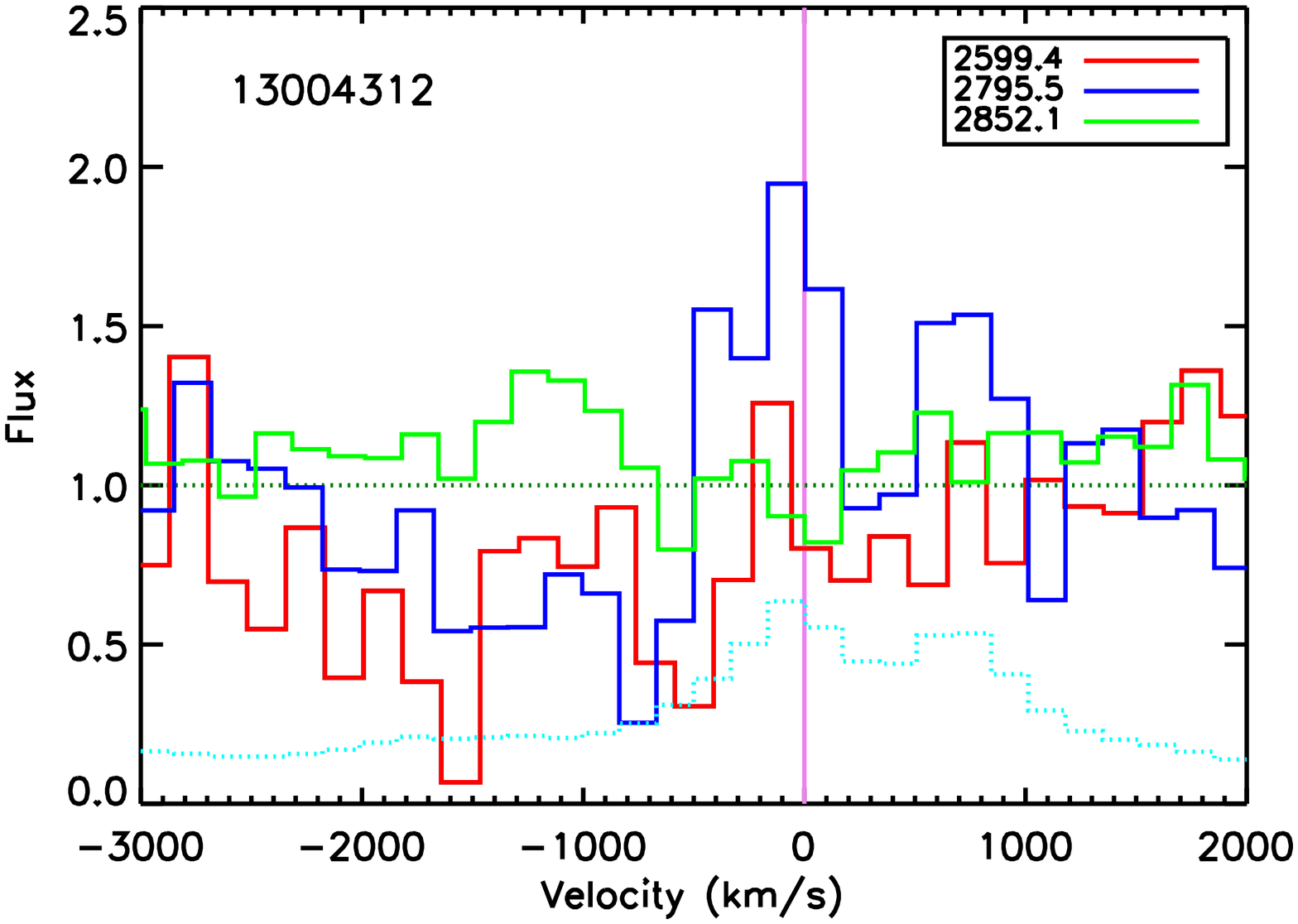}
\caption{\label{fig:extreme}
\small Absorption velocity profiles of \fetwo \ 2599 \AA \ (red), 
\mgtwo \ 2796 \AA \ (blue), and \mgone \ 2582 \AA \ (green) for object 13004312,
a red X-ray AGN host galaxy that may have a high velocity outflow.
The error spectrum for \mgtwo \ 2796 \AA \ is shown in cyan.
}
\end{figure}

One galaxy in our X-ray AGN host galaxy sample, 13004312, shows tentative 
evidence for a much higher velocity outflow.
The bluer \mgtwo \ line at 2796 \AA \ has a Gaussian fit with a 
velocity centroid of $-1225$ \kms, with
a width of $\sim500$ \kms. 
The velocity profiles for 
\fetwo \ 2599 \AA, \mgtwo \ (the bluer 2796 \AA \ line only), and \mgone \ 
for this object 
are shown in Figure~\ref{fig:extreme}.  
In this object \mgone \ does not have strong blueshifted absorption 
indicative of an outflow. 
The redder \mgtwo \ line at 2803 \AA \ does not show absorption, due to \mgtwo \
emission and the presence of the bluer \mgtwo \ line.
The absorption line fit at \fetwo \ 2599 \AA \ indicates a central velocity
of $-$550 \kms \ with a width of 157 \kms, but the EW measured is 
significant at only the 1.7$\sigma$ level.
The absorption at $\sim-$1600 \kms \ seen in the \fetwo \ line 
is due to \fetwo \ absorption at 2587 \AA. 
However, the \mgtwo \ 2796 \AA \ line shows a wide blueshifted absorption 
trough, 
extending out to $\sim-$2000 \kms \ before reaching the continuum level.  
We note that this is a red galaxy with an older stellar
population (see Figure~2) with significant, broad stellar \mgtwo \ absorption. 
The absorption feature in \mgtwo \ shown in Figure~\ref{fig:extreme} is 
apparent only after dividing out the stellar component (see Figure~6).  
This object also has clear \mgtwo \ emission 
(\mgtwo \ emission in the sample as a whole is discussed in detail
in Section 4.3 below). 
 The fractional error on the \mgtwo \ stellar absorption EW for this
object is $<$1\%, indicating that the profile seen after
removing stellar absorption is robust, in which case the extreme outflow 
detected in \mgtwo \ in this galaxy is also likely robust.
As the \fetwo \ 2599 \AA \ line, which is less affected by stellar 
absorption, shows that there is likely a wind at $\sim-$600 \kms, this 
suggests that the \mgtwo \ profile is valid.
We conclude that this object likely has an outflow with a
central velocity of at least $\sim-$600 \kms \ and may have a more extreme
outflow in \mgtwo.  

The typical velocity widths observed in the X-ray AGN sample are $\sim$100 -- 
300 \kms.  We note that given the resolution of these data, the absorption lines
are not resolved.  The maximum velocity at which 
we detect blueshifted absorption
is typically $\sim$500--800 \kms.  The covering fractions that we measure are 
lower limits, given the resolution of our data, and are generally within
the range of 0.1 -- 0.5.  The EWs span a range of values from 0.2 -- 2.5 \AA.

For our K+A samples, we detect outflowing winds in absorption 
in two of the seven DEEP2 
objects and two of six SDSS objects.  The velocity centroids of these four 
objects are between $-$130 -- $-$500 \kms, with widths of 65 -- 371 \kms. 
Given the resolution of our data, the lines are not resolved.  The maximum
velocities range from $-$225 -- $-$1192 \kms.  The covering fractions 
for the two SDSS K+A galaxies range from 0.1 -- 0.6.  For the DEEP2 K+A 
galaxies, the spectra are noisier (as the objects are much fainter) and 
the derived covering fraction can be greater than unity, though generally within
the error bars they are consistent with unity.  The EWs vary from 0.1 -- 2.0 
for the SDSS objects and 1.0 -- 7.6 for the DEEP2 objects.
We note that the presence of \mgtwo \ emission in many of these objects 
(discussed further in Section 4.3 below) may systematically affect our 
measurements of \mgtwo \ absorption \citep[see also][]{Prochaska11}.

On the whole, we detect winds via blueshifted
 \fetwo, \mgtwo, and \mgone \ 
absorption for 43\% of our sample.  
The velocity centroids of the winds are 
a few hundred \kms, much lower than the extreme winds seen by \cite{Tremonti07},
with the possible exception of one red X-ray AGN host galaxy  which
shows tentative evidence for an outflow with a centroid of $\sim-$600 -- $-$1200 \kms.

Two K+A galaxies in our sample (42021012 and J215518) show possible evidence 
for gas {\it inflow} along the line of sight.  Absorption line measures
for these two sources are given in Table~5.  For these two objects we
have {\it not} corrected for systemic absorption, as the velocity centroids 
are often measured to be positive (redward of systemic), such that the 
absorption is not symmetric about the systemic velocity.
Unlike for the other sources in this table, for these two objects we list the
maximum {\it redshifted} velocity of the absorption for each line.
For object 42021012, redshifted absorption is detected in both \mgone \ and
 \mgtwo, with a velocity centroid of 115 and 46 \kms, respectively.  The 
absorption detected for the various \fetwo \ lines is consistent with systemic.
 For object J215518 redshifted absorption is detected in \mgtwo, while 
blueshifted absorption is detected in a single \fetwo \ line.  
We further note that object 32008909 (not included in this table) has a 
complicated \mgtwo \ absorption profile that may indicate both blueshifted 
and redshifted gas (see Figure 7).  
The redshifted gas seen in absorption in these sources 
could be inflowing cool gas from a 
galactic fountain, gas accretion due to minor mergers, or gas cooling from 
the halo of the galaxy. The relatively high EW ($\sim$1-2 \AA) of the
redshifted gas should constrain its origin; in particular, this may be
higher than what is expected for gas cooling from the halo.

\subsection{Dependence of \mgtwo \ Absorption on Galaxy Properties}

\begin{figure*}[tbp]
\plottwo{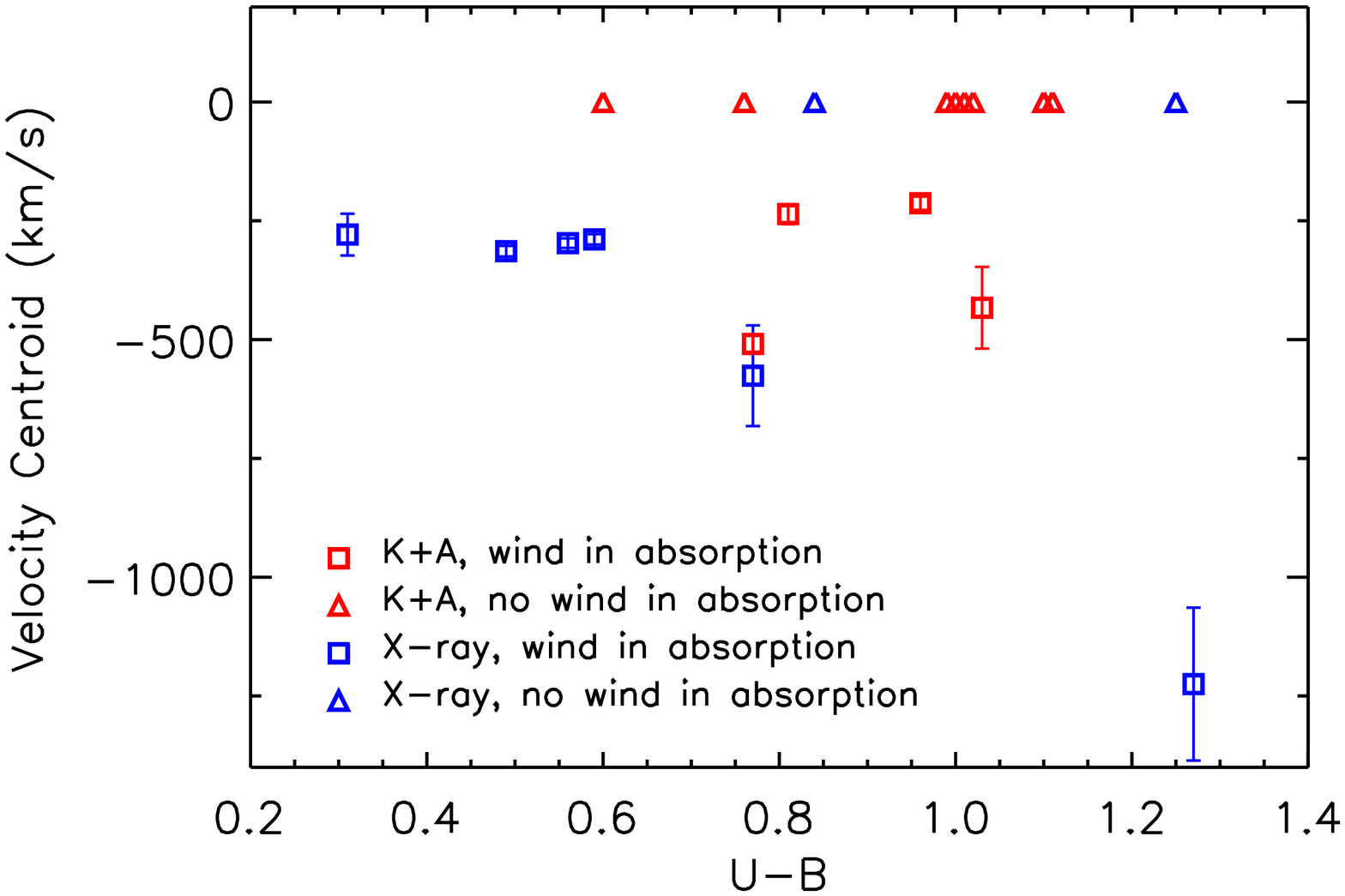}{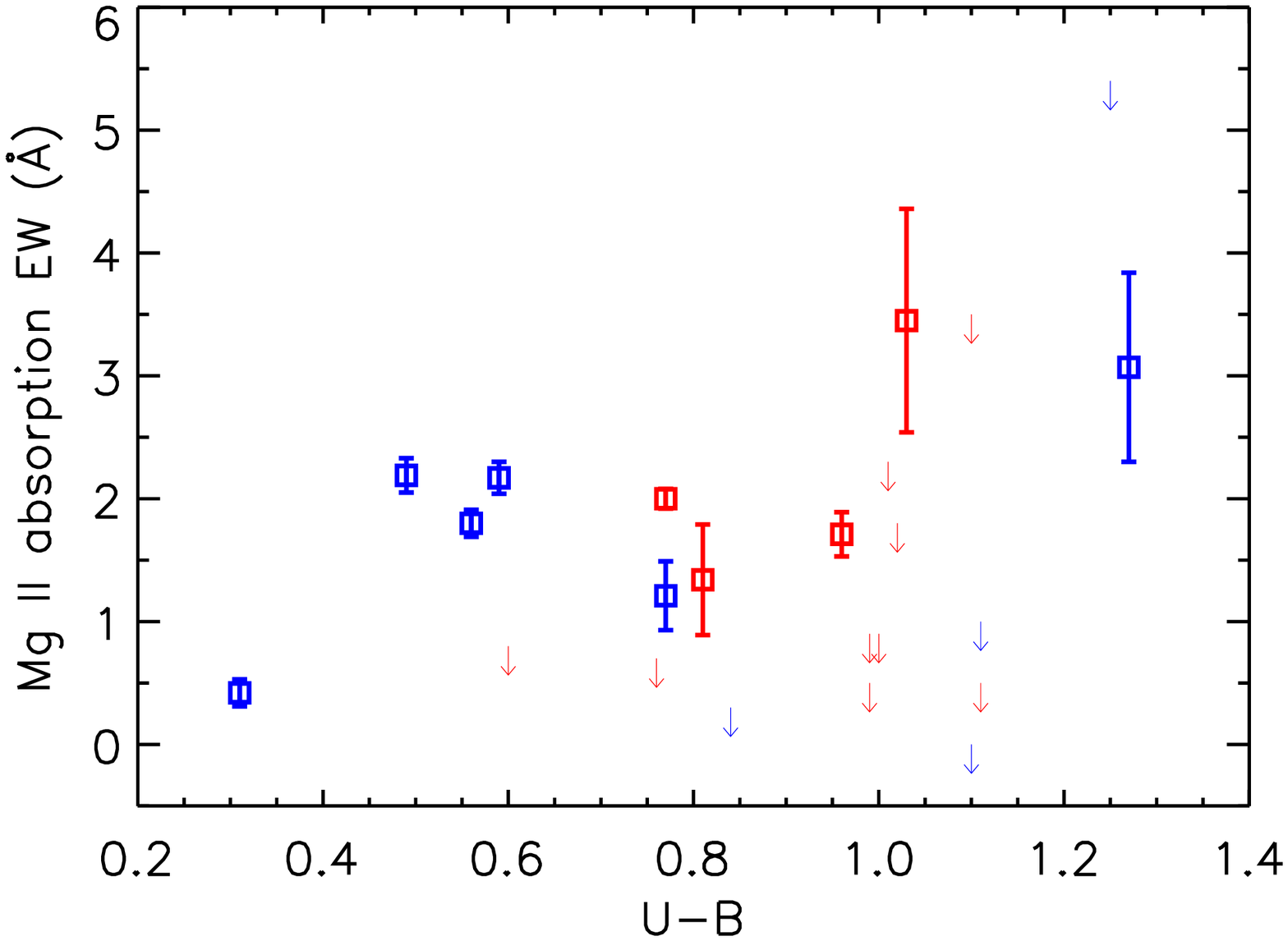}
\caption{\label{fig:colorew}
\small Velocity centroid (left) and equivalent width (right) of the \mgtwo \ 
2795.5 \AA \ absorption line versus restframe $U-B$ color for each object in 
our sample. 
 The equivalent width shown is for the outflow component of the absorption
line only, where any systemic contribution has been removed (see Section 3.2 
for details).
K+A galaxies are shown in red and X-ray AGN host galaxies are shown
in blue. Galaxies without blueshifted \mgtwo \ absorption are shown as 
triangles.
We have subtracted $U-B=$0.14 from the colors of the 
SDSS K+A galaxies to match the colors of the DEEP2 galaxies and remove the
effects of passive evolution.
In the right panel 2$\sigma$ upper limits on the EW are given for objects
without detected blueshifted absorption.}
\end{figure*}

In Figure~\ref{fig:colorew} we compare the \mgtwo \ 2796 \AA \ velocity
centroid (left) and EW (right) with the galaxy $U-B$ color.
Following \citet{Yan09}, who study K+A samples selected in both DEEP2
and SDSS, in this figure we put both samples on common ground by shifting 
the $U-B$ colors of the SDSS objects 
blueward by $\delta (U-B)=$0.14 to account for passive evolution between 
the SDSS and DEEP2 samples.  In the left panel, 
galaxies without detected blueshifted \mgtwo \ absorption are shown 
as triangles with velocities of 0 \kms, and in the right panel 2$\sigma$ 
EW upper limits are shown for these objects.


\begin{figure*}[tbp]
  \plottwo{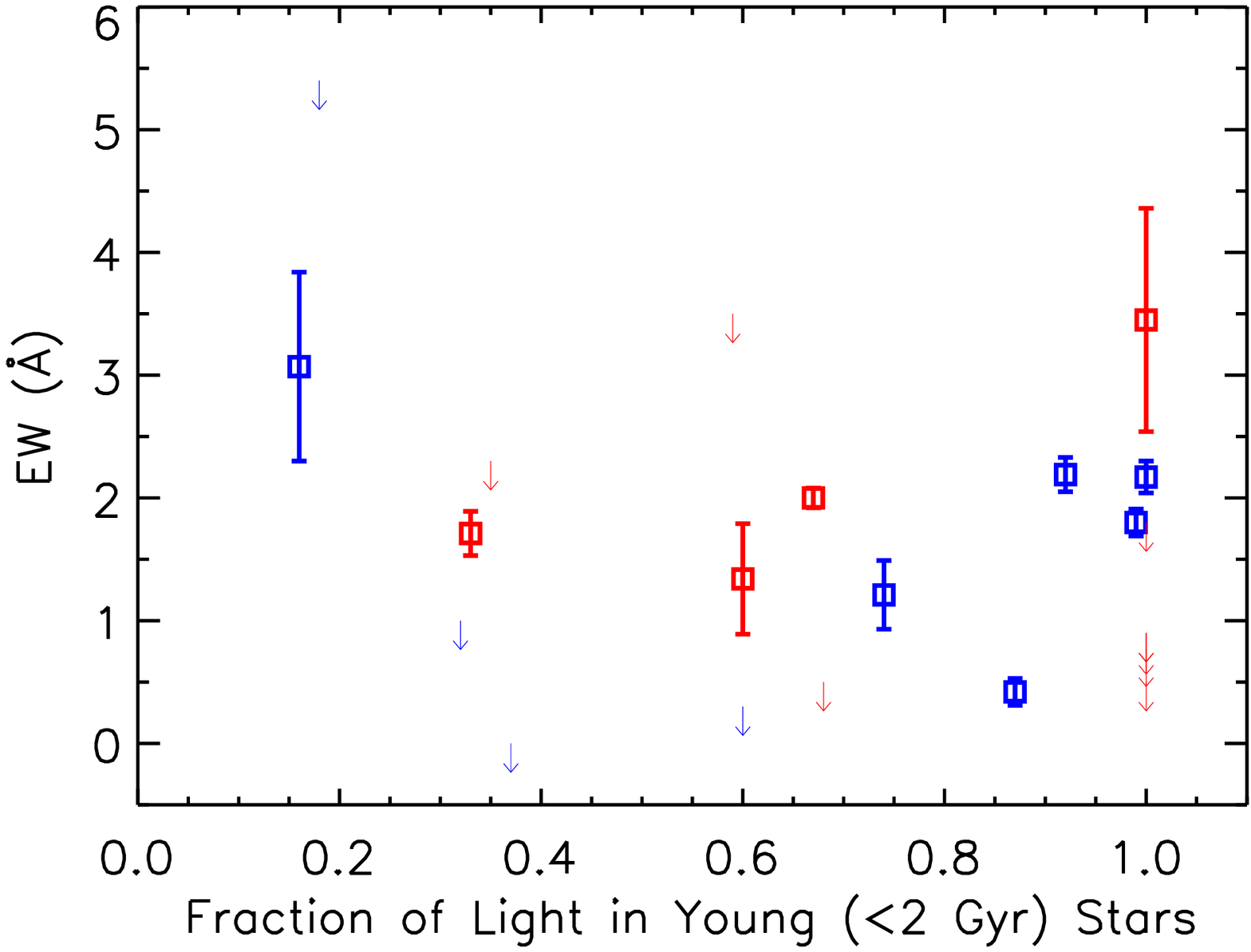}{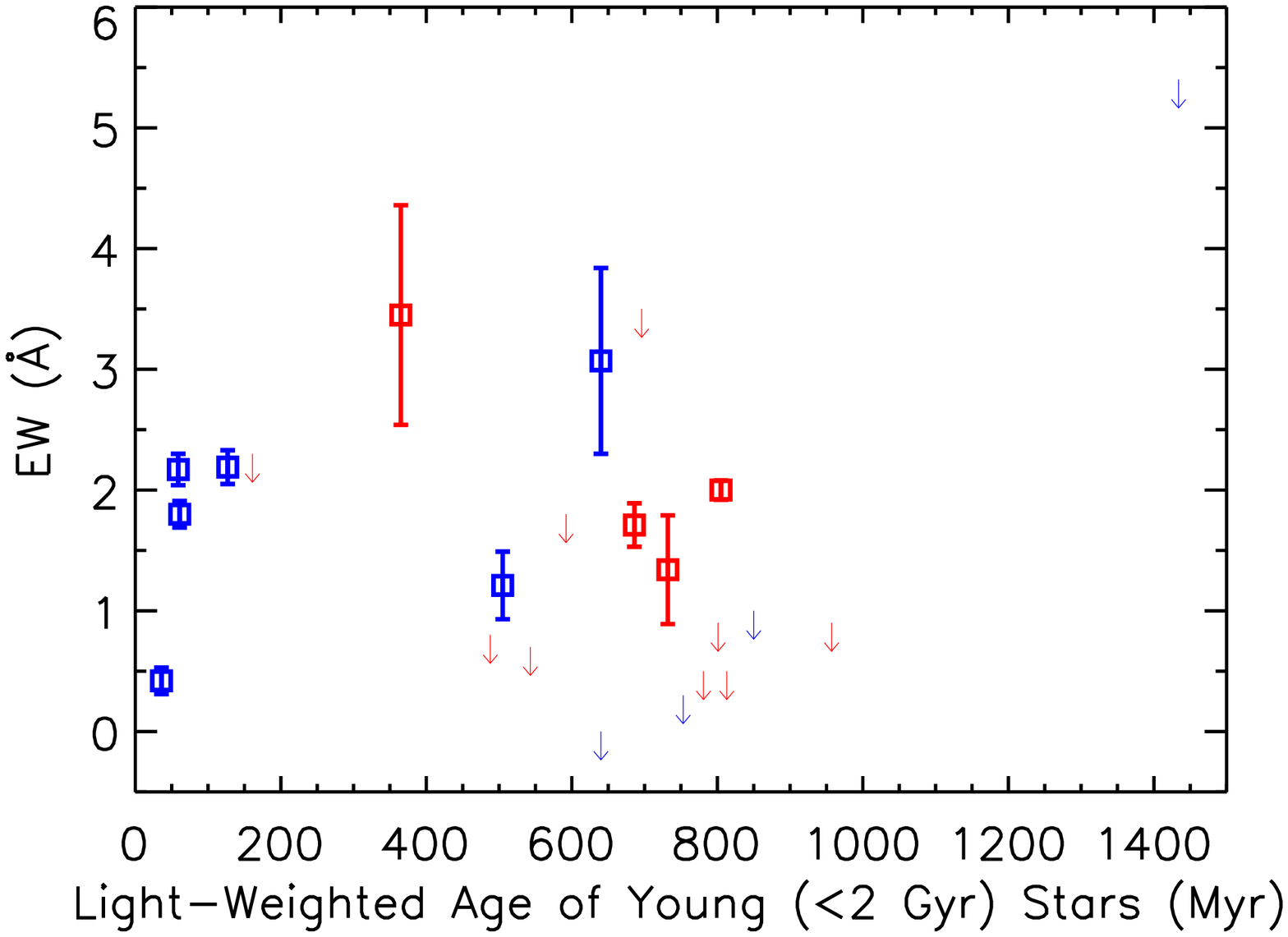}
\caption{\label{fig:ageew}
\small EW of the \mgtwo \ 2795.5 \AA \ absorption line versus the fraction
of light in young ($<$2 Gyr) stars (left) and the light-weighted age of 
the young stars (right).  
 As in the previous figure, the equivalent width shown is for the outflow component of the absorption
line only, where any systemic contribution has been removed.
K+A galaxies are shown in red and X-ray AGN host 
galaxies are shown in blue. 2$\sigma$ upper limits on the EW are given for 
objects without detected blueshifted absorption.}
\end{figure*}

Figure~\ref{fig:colorew} shows that the 
detection rate of winds via blueshifted \mgtwo \ absorption is higher 
for the blue galaxies with $U-B<1.0$.  
At $U-B\ge1.0$ only two of the ten galaxies
have winds detected in absorption.  However, the spectral S/N is lower for the
red galaxies, on the whole, such that some of the EW upper limits are not 
very constraining. Of the galaxies with winds detected in absorption, we do
not find a strong correlation between the velocity centroid and the galaxy
color, with the exception of the single red galaxy with a high velocity
outflow.  
The apparent trend in the right panel that the EW is higher for
redder galaxies may be a selection effect, as due to the lower S/N of the
red galaxy spectra we would not be able to detect low EW outflows.

We find that the velocity centroids and EWs of the \mgtwo \ absorption 
are similar for the AGN host galaxies and the K+A galaxies; within our small
samples, neither population appears to have faster winds or higher EW winds.
This is discussed further in Section 6.3 and 6.4.

In Figure~\ref{fig:ageew} we investigate the correlation of the 
\mgtwo \ 2796 \AA \ absorption EW with the light-weighted 
age of the stars in the galaxy,
comparing with the fraction of light due to 
young ($<$2 Gyr) stars on the left and the
light-weighted age of the young stars on the right.  
Most (7/9) detected winds in absorption are seen in galaxies with more 
than 50\% of their light due to young stellar populations.  Within that
population, however, we do not find a correlation between EW and the 
fraction of light in young stars.
Similarly, all detections are in galaxies in which the young stars have an
 age less than 800 Myr, though we do not find a correlation between EW and age 
of the young stars.  We do find that the detection fraction is a 
function of the light-weighted age of the young stars, with higher
detection fractions found for younger stellar populations.  For the K+A 
galaxies in our sample, the light-weighted age of young stars may be
interpreted roughly as the time since the last episode of star formation, 
and the observed trend is that the likelihood of detecting a wind lowers
once the time since the last burst is $\gtrsim$500 Myr.  
The five X-ray AGN host galaxies in this figure with light-weighted
young ages $<$500 Myr have blue colors and emission lines indicative of 
on-going star formation, however, such that the light-weighted age is not 
strictly the time since the last episode of star formation.

\begin{figure*}[tbp]
\epsscale{1.05}
\plottwo{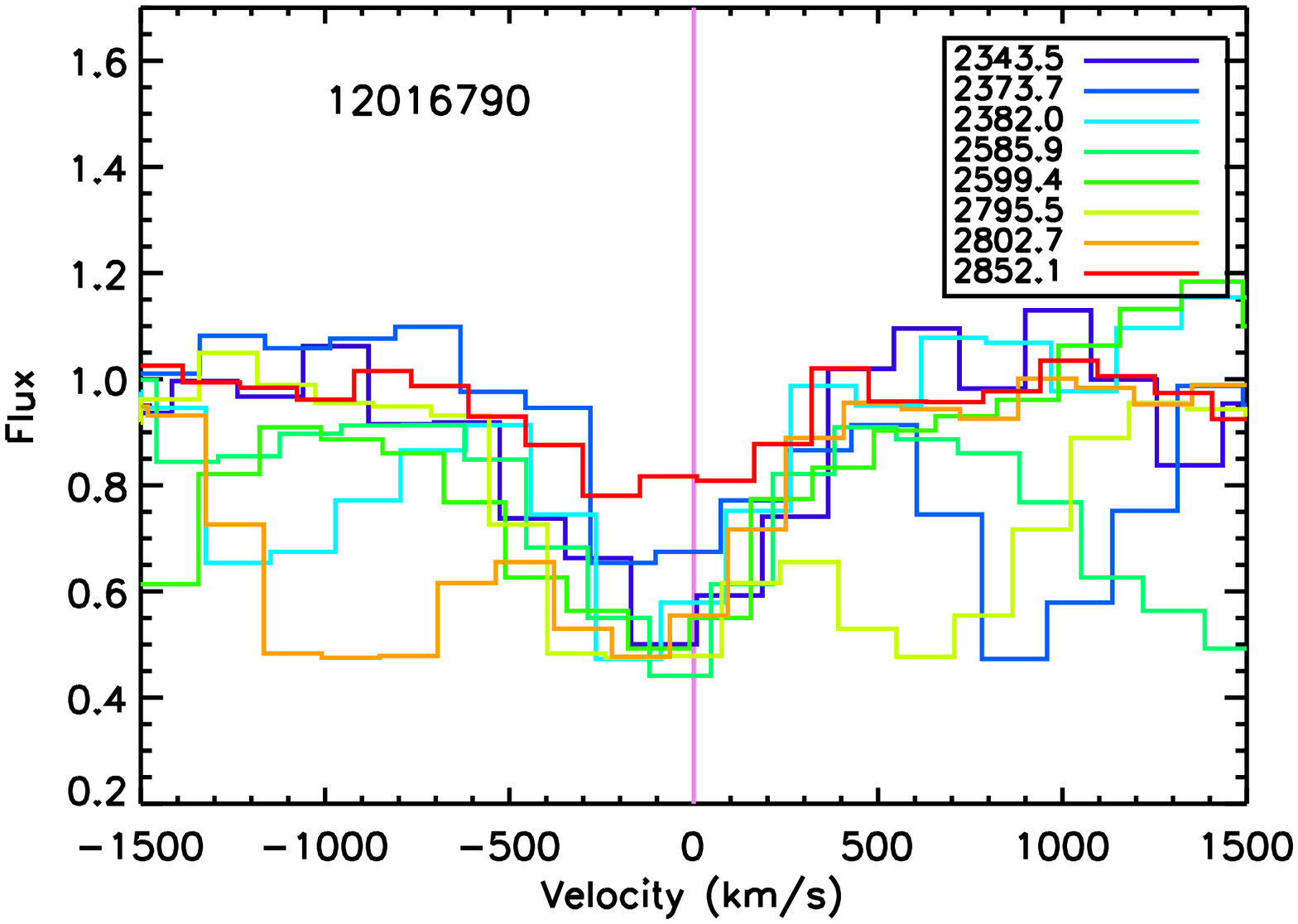}{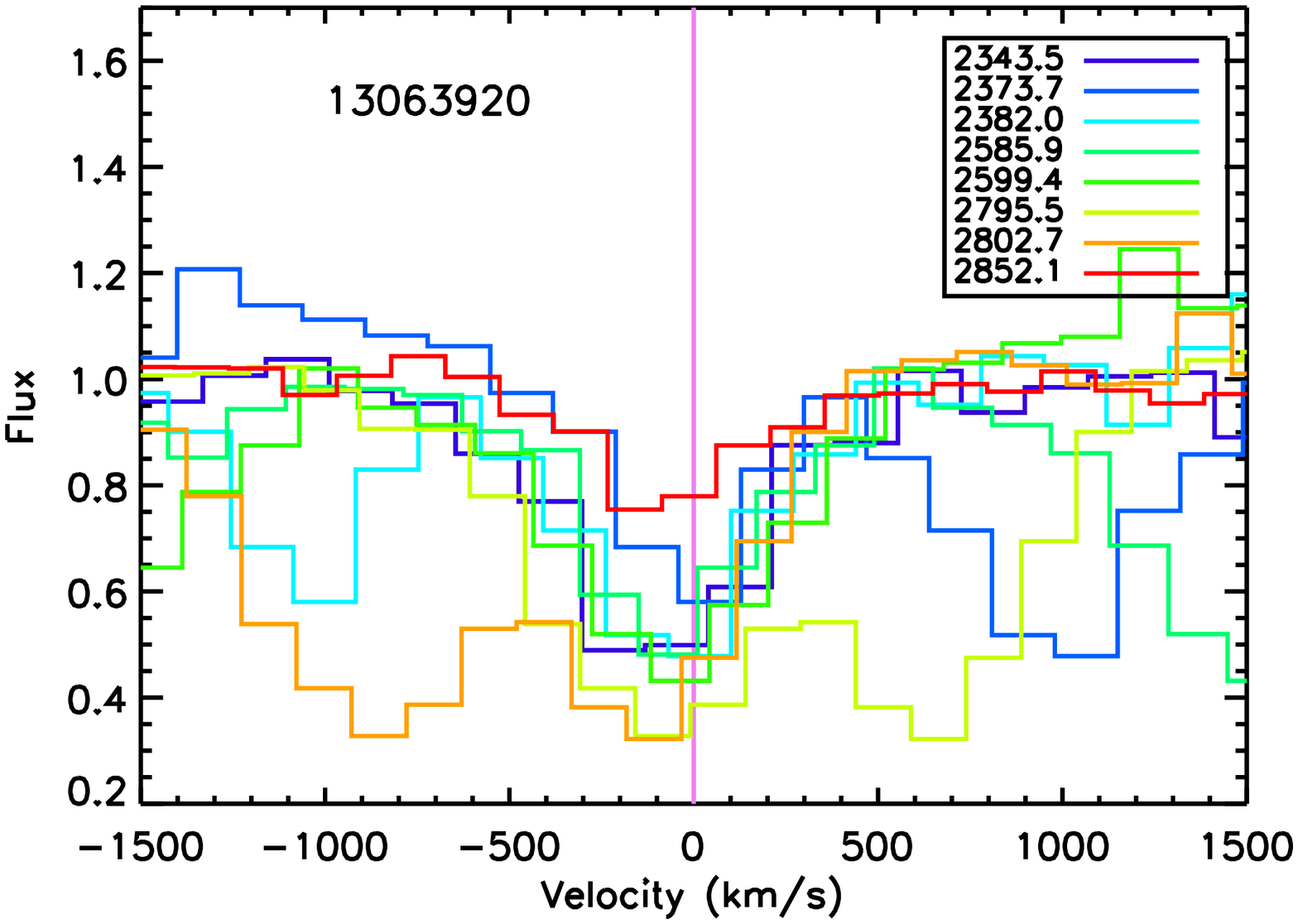}
\caption{\label{fig:velprof}
\small Comparison of the absorption profiles for all \fetwo, \mgtwo, and 
\mgone \ lines covered for two objects with high S/N 
in our X-ray AGN host galaxy sample. We show the observed spectra here, without
removing the systemic component.}
\end{figure*}

\subsection{Comparison of Absorption Lines}

We next study the absorption seen at 
\fetwo, \mgtwo, and \mgone \ 
on an object-by-object basis, 
comparing the absorption profiles and Gaussian fit 
results from the different ions and ionization states.  
Figure~\ref{fig:velprof} shows the velocity profile of 
each line for two of the X-ray AGN host galaxies in our sample.
Each line of the \mgtwo \ doublet (which has a velocity separation of 770 \kms) 
 is shown in green-yellow and
orange colors, which causes the absorption at $\sim-$900 \kms \ in
orange and $+$600 \kms \ in green-yellow.
The two spectra shown here have relatively high S/N, which facilitates a 
by-eye comparison of their absorption troughs.  Clearly at this resolution the 
general shape is very similar between 
\fetwo, \mgtwo, and \mgone,
 though the minimum line depth can vary between the lines, with
\mgone \ having less absorption in both of the objects shown.
For \mgone \ the lower absorption could be due either to a lower covering fraction 
or a lower optical depth or both.  As \mgone \ is a singlet we can not distinguish 
between these scenarios, unlike for the \mgtwo \ doublet, which is saturated 
and therefore optically thick. The line profiles for the \fetwo \ transitions 
are similar with each other and with \mgtwo, indicating that the lines are either 
saturated or nearly so \citep[e.g.,][]{Martin09}.

\begin{figure*}[tbp]
\epsscale{1.05}
\plottwo{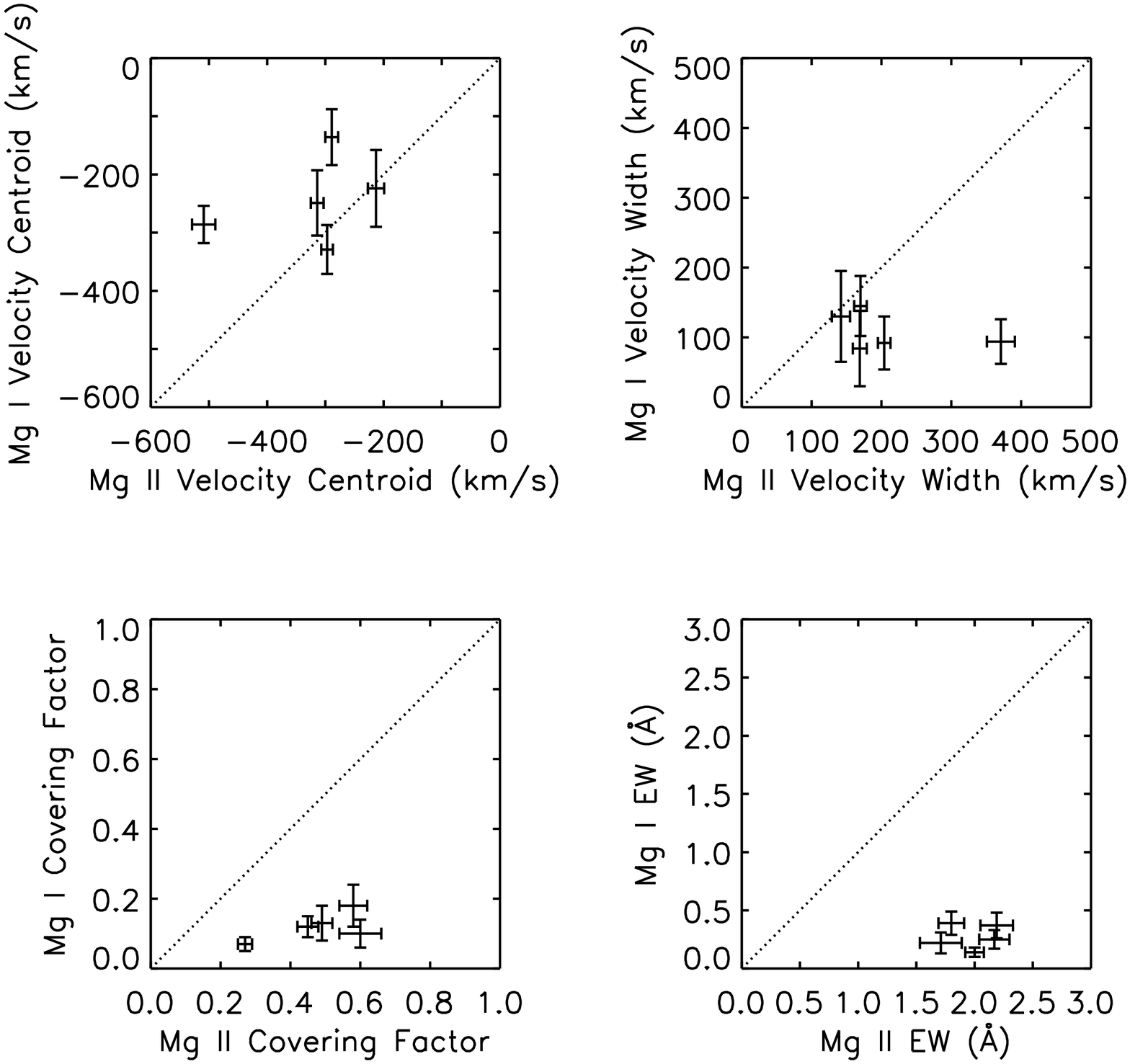}{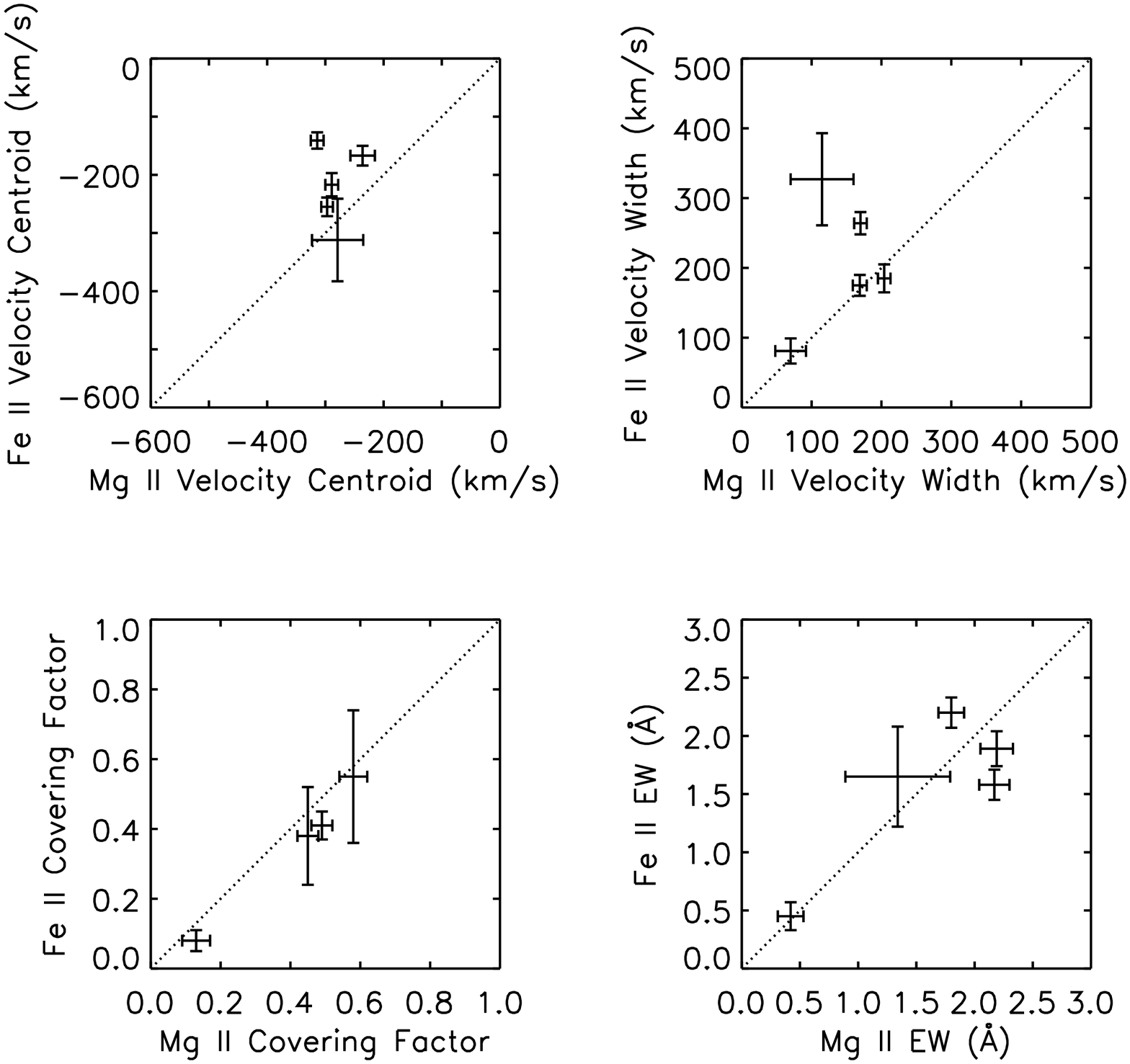}
\caption{\label{fig:linecomp}
\small Comparison of absorption line fits for \mgone \ 2852.1 \AA \ 
versus \mgtwo \ 2795.5 \AA \ (left) and \fetwo \ 2599.4 \AA \ 
versus \mgtwo \ 2795.5 \AA \ (right).}
\end{figure*}

Figure~\ref{fig:linecomp} quantitatively compares the results 
in Table~5 
 for 
the absorption line measurements---central velocity, velocity
width, covering fraction, and EW---among the different lines for 
individual objects.  
Here we treat the $A_{flow}$ values listed in Table~5 as the covering
fraction.
The left panel compares \mgone \ with the 
bluer \mgtwo \ line, while the right panel compares \fetwo \ 2599.4 \AA \
with the bluer \mgtwo \ line.  We have also compared (but do not show)
the various \fetwo \ lines against each other.
We find that
the velocity centroids, velocity widths, covering fractions and EWs all agree
reasonably well both between the various \fetwo \  lines and between the \fetwo \
lines and the bluer \mgtwo \ line, within the measured errors.  The covering
fraction of \fetwo \ 2599.4 \AA \ appears to be systematically slightly lower
than for \mgtwo.    We do not 
consider the redder \mgtwo \ line in this comparison as the absorption trough 
may be affected by the bluer \mgtwo \ line.  

In comparing \mgone \ and \mgtwo, we find that 
\mgone \ has a significantly lower covering fraction and/or optical depth 
and a lower EW  (which is closely tied to
the covering fraction in our unresolved data) 
than either \mgtwo \ or \fetwo.  
Additionally, \mgone \ often has a lower velocity centroid and width than 
either \mgtwo \ or \fetwo.  
As \mgone \ has a lower ionization potential than \mgtwo, 7.6 eV compared 
to 15.0 eV, it is likely that \mgone \ traces denser gas in smaller clumps,
which have a lower covering fraction.
\citet{Weiner09} also found that the EW and covering fraction of \mgone \ 
was smaller than \mgtwo \ in DEEP2 star-forming galaxies at $z\sim1.4$.

Comparing the various \fetwo \ lines with each other, we find that 
2599.4 \AA \ and 2585.9 \AA \ agree well in terms of velocity centroid, 
velocity width, and 
covering fraction, while the EW of 2599.4 \AA \ is generally slightly
higher than 2585.9 \AA.
Comparing \fetwo \ 2599.4 \AA \ with 2343.5 \AA, we find that both lines 
agree well in terms of velocity centroid and covering fraction, while for 
velocity width and EW 2599.4 \AA \ is a bit higher than 2343.5 \AA.  
However, our sample
size for this comparison is only a few data points, making robust conclusions
difficult.  We stress again that these data are not high resolution, which one
would ideally want for this kind of comparison.  Additionally, 
there can be emission systematically affecting our measurements, 
for \mgtwo \ in particular (discussed below).

\subsection{Na D}

The LRIS spectra for our SDSS K+A sample extend 
far enough in the red 
to cover the Na D 5889.95, 5898.92 \AA \ 
doublet, which is often used in low redshift galaxies to search
for winds.  In our two SDSS K+A galaxies in which we detect winds in \mgtwo \
absorption, we find that in one object 
(J022743) there is strong emission at Na D, centered on the systemic velocity.
In the other object (J225656) we find evidence for an outflowing wind in 
the Na D absorption profile.  In our other four SDSS K+A galaxies Na D 
either is seen in emission at the systemic velocity or has no detected 
emission or absorption.

 As Na D is a resonance absorption transition, if 
Na D is absorbed along certain sight lines due to an anisotropic wind 
(e.g., a biconical outflow), 
this can lead to isotropic Na D emission.
Na D emission is observed by \citet{Chen10} in their SDSS sample of
star-forming galaxies, primarily in face-on galaxies with little dust.
They suggest that the Na D emission is from the back side of an 
expanding bipolar outflow.  They also note that for some objects the 
emission is observed only after dividing out the stellar continuum; this 
is also the case for our object, J022743.  As discussed in detail in 
\citet{Chen10}, while one cannot rule out continuum fitting errors,
the emission may be real and from the wind itself.

In object J225656 the Na D doublet is consistent with a wind with the same 
velocity parameters found from the \mgtwo \ and \mgone \ lines: the velocity 
centroid is $-$158 \kms \ 
with a width of 81 \kms.  
The covering fraction is 0.07 $\pm$0.01, 
similar to what is found for \mgone \ (0.10 $\pm$0.04) and lower than 
 \mgtwo \ (0.53-0.60).

\section{Results on Winds Detected in Emission} \label{sec:resultsemis}

We now turn to emission line detections of outflowing winds in our sample, 
studying detections of both \mgtwo \ and \fetwo* emission in our spectra.

\subsection{\mgtwo \ Emission}

To observe \mgtwo \ emission, there must be scattered emission that is 
not re-absorbed; to produce such emission 
it therefore helps if there is a velocity gradient and 
non-isotropic illumination of the gas (e.g., outflowing material 
illuminated by a central source, observed along a sight line to the 
gas but not the central source).
\mgtwo \ emission is not generally seen in \ion{H}{2} \ regions, 
as it is disfavored by a factor of the abundance. 
As discussed in \cite{Weiner09} and \cite{Rubin11}, \mgtwo \ can often be seen in both
emission and absorption with a P-Cyngi-type profile in objects with outflowing 
winds, where the emission component is presumably from the back side of the 
wind \citep[see also ][for an example in Na I]{Phillips93}.  
This idea is modeled by \cite{Prochaska11} using radiative transfer 
calculations for a simple wind model.  
 Using this model they predict that the emission EW should equal the 
absorption EW for a galaxy with little dust and an isotropic outflow 
which is fully covered by the spectrograph slit; in the presence of 
dust the emission EW decreases.

\begin{table}[]
\tablewidth{0pt}
\begin{center}
\label{tab:mgews}
\small
\footnotesize
\caption{\small \mgtwo \ Emission Line Measurements and Equivalent Widths}
\begin{tabular}{lrrrr}
\colrule
\colrule
\vspace{-3 mm} \cr
Object & \ Velocity &  Velocity  & 2795.5 \AA \ \ & 2802.7 \AA \ \ \cr
       &  centroid  &  width \    & EW    \ \         & EW   \ \           \cr
       & (km/s)     &  (km/s)    & (\AA)  \ \         &  (\AA)  \ \      \cr
\vspace{-3 mm} \cr
\colrule
\colrule
\vspace{-3 mm} \cr
\multicolumn{5}{c}{DEEP2 X-ray AGN Host Galaxies} \cr
\vspace{-3 mm} \cr
\colrule
11046507 &    238 $\pm$100 &  174 $\pm$99  &  $-$0.43 $\pm$0.12 & $<$0.24\tablenotemark{1} \cr
13004312 &  $-$83 $\pm$76  &  151 $\pm$68  &  $-$3.83 $\pm$1.80 & $-$2.02 $\pm$1.51 \cr
13025528 &  $-$88 $\pm$40  &  235 $\pm$30  & $-$12.46 $\pm$1.99 & $-$8.84 $\pm$1.88 \cr
13043681 &     16 $\pm$126 &  107 $\pm$104 &  $-$5.09 $\pm$3.47 & $-$11.22 $\pm$3.90 \cr
13051909 & $-$155 $\pm$52  &  212 $\pm$40  &  $-$3.22 $\pm$0.73 & $-$2.57 $\pm$0.66 \cr
\vspace{-3 mm} \cr
\colrule
\vspace{-3 mm} \cr
\multicolumn{5}{c}{DEEP2 K+A Galaxies} \cr
\vspace{-3 mm} \cr
\colrule
32003698 & 290 $\pm$53 & 205 $\pm$42 & $-$7.70 $\pm$2.59 & $-$6.47 $\pm$1.58 \cr  
43030800 & 39 $\pm$25 & 68 $\pm$24 & $-$3.11 $\pm$0.87 & $<$1.88 \cr
\vspace{-3 mm} \cr
\colrule
\vspace{-3 mm} \cr
\multicolumn{5}{c}{SDSS K+A Galaxies} \cr
\vspace{-3 mm} \cr
\colrule
J022743 &     46 $\pm$13 &  51 $\pm$13 & $-$0.14 $\pm$0.04 & $-$0.17 $\pm$0.04 \cr
J210025 &  $-$70 $\pm$28 & 102 $\pm$27 & $-$1.17 $\pm$0.28 & $-$0.63 $\pm$0.22 \cr
J212043 &     19 $\pm$21 &  80 $\pm$22 & $-$0.98 $\pm$0.24 & $-$0.38 $\pm$0.19 \cr
J215518 & $-$260 $\pm$36 &  85 $\pm$34 & $-$0.95 $\pm$0.41 & $<$0.82  \cr
\vspace{-3 mm} \cr
\colrule
\colrule
\vspace{-8 mm} \cr
\tablenotetext{1}{2$\sigma$ upper limits are given for the few objects in which
the redder \mgtwo \ line was not detected.}
\end{tabular}
\end{center}
\end{table}

\begin{figure}[tbp]
\epsscale{1.2}
\plotone{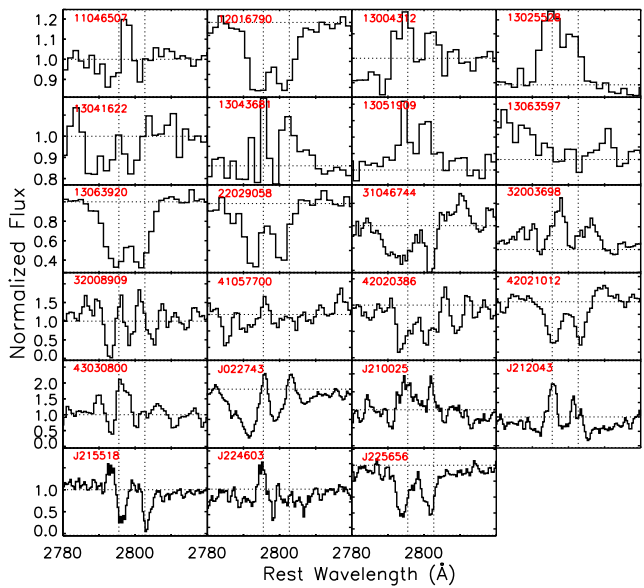}
\caption{\label{fig:mgemis}
\small Region around the continuum-normalized \mgtwo \ flux 
for each object in our sample. 
The y-axis range varies for each object.
There is a 
wide range of \mgtwo \ properties, ranging from strong absorption to strong 
emission. The K+A spectra have been smoothed by a boxcar of width three pixels
for this figure.}
\end{figure}

We show in Figure~\ref{fig:mgemis} the region around \mgtwo \ for each
of the objects in our sample.  As can be seen in the figure, there is a wide
range of \mgtwo \ profiles and properties, ranging from what appears to be
pure emission to pure absorption.  Table~6 
 lists emission
line measurements for those objects with significant ($>2\sigma$) 
\mgtwo \ emission
detected above the continuum level.  To determine these measurements we
fit double Gaussian line profiles in emission, where the two lines were 
constrained to have the same velocity width.  
 To obtain fits to the \mgtwo \ emission lines we use data 
within $\pm$40~\AA \ of the observed wavelength corresponding to 
the systemic velocity of each line in the doublet.  
This corresponds to roughly $\sim$3600 \kms \ for 
the $z\sim0.2$ objects and $\sim$2400 \kms \ for the $z\sim0.8$ 
objects.  EWs are 
measured within the velocity range where the Gaussian fit is within 
1\% of the continuum level of the spectrum.
Eight of the objects listed
have \mgtwo \ emission detected at $>3\sigma$, while another three objects
have emission detected at a significance between 2$\sigma$ and 3$\sigma$.
We detect \mgtwo \ emission
in half of our sample: five of the ten X-ray AGN host galaxies and six of the 
thirteen K+A galaxies.  The velocity centroids are consistent (less than 
3$\sigma$ deviant) with the systemic velocity for seven of the galaxies,
while two galaxies have blueshifted emission and two have redshifted emission.
 Presumably the redshifted emission is from the back side of the wind,
while the blueshifted emission is from the front side of the wind (possibly in
galaxies with dust that obscures the back side) and the emission at 
systemic is integrated over the entire geometry of the wind.

Of the eleven sources with emission detected in \mgtwo, three have detected 
blueshifted \mgtwo \ absorption as well and eight do not.  The eight 
galaxies with \mgtwo \ emission but no blueshifted absorption are all 
either in the green valley or on the red sequence; they are not star-forming
galaxies.  This implies that the \mgtwo \ emission is not from young 
stars.  All three of the galaxies in our sample that 
lie on the red sequence (all of which are X-ray AGN host galaxies) have \mgtwo 
\ emission detected (one at 2$\sigma$ and two at 3$\sigma$).

We looked in detail at all objects with \mgtwo \ emission that had
potentially large continuum errors to verify whether the emission
could be due to errors in the stellar continuum fit.  Object 13051909,
an X-ray AGN host galaxy on the red sequence, has a \mgtwo \ stellar
absorption EW error of 20\%.  Normalizing the observed spectrum by a
stellar continuum fit that has a \mgtwo \ EW that is low by
$\sim$1$\sigma$ results in \mgtwo \ emission in the
continuum-normalized spectrum significant at the 2.6$\sigma$ level; we
conclude that the \mgtwo \ emission in this object is likely real.
Within the DEEP2 K+A sample, object 43030800 (also on the red
sequence) has a 26\% error in the \mgtwo \ stellar absorption
EW. 
 Repeating the above test of renormalizing the observed spectrum by a
stellar continuum fit that is low by 
$\sim$1$\sigma$ results in \mgtwo \ emission significant at the 
$\sim$2$\sigma$ level. 
For the other DEEP2 K+A
galaxy with \mgtwo \ emission, the DEIMOS \mgtwo \ stellar EW is
within 6\% of the LRIS \mgtwo \ stellar EW, such that the continuum
error is subdominant.

For the SDSS sample, as discussed above the statistical errors in the
continuum fits are negligible.  However, as systematic errors in the
continuum fits to the LRIS spectra could lead to overestimated
emission, we checked all of these objects in detail.  In J022743, the
\mgtwo \ profile for this object shows a P-Cygni signature, indicating
that the \mgtwo \ emission in likely real.  For J210025, the \mgtwo
\ stellar EW in the SDSS fit is larger than in the LRIS fit, such that
using the SDSS stellar continuum fit would only lead to more emission
in the continuum-normalized spectrum.  For J212043 and J215518, the
fractional difference in the \mgtwo \ stellar EWs in the SDSS and LRIS
fits are well within the errors on the quoted \mgtwo \ emission in the
continuum-normalized spectra.  We therefore conclude that the \mgtwo
\ emission in all of these objects is likely real.

\cite{Weiner09} find that from their parent sample of $\sim$1500 star 
forming galaxies at $z=1.4$, $\sim$50 galaxies have detected \mgtwo \ emission.
These galaxies tend to be bright and blue, 
lying in the bluer half of the blue cloud with $U-B<0.6$ (see their Figure~7).
Their selection criteria of objects with \mgtwo \ emission 
may have selected the source with the highest emission EW, 
however, and so may not be directly comparable with the
selection used here.  We note that if the presence of \mgtwo \ emission 
is correlated with a lack of dust \citep{Prochaska11}, then it is 
plausible that both the bluest and reddest galaxies would show \mgtwo \ 
emission, as these are the galaxies expected to have less dust.
If this emission is indeed from the back side of the 
wind, including the objects in which \mgtwo \ is 
detected in emission raises the fraction of our sample with detected winds to 
83\% (19/23), including nine of the ten X-ray AGN host galaxies and nine of 
the thirteen K+A galaxies.  

As discussed in \cite{Prochaska11}, the presence of \mgtwo \ emission
can affect the \mgtwo \ absorption profile, which we use 
to measure kinematic properties of the outflow.  
The presence of emission
line filling can bias the absorption line fits in that it can lower
the measured EW, shift the velocity centroid to larger values, and lower 
the velocity width.  The effect is lessened or absent in the \fetwo \ lines,
however, and given the good agreement between the kinematics observed in
the different ions in section 4.1.2, it is likely not a major effect here.

\begin{figure}[tbp]
\epsscale{1.05}
\plottwo{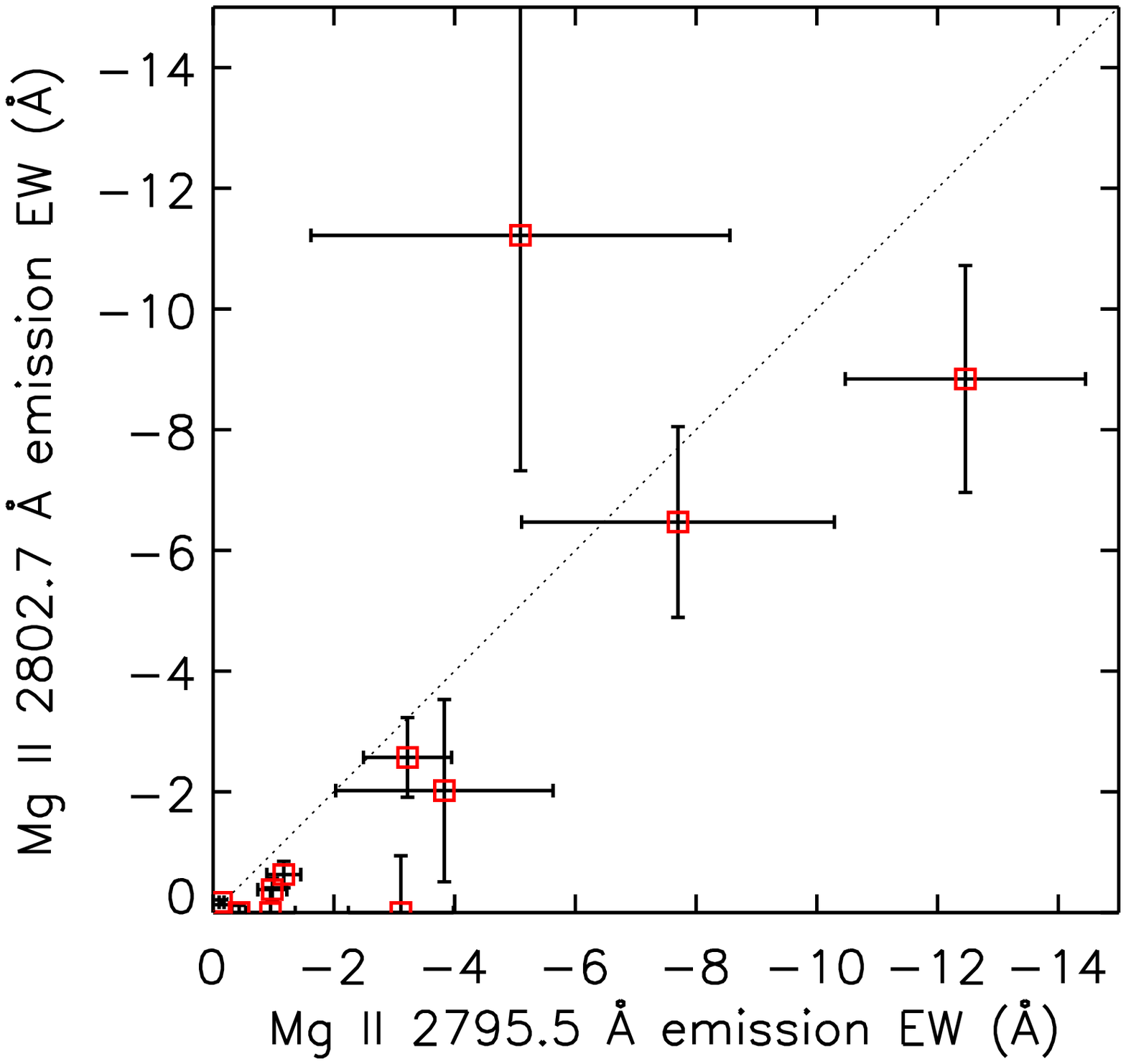}{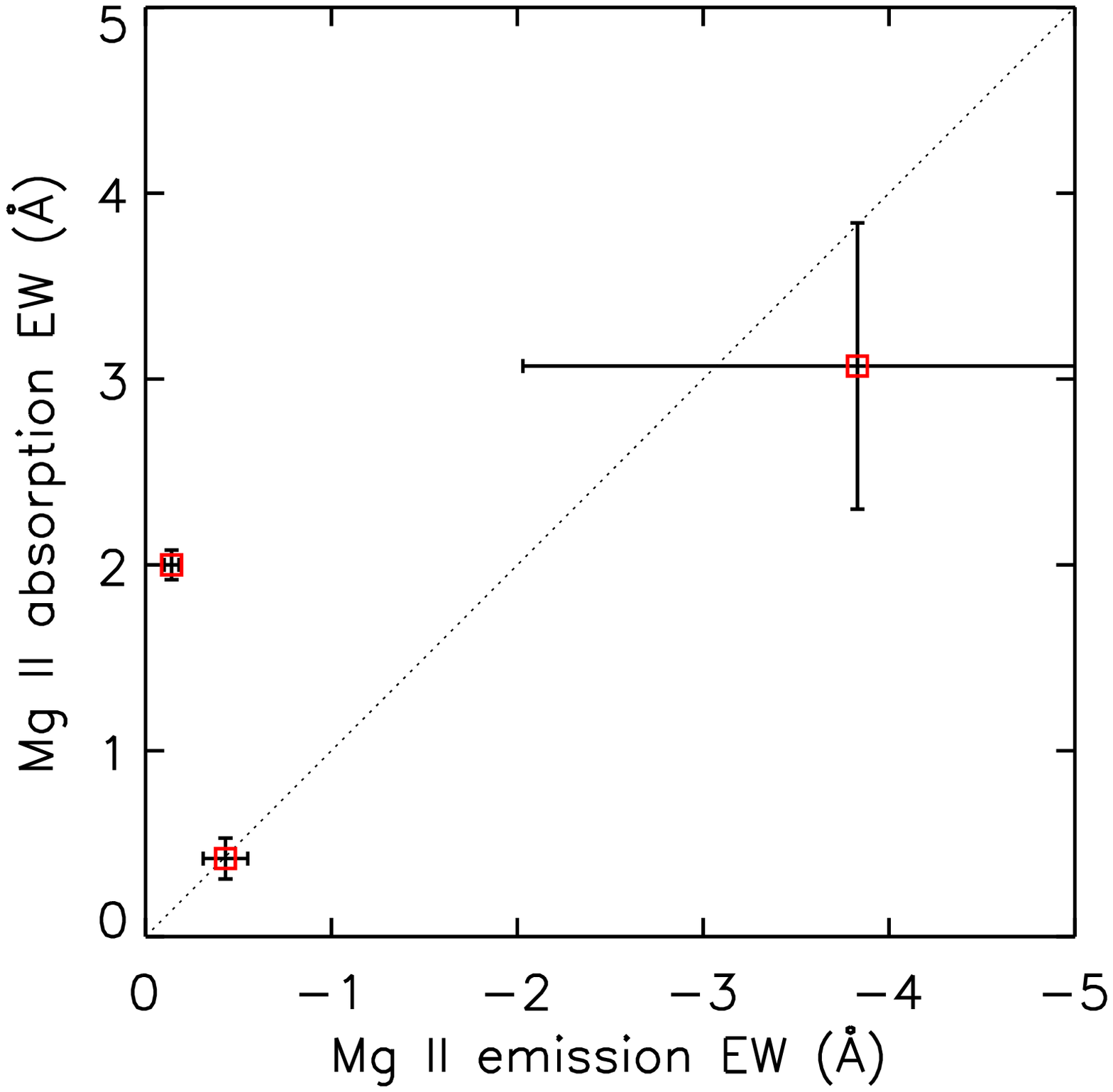}
\caption{\label{fig:mgemisabs}
\small 
Left: Comparison of \mgtwo \ emission EW in 
the 2802.7 \AA \ line with the 2795.5 \AA \ line.
Most objects have higher EW in the bluer \mgtwo \ line.
Right: Comparison of the EW of \mgtwo \ 2795.5 \AA \ 
in emission with the EW of \mgtwo \ 2795.5 \AA \ in
absorption for the three objects in which both are detected. Two of the
objects have equal EWs in emission and absorption, while one has much stronger
absorption than emission.}
\end{figure}

We note that the EW observed in the bluer \mgtwo \ 2795.5 \AA \ line in
our sample is 
generally larger than the EW observed in the redder 2802.7 \AA \ line.
This is quantified in the left panel of Figure~\ref{fig:mgemisabs},
which compares the EWs of the two emission lines. The median ratio of
the 2795.5 \AA \ EW relative to the 2802.7 \AA \ EW is 1.4.
  This difference between the EWs of the two \mgtwo \ emission lines 
is also seen in the coadded spectra of star-forming DEEP2 galaxies in 
\citet{Weiner09} and is not generally predicted by the models put forth in 
\citet{Prochaska11}.  The only model presented in which the 2795.5 \AA \ EW
is larger than the 2802.7 \AA \ line is their ``$\phi=0^\circ$ model'', in which
the wind is anisotropic and hemispherical, where half of the sphere has a wind
and half does not, and the wind is viewed from an angle such that the source is 
not covered ($\phi=180^\circ$ corresponds to the source being covered).  
However, in this model there is no \mgtwo \ absorption, which is 
seen in our data.  
This observed ratio of 2795.5 \AA \ EW to 2802.7 \AA \ EW 
in the data presented here and elsewhere 
is clearly a strong observational constraint on future, more sophisticated 
wind models.

We also note that several K+A galaxies in our sample (32003698, 43030800, 
J215518) show what appears to be emission blueward
of the absorption seen at \mgtwo \ 2795.5 \AA.  While such a signature 
could in theory originate as emission from the front side of an extended wind,
 for two of these sources -- 32003698 and J215518 -- this emission 
may be a residual from the stellar absorption fit, while in the remaining 
source (43030800) given the S/N of the data we conclude that this emission 
may not be significant.

\subsection{\mgtwo \ Emission Versus Absorption}

In the right panel of Figure~\ref{fig:mgemisabs} we compare the \mgtwo \ 
2795.5 \AA \ absorption and emission EWs for the three objects in which
both absorption and emission are detected at greater than 2$\sigma$ for this 
line.  We compare the bluer \mgtwo \ line as the absorption in the redder
line can be seriously affected by emission in the bluer line.  We find that 
two of the three objects have absorption and emission EWs that are consistent
with each other, while one object (40602938) has a much higher absorption
EW than emission EW.  As discussed in \citet{Prochaska11}, the ratio of 
the absorption and emission EW of \mgtwo \ can depend both on the opening angle
of the wind and the orientation of the wind relative to the observer, as well as
the amount of dust in the galaxy.  
 The two EWs will be roughly equal if the wind is isotropic or 
if an anisotropic, biconical wind with a large opening angle 
is observed 0$^\circ$ from the axis of rotational
symmetry (i.e. along the cone), in the absence of dust.
Obtaining absorption and emission EWs for
much larger samples of objects should therefore help to constrain the 
geometry of the outflowing wind.  Following the model of \citet{Prochaska11}, 
the one object observed here with a much higher absorption EW than emission EW 
should have $\tau_{dust}\gtrsim3$, where $\tau_{dust}$ is the integrated opacity 
of dust from the center of the system.
However, we note that this object (40602938) is fairly blue and is 
the bluest of the SDSS K+As, which, at least superficially, contradicts the 
idea of a high opacity dust screen.

\begin{figure}[tbp]
\epsscale{1.2}
\plotone{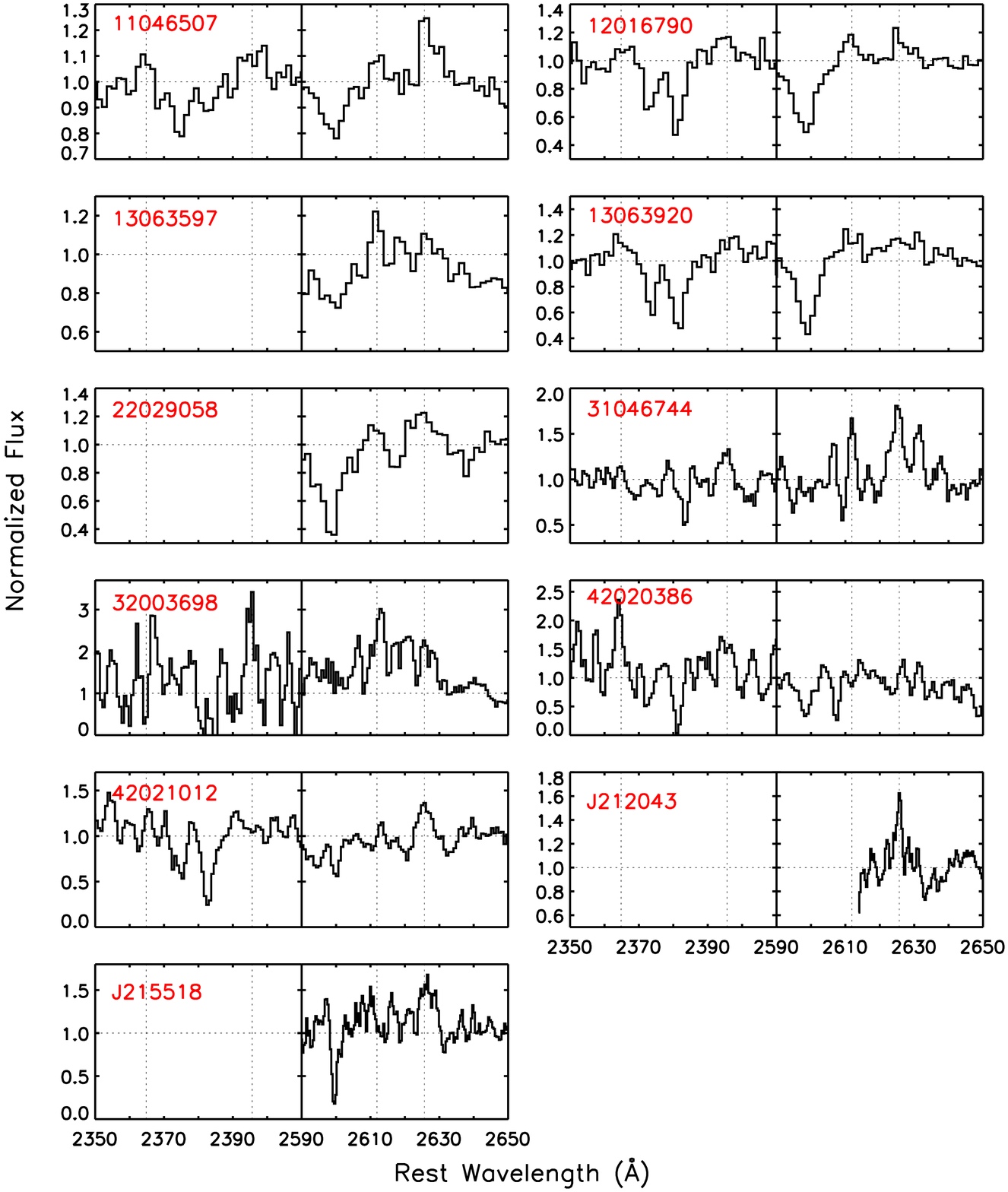}
\caption{\label{fig:feemis}
\small Region around \fetwo* lines for each object in our sample in 
which we detect significant \fetwo* emission. The wavelengths of the 
\fetwo* lines are shown with dotted vertical lines corresponding to 
2625.6679, 2611.8743, 2395.6263, and 2364.8292 \AA. Note that the y-axis 
range varies between objects.  
The K+A spectra have been smoothed by a boxcar of width three pixels
for this figure.}
\end{figure}

\subsection{\fetwo* Emission}

In addition to emission from \mgtwo, emission from \fetwo*  non-resonant 
fine structure 
lines has also been observed in galaxies at $z\sim0.5-1$ that 
have outflowing winds.  \cite{Rubin11} present observations of \fetwo* 
emission at 2364.8, 2395.6, 2611.9, and 2625.7 \AA \ in a bright starburst 
galaxy at $z=0.69$ that has a star formation rate $\sim$80 \msunyr.
The galaxy has a wind detected in \mgtwo \ and \fetwo \ absorption with
a central velocity of $\sim-$200 -- $-$300 \kms.  
They find that the \fetwo* emission is at or near the systemic velocity of
the galaxy, slightly redward of nebular lines such as [Ne III], H$\delta$, 
and H$\gamma$, and propose that the emission originates from photon 
scattering in the outflowing wind.  Unlike the \mgtwo \ resonant line,
blueshifted \fetwo* line emission is not absorbed and therefore the line is 
expected to be at the systemic velocity if the wind is symmetric and dust in
the galaxy is not obscuring the back side of the wind.
They present alternative possible origins,
including emission from gas in the galaxy disk as opposed to the gas in
the outflowing wind, but conclude that the emission is most likely from
the wind, due to differences in the emission line profile of \fetwo* 
compared to nebular lines.

Here we investigate \fetwo* emission in our sample for these same 
transitions. 
As with the \mgtwo \ emission lines, to obtain Gaussian fits to the 
\fetwo* emission lines we use data 
within $\pm$40 \AA \ of the observed wavelength corresponding to 
the systemic velocity.  This corresponds to roughly $\sim$4000 \kms \ for 
the $z\sim0.2$ objects and $\sim$2500 \kms \ for the $z\sim0.8$ objects.  
EWs are 
measured within the velocity range where the Gaussian fit is within 
1\% of the continuum level of the spectrum.
 Table~7 
 lists emission line measurements and EWs for
\fetwo* as well as [Ne III] and H$\zeta$.  Not all of our spectra have
the wavelength coverage to include all of these transitions; these are
marked with ellipses in the table.  Transitions where no significant
emission is observed are marked with an `X'.  We include [Ne III] and 
H$\zeta$ as nebular comparison lines.  By definition, the K+A galaxies will 
not have detected H$\zeta$ emission.  Figure~\ref{fig:feemis} shows the
region around the various \fetwo* lines for each object in which it is
detected.  Note that not all objects include spectral coverage of
 the bluest lines.

\begin{table*}[]
\tablewidth{0pt}
\begin{center}
\label{tab:feews}
\footnotesize
\caption{\small \fetwo*, \nethree, and H$\zeta$ Emission Line Measurements and Equivalent Widths}
\resizebox{\columnwidth}{!}{
  \begin{tabular*}{0.8\columnwidth}{crrr}
    \colrule
    \colrule
    \vspace{-3 mm} \cr
    Line \ \ &  Velocity  &  Velocity  &  EW    \ \ \ \  \cr
             &  centroid    &     width \   &            \cr
      (\AA)  & (km/s)  \     &    (km/s)    & (\AA) \ \ \ \   \cr
    \vspace{-3 mm} \cr
    \colrule
    \colrule
    \vspace{-3 mm} \cr
    \multicolumn{4}{c}{DEEP2 X-ray AGN Host Galaxies} \cr
    \vspace{-3 mm} \cr
    \colrule
    \vspace{-3 mm} \cr
    \multicolumn{4}{c}{11046507} \cr
    \vspace{-3 mm} \cr
    2364.8 & \ $-$81 $\pm$92  & 161 $\pm$90  & \ $-$0.39 $\pm$0.16  \cr
    2395.6 &    99  $\pm$130 & \ 373 $\pm$128 & $-$0.79 $\pm$0.23  \cr
    2611.9 &    40  $\pm$67  & 168 $\pm$67  & $-$0.39 $\pm$0.12  \cr
    2625.7 &   126  $\pm$39  & 255 $\pm$39  & $-$1.26 $\pm$0.16  \cr
    3868.8 &    52  $\pm$12  & 165 $\pm$12  & $-$6.00 $\pm$0.41  \cr
    3889.1 & $-$12  $\pm$13  & 166 $\pm$12  & $-$6.19 $\pm$0.44  \cr
    \vspace{-3 mm} \cr
    \colrule
    \vspace{-3 mm} \cr
    \multicolumn{4}{c}{12016790} \cr
    \vspace{-3 mm} \cr
    2364.8 &     X\tablenotemark{1}  & X  & $<$0.44  \cr
    2395.6 & $-$112 $\pm$86  & 299 $\pm$86  & $-$0.99 $\pm$0.25  \cr
    2611.9 & $-$54  $\pm$43  & 185 $\pm$43  & $-$0.75 $\pm$0.14  \cr
    2625.7 & $-$29 $\pm$38  & 169 $\pm$37  & $-$0.74 $\pm$0.13  \cr
    3868.8 & X  & X         & X                  \cr
    3889.1 &     11 $\pm$10  & 105 $\pm$10  & $-$1.93 $\pm$0.17  \cr
    \vspace{-3 mm} \cr
    \colrule
    \vspace{-3 mm} \cr
    \multicolumn{4}{c}{13063597} \cr
    \vspace{-3 mm} \cr
    2364.8 & ...\tablenotemark{2} &...  & ... \cr                                  
    2395.6 & ... &...  & ... \cr                                  
    2611.9 & $-$11 $\pm$38 & 117 $\pm$32  & $-$0.63 $\pm$0.14 \cr
    2625.7 & X & X & $<$0.32 \cr
    3868.8 & $-$144 $\pm$53 & 199 $\pm$53  & $-$0.30 $\pm$0.06 \cr
    3889.1 &     11 $\pm$10 & 226 $\pm$10  & $-$2.58 $\pm$0.10 \cr
    \vspace{-3 mm} \cr
    \colrule
    \vspace{-3 mm} \cr
    \multicolumn{4}{c}{13063920} \cr
    \vspace{-3 mm} \cr
    2364.8 & $-$60 $\pm$86  & 275 $\pm$86  & $-$0.92 $\pm$0.24  \cr
    2395.6 &   184 $\pm$96  & 318 $\pm$95  & $-$0.98 $\pm$0.27  \cr
    2611.9 & $-$10 $\pm$69  & 342 $\pm$69  & $-$1.46 $\pm$0.24  \cr
    2625.7 &    47  $\pm$116 & 654 $\pm$146 & $-$2.02 $\pm$0.25  \cr   
    3868.8 & $-$13  $\pm$38  & 128 $\pm$37  & $-$0.56 $\pm$0.13  \cr  
    3889.1 & $-$13  $\pm$13  & 129 $\pm$13  & $-$2.37 $\pm$0.21  \cr   
    \vspace{-3 mm} \cr
    \colrule
    \vspace{-3 mm} \cr
    \multicolumn{4}{c}{22029058} \cr
    \vspace{-3 mm} \cr
    2364.8 & ...             &...           & ...                \cr
    2395.6 & ...             &...           & ...                \cr
    2611.9 & $-$105 $\pm$77  & 165 $\pm$78  & $-$0.53 $\pm$0.18  \cr
    2625.7 & $-$63 $\pm$61  & 302 $\pm$61  & $-$1.60 $\pm$0.26  \cr
    3868.8 & $-$86  $\pm$43  & 304 $\pm$58  & $-$1.55 $\pm$0.15  \cr
    3889.1 & $-$32  $\pm$14  & 171 $\pm$15  & $-$2.53 $\pm$0.17  \cr
    \vspace{-3 mm} \cr
    \colrule 
    \colrule 
    \vspace{1.04in}
  \end{tabular*}
}	
\resizebox{\columnwidth}{!}{
  \begin{tabular*}{0.8\columnwidth}{crrr}
    \colrule
    \colrule
    \vspace{-3 mm} \cr
    Line \ \ &  Velocity  &  Velocity  &  EW    \ \ \ \  \cr
             &  centroid    &     width \   &            \cr
      (\AA)  & (km/s)  \     &    (km/s)    & (\AA) \ \ \ \   \cr
    \vspace{-3 mm} \cr
    \colrule
    \colrule
    \vspace{-3 mm} \cr
    \multicolumn{4}{c}{DEEP2 K+A Galaxies} \cr
    \vspace{-3 mm} \cr
    \colrule
    \vspace{-3 mm} \cr
    \multicolumn{4}{c}{31046744} \cr
    \vspace{-3 mm} \cr
    2364.8 & X              & X           & $<$1.42             \cr
    2395.6 & \ $-$49 $\pm$97 & \ 163 $\pm$97 & \ $-$1.13 $\pm$0.56  \cr
    2611.9  & X              & X           & $<$1.95            \cr
    2625.7 & $-$61 $\pm$47 & 182 $\pm$47 & $-$3.12 $\pm$0.76  \cr
    3868.8 &      0 $\pm$22 &  61 $\pm$22 & $-$0.93 $\pm$0.32  \cr
    3889.1 & X              & X           & X                  \cr
    \vspace{-3 mm} \cr
    \colrule
    \vspace{-3 mm} \cr
    \multicolumn{4}{c}{32003698} \cr
    \vspace{-3 mm} \cr
    2364.8 & X              & X           & $<$3.17            \cr
    2395.6 & $-$60 $\pm$32 & 120 $\pm$32 & $-$6.83 $\pm$1.60  \cr
    2611.9 & 89  $\pm$44 & 225 $\pm$43 & $-$10.05 $\pm$1.59 \cr
    2625.7 & X              & X           & $<$1.82            \cr
    3868.8 & X              & X           & X                  \cr
    3889.1 & X              & X           & X                  \cr
    \vspace{-3 mm} \cr
    \colrule
    \vspace{-3 mm} \cr
    \multicolumn{4}{c}{42020386} \cr
    \vspace{-3 mm} \cr
    2364.8 & $-$81 $\pm$23  & 79 $\pm$22  & $-$3.36 $\pm$0.87 \cr 
    2395.6 & $-$67 $\pm$88  & 208 $\pm$88 & $-$2.47 $\pm$1.11 \cr 
    2611.9 & X & X & $<$1.30 \cr                                         
    2625.7 & X & X & $<$1.24 \cr                                         
    3868.8 & X & X & X \cr                                        
    3889.1 & X              & X           & X                  \cr 
    \vspace{-3 mm} \cr
    \colrule
    \vspace{-3 mm} \cr
    \multicolumn{4}{c}{42021012} \cr
    \vspace{-3 mm} \cr
    2364.8 & X & X & $<$0.99 \cr
    2395.6 & X & X & $<$0.91 \cr
    2611.9 & X & X & $<$0.81 \cr
    2625.7 & 5 $\pm$60 & 168 $\pm$60 & $-$1.38 $\pm$0.44 \cr
    3868.8 &   112 $\pm$34 &  79 $\pm$34 & $-$0.92 $\pm$0.33 \cr
    3889.1 & X              & X           & X                  \cr
    \vspace{-3 mm} \cr
    \colrule
    \colrule
    \vspace{-3 mm} \cr
    \multicolumn{4}{c}{SDSS K+A Galaxies} \cr
    \vspace{-3 mm} \cr
    \colrule
    \vspace{-3 mm} \cr
    \multicolumn{4}{c}{J212043} \cr
    \vspace{-3 mm} \cr
    2364.8 & ... & ... & ... \cr
    2395.6 & ... & ... & ... \cr
    2611.9 & ... & ... & ... \cr
    2625.7 & $-$8 $\pm$ 20 & 83 $\pm$20 & $-$1.19 $\pm$0.26  \cr
    3868.8 & ... & ... & ... \cr
    3889.1 & ... & ... & ... \cr
    \vspace{-3 mm} \cr
    \colrule
    \vspace{-3 mm} \cr
    \multicolumn{4}{c}{J215518} \cr
    \vspace{-3 mm} \cr
    2364.8 & ... & ... & ... \cr
    2395.6 & ... & ... & ... \cr
    2611.9 & X   & X   & $<$0.97  \cr
    2625.7 & 85 $\pm$ 54 & 224 $\pm$51 & $-$2.82 $\pm$0.57  \cr
    3868.8 & ... & ... & ... \cr
    3889.1 & ... & ... & ... \cr
    \vspace{-3 mm} \cr
    \colrule
    \colrule
  \end{tabular*}
}
  \tablenotetext{1}{'X' indicates spectral coverage of this line but no significant emission.  2$\sigma$ upper limits are given for the EW in these cases.}
  \tablenotetext{2}{'...' indicates no spectral coverage of this line.}
\end{center}
\end{table*}

\begin{figure}[tbp]
\epsscale{1.05}
\plotone{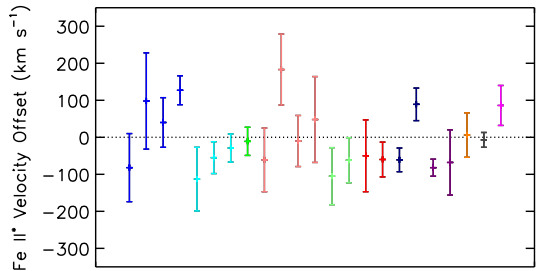}
\caption{\label{fig:feveloff}
\small Velocity offset from the systemic velocity of the galaxy 
for each detected \fetwo* emission line in our spectra.  Different 
lines observed in the same object are shown with the same color.
All but two of the observed \fetwo* emission lines are within 
2$\sigma$ of the systemic velocity of the galaxy.
}
\end{figure}

\begin{figure}[tbp]
\plotone{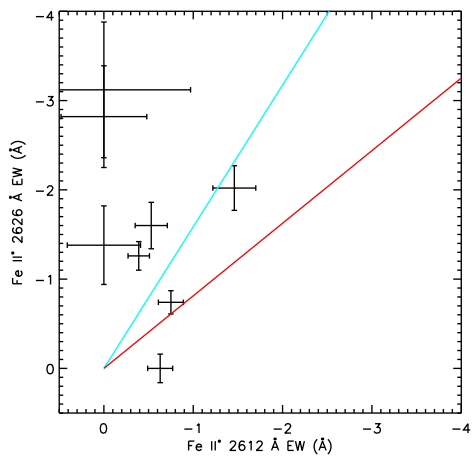}
\caption{\label{fig:fecompew}
\small Comparison of the \fetwo* 2611.9 \AA \ EW to
the \fetwo* 2625.7 \AA \ EW.  Data points from our
spectra are shown with error bars.  The red and cyan lines show
the extreme ratios predicted by the models of \citet{Prochaska11}.
}
\end{figure}

We detect significant \fetwo* emission in five of the ten X-ray AGN host
galaxies and six of the thirteen K+A galaxies.  However, 
only two of the six SDSS K+A galaxies have blue enough spectral 
coverage to observe the \fetwo* 2625.7 \AA \ line, and \fetwo* is detected
in both of them, so it is entirely possible that 
the fraction of K+A galaxies with \fetwo* emission is higher.

In the objects where H$\zeta$ is detected, it is always consistent with
the systemic velocity.  [Ne III] is at the systemic velocity in three out
of six objects, blueshifted in one object, and redshifted in two objects.
Similarly, each of the observed \fetwo* lines 
is consistent at 2$\sigma$ with the systemic velocity in all
but two lines.  The median velocity centroid observed among
all of the \fetwo* lines in galaxies in our sample is $-29$ \kms, 
while the median error of 61 \kms. 
Figure~\ref{fig:feveloff} shows the velocity offset from systemic of
the velocity centroids for all measured \fetwo* lines 
in Table~7. 
The velocity centroids measured from different \fetwo* lines in the same
object are generally consistent with each other.
Our finding that the \fetwo* emission is centered on the systemic velocity 
of the galaxy is consistent with the results of \citet{Rubin11} and the models 
of \citet{Prochaska11}.

Additionally, we detect \fetwo* emission in many of
our K+A galaxies, which have little to no nebular emission and are not 
currently forming stars.  It is therefore unlikely that the \fetwo* 
emission in these galaxies arises from gas in the disk of the galaxy itself.

We compare the ratio of EW in the \fetwo* 2611.9 \AA \ and 2625.7 \AA \ 
lines in our objects in Figure~\ref{fig:fecompew}.  Our data points with errors
are shown in black, along with two models from \citet{Prochaska11} that
illustrate the minimum and maximum ratios found for these lines in their 
model.  The maximum ratio predicted in their models is 1.23, which is for an 
anisotropic wind without dust (shown as a red line).  The minimum ratio predicted
is 0.63, which is for a model with an ISM component included (shown as a cyan
line).  Most of our objects have a ratio less than 0.5, where we have included
objects that do not have emission detected at \fetwo* 2611.9 \AA.  We have not 
included object 32003698 in this figure, as the emission detected at 2611.9 \AA \ 
has a broad emission component which leads to a stronger observed EW but 
is not consistent with the simple wind models from \citet{Prochaska11}.
We note that many of our objects do not fall within the allowed bounds of 
the models
in \cite{Prochaska11}.  Data are needed (at higher S/N) on many more objects, 
but these few examples show that measurements of this kind have the 
potential to be strongly constraining for theoretical wind models.

\subsection{\fetwo* Emission Versus \mgtwo \ Emission}

Three objects in our sample have detected emission in both \mgtwo \ and \fetwo*.
We calculate the ratio of the emission line EW in \mgtwo \ 2795.5 \AA \ to 
either the \fetwo* 2611.9 or 2625.7 \AA \ lines in two of these objects, as these
line ratios should contain information about the physical conditions in 
the outflowing gas.  In object 11046507 the ratio of 2795.5 \AA \ EW to
 2611.9 \AA \ EW is 1.10 $\pm$0.17, while the ratio of 2795.5 \AA \ EW to
 2625.7 \AA \ EW is 0.34 $\pm$0.20. 
In object 
71633703 the ratio of 2795.5 \AA \ EW to 2625.7 \AA \ EW is 0.34 $\pm$0.70.
In most of the models (fiducial model and variants) 
presented in \citet{Prochaska11} the ratio of 2795.5 \AA \ to 2611.9 \AA \
varies from $\sim2.9-14.6$, all higher than what we find here. 
 The `ISM' wind model predicts a ratio of 1.5, closer to what is found here. 
The models predict a ratio of  2795.5 \AA \ to 2625.7 \AA \
of $\sim0.9-18$, significantly higher than what is found for our two objects.  
As in the case of the ratio of \fetwo* EWs discussed above, 
these line ratios have the potential to constrain future theoretical models.

\section{Discussion and Conclusions} \label{sec:discussion}

In this section we discuss our results, including the prevelance and 
properties of 
outflowing galactic winds in AGN host galaxies and post-starburst galaxies
at intermediate redshift, the driving mechanism of the winds, and the 
implications for star formation quenching.

\subsection{Prevalence of Winds}

Table~8 
 provides a summary of the objects in our sample, their
location in the restframe optical color-magnitude diagram,  and an
indication of which objects have outflowing winds detected in blueshifted
\mgtwo \ and \fetwo \ absorption and detections of 
\mgtwo \ emission and/or \fetwo*
emission.  We find blueshifted absorption indicative of an outflow 
in six of our ten X-ray AGN host galaxies and four of our thirteen K+A galaxies,
43\% of our full sample.  For all of these
galaxies, blueshifted absorption is detected in \mgtwo; some galaxies
additionally show blueshifted absorption in either \fetwo \ or \mgone.
We find evidence for winds in galaxies across the color-magnitude 
diagram, including one X-ray AGN host galaxy on the red sequence.
These results are shown graphically in Figure~1.
We note that 
these are galactic-scale winds, as the observed covering fraction is high and 
for the galaxies with AGN, the AGN does not dominate the galaxy continuum 
light; therefore the background light source is extended.

We additionally find that most of 
the galaxies in our sample show line emission in either \mgtwo \ or \fetwo*. 
 While this emission may be from the wind itself, it is difficult 
to extract kinematic information about possible outflows as the emission is 
isotropic and is observed 
summed over the entire galaxy.  Therefore symmetry washes out kinematic 
information along a given sightline,
unlike in the case of absorption.
 Five of the ten X-ray AGN host galaxies and 
six of the thirteen K+A galaxies have \mgtwo \ emission.  
 If the emission is from the wind itself, then taken together,
nine of the ten X-ray AGN host galaxies and nine of the thirteen K+A galaxies,
or 78\% of the full sample, have winds detected either in \mgtwo \ absorption
or emission. All three of the X-ray AGN host galaxies that are on the red 
sequence show \mgtwo \ emission, though this emission is detected only 
after removing the absorption due to stellar continuum in the observed spectra 
 and therefore should be treated with caution.
We do not find a strong correlation between blueshifted
absorption and \mgtwo \ emission; while three galaxies in our sample have both,
the rest do not.  The presence of emission can fill in absorption features, however,
and we emphasize that we are reporting only absorption and emission that are
detected with high significance.  Object 32003698, for example, has narrow 
blueshifted absorption in both \mgtwo \ lines, but the EW of the absorption 
is not significant in the presence of the strong \mgtwo \ emission in 
this galaxy.

\fetwo* emission is detected in 50\% of the X-ray AGN host galaxies and 46\% 
of the K+A galaxies.  
If the \fetwo* emission also originates in the wind, then a total of 100\% of 
our X-ray AGN host sample and 77\% of our K+A sample has 
either blueshifted absorption or line emission from a wind.  As seen in 
Table~8,  
roughly half of the galaxies with blueshifted absorption
also have \fetwo* emission, but some objects with \fetwo* emission do not
have blueshifted absorption.  There is also very little correlation between 
detection of \mgtwo \ emission and \fetwo* emission.  However, given that 
\fetwo* emission is detected in K+A galaxies that do not have on-going star 
formation or nebular lines, it is likely that the \fetwo* emission 
is from the wind itself.  
The fact that we do not often detect both \mgtwo \ and \fetwo* emission in 
the same galaxy may be due to  dust in the galaxy.  

We find that outflows are common both in galaxies that host low-luminosity
AGN and in post-starburst galaxies that are no longer forming stars.
The prevalence of winds as detected in blueshifted absorption can be used 
to constrain the opening angle of the outflows.  As we detect blueshifted
absorption in 60\% of the X-ray AGN host galaxies, this implies that at least
60\% of these galaxies have a wind (if it is isotropic) or 100\% of these 
galaxies could have a wind with a geometric covering fraction of 60\%, 
corresponding to an opening angle of 66 degrees  
for a biconical outflow.  In the post-starburst
sample we detect blueshifted absorption in 29\% of the DEEP2 K+A galaxies 
and 33\% of the SDSS K+A galaxies, implying a geometric covering fraction of 
at least $\sim$30\%, which corresponds to an opening angle of 46 degrees.
Larger sample sizes with higher S/N would, of course, be preferable for 
determining these fractions and constraining wind models, as the prevalence 
found here is a lower limit due to the low S/N of our data.

\begin{table}
\begin{center}
\label{tab:summary}
\footnotesize
\caption{\small Summary of Detections in Each Galaxy}
\begin{tabular}{llccc}
\colrule
\colrule
\vspace{-3 mm} \cr
\vspace{-3 mm} \cr
Object   &  \ Color \ & Blueshifted & \mgtwo \ & \fetwo*  \cr
         &          & absorption  & emission & emission \cr
\vspace{-3 mm} \cr
\vspace{-3 mm} \cr
\colrule 
\colrule 
\vspace{-3 mm} \cr
\vspace{-3 mm} \cr
\multicolumn{5}{c}{DEEP2 X-ray AGN Host Galaxies} \cr
\vspace{-3 mm} \cr
\colrule 
11046507 & \ \ blue  & O\tablenotemark{1} & O & O \cr
12016790 & \ \ blue  & O & X & O \cr
13004312 & \ \ red   & O & O & X \cr
13025528 & \ \ green & X & O & X \cr
13041622 & \ \ blue  & O & X & X \cr
13043681 & \ \ red   & X & O & X \cr
13051909 & \ \ red   & X & O & X \cr
13063597 & \ \ blue  & X & X & O \cr
13063920 & \ \ blue  & O & X & O \cr
22029058 & \ \ blue  & O & X & O \cr
\vspace{-3 mm} \cr
\colrule
\colrule
\vspace{-3 mm} \cr
\vspace{-3 mm} \cr
\multicolumn{5}{c}{DEEP2 K+A Galaxies} \cr
\vspace{-3 mm} \cr
\colrule 
31046744 & \ \ green & O & X & O \cr
32003698 & \ \ green & X & O & O \cr
32008909 & \ \ green & X & X & X \cr
41057700 & \ \ blue  & X & X & X \cr
42020386 & \ \ blue  & O & X & O \cr
42021012 & \ \ blue  & X & X & O \cr
43030800 & \ \ green & X & O & X \cr
\vspace{-3 mm} \cr
\colrule
\colrule
\vspace{-3 mm} \cr
\vspace{-3 mm} \cr
\multicolumn{5}{c}{SDSS K+A Galaxies} \cr
\vspace{-3 mm} \cr
\colrule 
J022743  & \ \ blue  & O & O & X \cr
J210025  & \ \ green & X & O & X \cr
J212043  & \ \ green & X & O & O \cr
J215518  & \ \ green & X & O & O \cr
J224603  & \ \ green & X & X & X \cr
J225656  & \ \ green & O & X & X \cr
\vspace{-3 mm} \cr
\colrule
\colrule
\end{tabular}
  \tablenotetext{1}{An 'O' indicates a detection while an 'X' indicates no detection.}
\end{center}
\end{table}

\vspace{1cm}

\subsection{Wind Properties}

From measurements of blueshifted absorption we can constrain
the minimum covering fraction and the velocities of the outflowing winds.
The maximum depth of the absorption reflects the covering fraction for 
optically thick lines, in that
it is an indication of the fraction of photons that have been absorbed 
and therefore the fraction of the galaxy that is covered by the wind.
However, as our spectral resolution is not high, an absorption feature that 
is `black' (fully absorbed) may appear to have a lower covering fraction due
to the poor velocity resolution.  Therefore we can measure only a lower limit
on the covering fraction.
As indicated in Table~5, 
the minimum 
covering fraction (equal to $A_{flow}$) of \mgtwo \ ranges from $\sim$0.1--0.5 in the 
X-ray AGN host galaxy sample and $\sim$0.2--1.0 in the K+A sample.  
Given the lower S/N of the K+A sample, we may not have been able to detect 
weaker absorption features in our data.  As discussed in Section 4.1, the 
covering fraction of \fetwo \ is similar to that of \mgtwo.
The minimum line depth of \mgone \ is higher, which would indicate a lower
covering fraction ($\sim$0.1--0.2) if the line is saturated.  However, for \mgone \
we do not know if this is the case; therefore the difference in line depth may 
reflect either a lower covering fraction or a lower optical depth.  
As the ionization potential of
\mgone \ is 7.6 eV, much of the Mg in the wind is likely ionized, which 
would lead to a lower covering fraction of \mgone \ compared to \mgtwo.
In that case, the \mgone \ that survives would likely be 
in dense clumps with a lower covering fraction.

The velocity centroids of the outflows detected in blueshifted
absorption are typically $\sim-$200 -- $-$300 \kms.  The velocity widths are
generally unresolved, $\sim$100--300 \kms.  One AGN host galaxy has a
velocity centroid of $\sim-$600 \kms \ and a width of $\sim$270 \kms,
and another AGN host galaxy has a centroid of $\sim-$600 to $-$1200
\kms \ with a width of $\sim$500 \kms.  The absorption EWs vary from
$\sim$0.2 -- 3 \AA.  The maximum velocity to which we detect
absorption is typically $\sim$-400 to -800 \kms.  We do not find a
correlation between the outflow properties (velocity centroid,
velocity width, EW, covering fraction) and the optical color of the
galaxy or the fraction of young ($<$2 Gyr) stars.  We detect outflows
only in galaxies in which the younger stars ($<$2 Gyr) have ages less
than 700 Myr, but given the upper limits on the non-detections this is
not a particularly strong constraint.  
Our sample does not cover a wide enough range in X-ray luminosity to test
for a dependence between outflow properties and X-ray luminosity.

\subsection{Comparison With Other Samples}

The sample we present here is quite complementary to other samples with
measured wind properties from \mgtwo \ absorption 
at $z\sim0.5-1$.  \citet{Weiner09} 
and \citet{Rubin10a} present results for typical star-forming galaxies 
at $z\sim1$, selected from flux-limited galaxy samples.  Both studies find that 
star-forming galaxies at these redshifts commonly have outflows with velocity 
centroids of $\sim200-300$ \kms, with absorption seen out to $\sim800$ \kms.
Our results here for winds in X-ray AGN host galaxies and K+A galaxies are
very similar, even though our galaxy samples are selected differently. 

Our detection of \mgtwo \ winds in X-ray AGN host galaxies that are on the 
red sequence is somewhat analogous to results from \citet{Sato09}, who study 
Na~D absorption in lower redshift galaxies in AEGIS, at $z\sim0.1-0.5$.  They 
find that many red sequence galaxies have winds in Na D (see their Fig. 10) 
with velocities $\lesssim200$ \kms \ (though stellar absorption was not removed).
They note that many of the red sequence galaxies with winds show signs of
recent star formation, however, which may indicate a correlation between
the wind and the quenching of star formation in these systems.
More recently, \citet{Bowen11} find that \mgtwo \ absorption along QSO lines of 
sight around massive
luminous red galaxies (LRGs) at similar redshifts is very rare, $\lesssim$10\%.
LRGs are expected to be at a later evolutionary stage than the red sequence
galaxies studied here, however, and it is entirely possible that their assembly
history and/or star formation quenching mechanism(s) are quite different than
for typical red galaxies, or that the winds do not reach the combination of 
impact parameter and density needed for detection against background QSOs.

Our sample does not show the extreme velocity 
outflows seen by \citet{Tremonti07} in their SDSS post-starburst sample at 
$z\sim0.6$, except for one X-ray AGN host galaxy that appears to have an 
outflow with a velocity centroid $\gtrsim$600 \kms.  Compared to the 
\citet{Tremonti07} K+A galaxies, the post-starburst galaxies that we study 
are more common and less luminous and 
are therefore likely to reflect a more frequent path to the red sequence.  
Our SDSS K+A galaxies are also redder, which may indicate that 
they are at a later evolutionary stage.
Additionally, 
the absorption profile that we observe in our sample is similar to
that seen in the coadded spectra 
of star-forming galaxies in \cite{Weiner09} and 
\citet{Rubin10a}, with absorption extending from systemic out to $\gtrsim-600$ 
\kms, often with a sawtooth-like profile.  We do not observe profiles akin to 
what is seen by \citet{Tremonti07}, in which the absorption profile is narrower
and offset from systemic, with very little to no absorption at systemic.
It appears that either the driving mechanism or duration of the gas ejection 
is different between our sample and that of \citet{Tremonti07}.

Our results regarding winds in AGN host galaxies can be compared with 
work by \citet{Rupke05} at lower redshift and \citet{Hainline11} at higher
redshift.  
At $z\sim0.2$, \citet{Rupke05} find that starburst, IR-bright galaxies
with LINER emission have marginally higher velocity winds ($\sim$100 \kms, 
measured with 95\% confidence) than non-LINER galaxies.  They find that 
the detection rate of winds in LINER galaxies is the same as in non-LINER
galaxies and note that the LINER emission may come from shocks in the 
gas of the galaxy or wind, as opposed to weak AGN emission.
At $z\sim2-3$, \citet{Hainline11} find that in UV-selected galaxies with
optically-identified, narrow-line AGN, the \ion{Si}{4} absorption line
in a coadded spectrum of 33 AGN host galaxies shows absorption at 
higher velocities than in a coadded spectrum of non-AGN host galaxies.
However, the interpretation is complicated by the presence of 
\ion{Si}{4} emission, which is stronger in the AGN host galaxies. Emission-line
filling may therefore account for the bulk of the difference observed.
We conclude, therefore, that our finding that X-ray selected AGN host
galaxies at $z\sim0.2-0.5$ do not have significantly faster winds
than star-forming galaxies at similar redshifts is not strongly at odds 
with results from lower and higher redshift.

\subsection{Implications for Wind Models and Star Formation Quenching}

What is the physical driver of the winds detected here?  For sources
in our AGN host galaxy sample, given that their wind properties are
similar to those in our K+A sample and in the star-forming samples of
\citet{Weiner09} and \citet{Rubin10a}, we conclude that the winds 
are not primarily AGN-driven.  
The fact that whether or not a galaxy has a low luminosity
AGN does not change its wind properties suggests that
the winds are driven by supernovae.  
The relatively low outflow velocities observed are consistent with this
scenario, though it is possible that low luminosity AGN drive winds with 
comparable velocities as those observed in star-forming galaxies.
However, for the X-ray
AGN host galaxies that have ongoing star formation, SNe-driven winds
seem quite plausible.  We also note that as these galaxies are continuing
to form stars, the AGN in them can not have driven such strong winds so 
as to have quenched star formation. 
 Four of the six blue X-ray AGN host galaxies
studied here also show \mgtwo \ and/or \fetwo \ 
absorption at the systemic velocity of the galaxy, indicating that the 
winds in these galaxies have not fully expelled the cool gas present.
 The LIRG in our sample lies on the
lower-redshift relation of \citet{Martin05} for wind velocity versus
SFR (the LIRG has a SFR of 45 \msun $yr^{-1}$, estimated from the
24\micron \ flux), implying consistency with a SNe-driven wind.  For the
red X-ray AGN host galaxies in our sample 
with winds, the winds may be relic winds
that were driven by SNe during the quenching of star formation in
these galaxies.

If the winds in the X-ray AGN host galaxies are nonetheless driven by
the AGN itself and not SNe, we note that the wind velocities are not
particularly high, indicating that low luminosity AGN are unlikely to
drive strong enough winds to quench star formation.  Whether the winds
are due to either star formation (at least in the blue galaxies) or
low luminosity AGN activity, the outflows observed do not appear to be
fast enough to clear the ISM of the galaxy and lead to migration to
the red sequence.
\citet{Weiner09} estimate the escape velocities of their $z\sim1.4$ 
star-forming DEEP2 galaxies using the \ion{O}{2} 3727 \AA \ 
linewidth and find that the
median escape velocity is $\sim$400-450 \kms \ for galaxies with stellar
masses log(M)$\sim$9.5-10.5 \msun \ and SFR$\sim$10-40 \msunyr.  The velocity centroids
of the winds we detect in our sample are generally lower than this estimated
escape velocity,
though the maximum velocity observed for many of our sources (especially
the AGN hosts) does exceed this.  This implies, similarly to the 
\citet{Weiner09} sample, that while some of the gas may be travelling at 
high enough speeds to escape the halo of the galaxy, the bulk of the gas
remains bound.

Winds in the post-starburst galaxies studied here 
may be driven by SNe resulting from the most recent starburst. 
It is worth noting that several of our K+A galaxies have \mgtwo \ and/or 
\fetwo \ absorption at the systemic velocity.  These galaxies, which are 
either in the blue cloud or green valley, apparently have cool gas 
that was not expelled either during the recent starburst or by the current 
outflowing wind.
Why do we not observe extreme outflows similar to those observed 
by \citet{Tremonti07}?
Our K+A galaxies may be at a later evolutionary stage,
which could suggest that the extreme wind phase, if present, does not
last long.  However, if our K+A galaxies are at a later evolutionary
stage and further on their way to the red sequence, and if the extreme
outflows only exist in the youngest objects, our objects should have already
gone through the extreme wind phase.  
As our objects are still in the process of ejecting gas, 
this scenario would require that they had a wind with a central 
velocity $>1000$ \kms \ which nonetheless did not clear the ISM from the galaxy.
A more plausible explanation is that these sources have not passed through the 
extreme wind phase 
observed by \citet{Tremonti07}.  Either these galaxies have not had a 
violent AGN outburst, or if the winds in the \citet{Tremonti07} sources 
are SNe-driven and not AGN-driven, then perhaps the recent starburst 
in the K+A galaxies in our sample did not form as many stars as those in 
the \citet{Tremonti07} sample and as such there are fewer SNe
driving winds in our galaxies.
 Preliminary
evidence from the \citet{Tremonti07} sample indicates that the starburst
decay time -- or timescale
for star formation quenching -- is extremely short in those galaxies 
($\sim$25 Myr, Tremonti, private communication), which implies that 
they are not `typical' post-starburst galaxies.

It is notable that the one object in our sample that may have an extreme
wind with a velocity $>$1000 \kms \ is an X-ray AGN host galaxy on the red
sequence.  
We note that if the wind was launched 640 Myr ago (corresponding to the 
age of the young stellar population in this galaxy) and has been traveling 
at a constant speed of $\sim$1,000 \kms \ since, it would now be at a 
radius of $\sim$650 kpc and 
would likely not be observable in absorption.  This argues that the wind 
was either launched more recently or (possibly more likely) has not been 
traveling at this high constant velocity since its inception.
If this wind is a relic AGN wind, then it may be that the AGN
was powerful enough to drive a wind that shut off star formation in this
red galaxy.
If this is the case and is common for AGN in red host galaxies, 
then the AGN observed in blue host 
galaxies may be intrinsically different (e.g., have lower mass black holes,  
lower accretion rates, and/or be at an earlier evolutionary stage) 
and therefore drive lower velocity winds.  We note, however, that 
only one of the three red AGN host galaxies studied here has a wind 
detected in absorption for which we can derive kinematic information.

In summary, we find in our sample targeting low-luminosity AGN host galaxies
and post-starburst galaxies at intermediate redshift that outflowing galactic
winds are prevalent in these galaxies, as detected either by \mgtwo \ and 
\fetwo \ absorption or \mgtwo \ and \fetwo* emission.  The velocities of 
the winds are roughly consistent with those observed in star-forming galaxies 
at similar redshifts.  Further, the presence of a low-luminosity AGN does 
not appear to commonly drive faster winds, such that these winds are likely
SNe-driven. 
In our post-starburst sample we do not detect the extreme winds
observed by \citet{Tremonti07} and conclude that most post-starburst 
galaxies likely do not host such high velocity winds. We therefore find that
while galactic winds are common in both low-luminosity AGN host 
galaxies and post-starburst galaxies, they 
do not appear to play a major role in quenching star formation.

\acknowledgements

The authors would like to thank 
Aleks Diamond-Stanic, 
Katherine Kornei, 
Crystal Martin, 
John Moustakas, 
Jason X. Prochaska, 
Kate Rubin, 
Alice Shapley, and Christy Tremonti for useful discussions.  We also thank
Christy Tremonti for sharing her stellar continuum fitting code.
The data presented herein were obtained at the W.M. Keck Observatory, which is 
operated as a scientific partnership among the California Institute of 
Technology, the University of California and NASA. 
The Observatory was made possible by the generous financial 
support of the W.M. Keck Foundation.
The authors wish to recognize and acknowledge the very significant cultural role 
and reverence that the summit of Mauna Kea has always had within the indigenous 
Hawaiian community.  
The Keck access used for these observations is from NOAO and the University
of California.
This study uses data from both the AEGIS survey, which is supported in part
by the NSF, NASA, and the STFC, and the SDSS survey, which is supported in part
by the Alfred P. Sloan Foundation, the participating institutions, the
NSF, the U.S. Department of Energy, and NASA.


\end{document}